\documentclass[letterpaper]{article}
\usepackage{authblk}
\usepackage{url}
\usepackage[margin=1in,footskip=0.25in]{geometry}
\usepackage[bbgreekl]{mathbbol}
\usepackage{graphicx}
\usepackage{amsfonts, amsmath, amssymb}
\usepackage{booktabs,siunitx}
\usepackage[tableposition=above]{caption}
\usepackage{pifont}
\usepackage{bm}
\usepackage{cancel}
\usepackage{caption}
\usepackage{color}
\usepackage{enumerate}
\usepackage{float}
\usepackage{hyperref}
\usepackage[english]{babel}
\usepackage{mathtools}
\usepackage{longtable}
\usepackage{setspace}
\usepackage{subfigure}
\usepackage{lineno}

\renewcommand \d [2]{\frac{{\rm d} #1}{{\rm d} #2}}
\newcommand \D [2]{\frac{\partial #1}{\partial #2}}

\renewcommand{\vec}[1]{\bm{#1}}
\newcommand{\V}[1]{\bm{#1}}

\newcommand{\Lim}[1]{\raisebox{0.5ex}{\scalebox{0.8}{$\displaystyle \lim_{#1}\;$}}}

\def \div{\nabla \cdot \mbox{}}
\def \grad{\nabla}

\def \x{\vec{x}}

\def \n{\vec{n}}

\def \u{\vec{u}}

\def \I{\vec{I}}
\def \F{\vec{F}}

\def \U{\vec{U}}
\def \L{\vec{L}}

\def \cM{\vec{\mathcal{M}}}

\def \Sb{S_{\rm b}}

\def \Vb{V_{\rm b}}

\def \A{\vec{A}}

\def \B{\vec{B}}
\def \C{\vec{C}}

\def \cF{\vec{\mathcal{F}}}

\def \F{\vec{F}}

\def \g{\vec{g}}

\def \I{\vec{I}}
\def \Ib{\I_{\text{b}}}

\def \M{\vec{M}}
\def \Mb{\text{M}_{\text{b}}}

\def \Nx{N_x}
\def \Ny{N_y}

\def \Omegal{\Omega_{\text{l}}}
\def \Omegag{\Omega_{\text{g}}}

\def \U{\vec{U}}

\def \Ur{\U_{\text{r}}}
\def \W{\vec{W}}

\def \Wr{\W_{\text{r}}}

\def \X{\vec{X}}
\def \Xcom{\X_{\text{COM}}}

\def \cO{\mathcal{O}}

\def \f{\vec{f}}

\def \fc{\f_{\text{c}}}

\def \half{\frac{1}{2}}
\def \3half{\frac{3}{2}}
\def \5half{\frac{5}{2}}

\def \n{\vec{n}}

\def \ncells{n_{\text{cells}}}
\def \ncycles{n_{\text{cycles}}}

\def \sgn{\textrm{sgn}}
\def \u{\vec{u}}

\def \ub{\u_{\text{b}}}

\def \x{\vec{x}}

\def \div{\nabla \cdot \mbox{}}
\def \grad{\nabla}

\def \dt{\Delta t}
\def \dx{\Delta x}
\def \dy{\Delta y}

\def \dt{\Delta t}
\def \dx{\Delta x}

\def \phii{\Phi_{\text{I}}}
\def \phid{\Phi_{\text{D}}}
\def \phir{\Phi_{\text{R}}}
\def \xi{\x_{\text{i}}}

\newcounter{subsubsubsection}[subsubsection]
\def\subsubsubsectionmark#1{}

\def\subsubsubsection{\@startsection
      {subsubsubsection}{4}{\z@} {-3.25ex plus -1
      ex minus -.2ex}{1.5ex plus .2ex}{\normalsize\bf}}
\def\l@subsubsubsection{\@dottedtocline{4}{4.8em}
      {4.2em}}

\newcommand{\upperRomannumeral}[1]{\uppercase\expandafter{\romannumeral#1}}

\newcommand{\REVIEW}[1]{#1}

\title{Comparison of wave-structure interaction dynamics of a submerged cylindrical point absorber with three degrees of freedom using potential flow and computational fluid dynamics models}

\author[1]{Panagiotis Dafnakis} %\corref{cor1}}
\author[2]{Amneet Pal Singh Bhalla$^*$}
\author[1]{Sergej Antonello Sirigu }
\author[1]{Mauro Bonfanti} %\corref{cor2}}
\author[1]{Giovanni Bracco} %\corref{cor2}}
\author[1]{Giuliana Mattiazzo} %\corref{cor2}}
\affil[1]{Department of Mechanical and Aerospace Engineering, Politecnico di Torino, Turin 10129, Italy}
\affil[2]{Department of Mechanical Engineering, San Diego State University, San Diego, California 92182, USA}
\affil[ ]{$^*$ Corresponding author: \texttt{asbhalla@sdsu.edu}}

\begin{document}

\date{}
\maketitle

\begin{abstract} 

In this paper we compare the  heave, surge, and pitch dynamics of a submerged cylindrical point absorber, simulated using potential flow and fully-resolved computational fluid dynamics (CFD) models. The potential flow model is based on the time-domain Cummins equation, whereas the CFD model uses the fictitious domain Brinkman penalization (FD/BP) technique. The submerged cylinder is tethered to the seabed using a power take-off (PTO) unit which restrains the heave, surge, and pitch motions of the converter, and absorbs energy from all three modes. It is demonstrated that the potential theory over-predicts the amplitudes of heave and surge motions, whereas it results in an insignificant pitch for a fully-submerged axisymmetric converter. It also under-estimates the slow drift of the buoy, which the CFD model is able to capture reliably.  Further, we use fully-resolved CFD simulations to study the performance of a three degrees of freedom (DOF) cylindrical buoy under varying PTO coefficients, mass density of the buoy, and incoming wave heights. It is demonstrated that the PTO coefficients predicted by the linear potential theory are sub-optimal for waves of moderate and high steepness. The wave absorption efficiency improves significantly when higher than the predicted value of the PTO damping is selected. Simulations with different mass densities of the buoy show that converters with low mass densities have an increased tension in their PTO and mooring lines. Moreover, the mass density also influences the 
range of resonance periods of the device. Finally, simulations with different wave heights show that at higher heights, the wave absorption 
efficiency of the converter decreases and a large portion of available wave power remains unabsorbed.

\end{abstract}

\noindent \textbf{Keywords:} incompressible Navier-Stokes equations, Brinkman penalization method, numerical wave tank, potential flow theory, Cummins equation

%%%%%%%%%%%%%%%%%%%%%%%%%%%%%

\section{Introduction}
Power production using wave energy gained momentum in the 1970s during the oil crisis. This field is regaining a 
renewed interest in the marine hydrokinetic research community that is aiming to reduce the current carbon footprint of power production.  
In spite of the abundantly available wave power in the oceans and seas worldwide~\cite{Gunn12}, and research 
efforts dating back since the seventies~\cite{Evans1979}, no 
commercial-scale wave power production operations exist today.  Consequently, various wave energy conversion (WEC) 
concepts have been proposed and implemented, yet no single device architecture has been recognized as the ultimate solution. 

Point absorber (PA) is a type of WEC system, for which the device characteristic dimensions are relatively small compared to the wave 
length of the site~\cite{Falnes2012,Soares2012}.  Depending upon the wave energy extraction mechanism and the power take-off (PTO) system employed,
PAs can be further categorized into different subtypes. To name a few, \emph{Inertial Sea Wave Energy Converter} (ISWEC) developed by the Polytechnic University of Turin is a floating point absorber (FPA) that converts the pitching motion of the hull to an electrical output using a gyroscopic PTO system~\cite{Sirigu2020a,Sirigu2020b,Vissio2017}.  \emph{PowerBuoy} is a two-body FPA developed by Ocean Power Technologies that uses 
heave mode to extract energy from the waves~\cite{Poullikkas2014,Van2017,Yu2013}. Although FPAs have the advantage of receiving 
a dense concentration of wave energy from the ocean or sea surface, they are also prone to extreme waves and other severe weather 
conditions that can limit their operability and long-term survivability. Fully submerged point absorbers (SPA) have been designed to overcome 
these issues. \emph{CETO} is a cylindrical shaped SPA developed by Carnegie Wave Energy that is able to absorb wave energy 
using multiple degrees of freedom~\cite{Sergiienko2018,Rafiee2015}. An added advantage of SPAs is their zero visual impact on the ocean or sea shorelines~\cite{Sergiienko2017}. Currently, efforts are underway that are testing point absorber devices at various locations around the world, including but not 
limited to, the Pacific Ocean~\cite{Paasch2012}, the Atlantic Ocean~\cite{Dacobem}, and the Mediterranean Sea~\cite{Dafnakis2019,Mattiazzo2019}. Some of us are also directly involved in testing and improving WEC devices at various sea locations~\cite{Sirigu2020a,Sirigu2020b,Vissio2017,Mattiazzo2019}.
        
Numerical models based on frequency- or time-domain methods are commonly used to study the performance of point absorbers~\cite{Tetu2018,Ding2016,Sirigu2018}. The hydrodynamic loads in these methods are calculated using the boundary element 
method (BEM) approach which is based on linear potential flow (LPF) formulation. The BEM approach to 
WEC modeling solves a radiation and a diffraction problem of the oscillating converter separately. The pressure solutions 
from the radiation and diffraction problems, and the pressure field of the undisturbed incident wave are superimposed to obtain the net hydrodynamic load on the wetted surface of the converter.  Frequency- or time-domain methods ignore the viscous phenomena and the nonlinear convective terms from the equations of motion. As a result, these methods cannot capture highly nonlinear phenomena like wave-breaking and wave-overtopping. Moreover, these methods over-predict the dynamics and the wave absorption efficiency of the WEC systems~\cite{Yu2013}.  An improvement over LPF based models is the fully nonlinear potential flow (FNPF) formulation, that permits 
large-amplitude displacements of the WECs and modeling of nonlinear free-surface~\cite{Davidson2020}. 
Furthermore, FNPF models impose body boundary conditions based on the instantaneous location of the WEC in the computational domain, rather than assuming the free-surface and WECs at their equilibrium positions.

A considerable amount of accuracy in WEC modeling is achieved by solving the nonlinear incompressible Navier-Stokes (INS) equations of motion~\cite{Agamloh08,Penalba2017,Yu2013,Ghasemi2017,Anbarsooz2014}, albeit 
at a higher computational cost compared to the BEM technique. Several approaches 
to fully-resolved wave-structure interaction (WSI) modeling have been adopted in the literature. The two main categories 
are (i) the \emph{overset} or the \emph{Chimera} grid-based methods, and (ii) the \emph{fictitious domain}-based methods. 
The overset method employs an unstructured mesh for the solid structure and a background fluid grid which is generally taken as block structured~\cite{Shen2015,Carrica2007,Hou18,Facci2016}. The fictitious domain (FD) approach to fluid-structure interaction (FSI) 
modeling is a Cartesian grid-based method in which the fluid equations are extended \emph{into} the solid domain, and a common set of 
equations are solved for the two domains. \REVIEW{Fictitious domain methods have been used to simulate FSI of porous structures~\cite{Hsieh2016}, elastic boundaries~\cite{Kim2007}, and rigid bodies of complex shapes~\cite{Kumar2019,Walayat2020}}. FD methods can be implemented in several 
ways, for example, by using the \emph{distributed Lagrange multiplier} (DLM) technique~\cite{Patankar2000,Sharma2005} or by employing the \emph{Brinkman penalization} (BP) approach~\cite{Angot99,Carbou03,Bergmann11}.  Ghasemi et al.~\cite{Ghasemi2017} and Anbarsooz et al.~\cite{Anbarsooz2014} have studied the performance of a submerged cylindrical shaped 
PA with two degrees of freedom using the FD/DLM approach. The cylindrical buoy was constrained to move in two orthogonal directions 
(heave and surge) in these studies. One of the limitations of the FD/DLM method is that any external force acting on the immersed 
object (e.g. via tethered PTO system) needs to be expressed as a \emph{distributed body force density} in the INS momentum equation. 
This is not a versatile approach, but it can work for simple scenarios~\footnote{\REVIEW{For example in cases where linear and angular momentum due to additional external force and torque can be included during the momentum conservation stage of the FD/DLM method.}}. However, applying 
external torque on the structure is not straightforward for FD/DLM method because velocity, and not vorticity, is generally solved for in the 
INS equations. In contrast, it is straightforward to include both 
external forces and torques on the immersed object using FD/BP methodology. \REVIEW{Moreover, the FD/BP method is a fully-Eulerian approach to FSI. This makes the parallel implementation of the technique relatively easier compared to the FD/DLM method, which is typically implemented using two (Eulerian and Lagrangian) grids.} Since, we are interested in studying the dynamics of a three degrees of freedom (3-DOF) buoy under the action of external forces and torques, we employ the more versatile FD/BP approach in our CFD model.  \REVIEW{To our knowledge, this works presents the first application of FD/BP method for simulating WEC devices.}

Using the FD/BP framework, we simulate the wave-structure interactions of a cylindrical buoy in one, two, and three degrees of freedom. The CFD solution is compared against the potential flow model. We find that the Cummins model over-predicts the heave and surge amplitude, and does not capture the slow drift in the surge dynamics. Moreover, the potential flow model results in an insignificant pitch of an axisymmetric converter. Finally, we study the wave absorption efficiency of a 3-DOF cylindrical buoy 
under varying PTO coefficients, mass density of the buoy, and incoming wave heights using the CFD method. The resolved simulations 
provide useful insights into an efficient design procedure for a simple WEC. 

The rest of the paper is organized as follows.  We first describe the potential flow formulation and the time-domain Cummins model 
in Sec.~\ref{sec_potential_eqs}. Next, we describe the continuous and discrete equations for the multiphase fluid-structure system in Secs.~\ref{sec_cont_eqs} and~\ref{sec_sol_method}, respectively. Validation cases for the FD/BP framework are presented in Sec.~\ref{sec_validation}. The tests also highlight the solver stability in presence of large density contrasts. 
Sec.~\ref{sec_model_compare} compares the dynamics of the cylindrical buoy using potential flow and CFD models. The performance of a submerged cylindrical buoy using the CFD model under various scenarios is presented in Sec.~\ref{sec_results}.

%%%%%%%%%%%%%%%%%%%%%%%%%%%%%%
\section{Numerical model based on the potential flow theory} \label{sec_potential_eqs}
\subsection{State-space fluid-structure interaction formulation} \label{sec_cummins}
Using the potential flow model, the velocity potential $\Phi$ of an inviscid and incompressible fluid, under the assumption of irrotational flow
is obtained by solving the Laplace equation in the water domain
\begin{equation}
\nabla^2 \Phi = 0,
\end{equation}
using suitable kinematic and dynamic boundary conditions~\cite{Holthuijsen2010,Dean1991}. The fluid velocity is expressed as  gradient 
of the velocity potential, $\u = (u,v) = \V{\nabla} \Phi$. \REVIEW{Once the solution to the Laplace equation is found, the fluid pressure $p$ is obtained from the linearized Bernoulli equation $\partial \Phi/\partial t + p/\rho_\text{w} + g y = 0$  as}
\begin{align}
  p(\x,t) &= p_{\text{dynamic}}(\x,t) +  p_{\text{hydrostatic}} (y)  & \nonumber \\
            &= -\rho_{\text{w}} \D{\Phi}{t}(\x,t) - \rho_{\text{w}} g y,  & \label{eqn_bernoulli}
\end{align}
in which $g$ is the acceleration due to gravity, $\rho_{\text{w}}$ is the density of water, and $y = 0$ represents the \REVIEW{undisturbed} free-water surface.
The hydrodynamic force on a submerged body is obtained by integrating pressure forces on the wetted surface of the body.
 
In the linear wave theory, the wave amplitude and the resulting body motions are assumed to be small compared to the wavelength 
of the incident wave. Under this assumption, the flow potential can be divided into three distinct parts~\cite{Giorgi2017}
\begin{align}
\Phi  = \phii+\phid+\phir,
\end{align}  
in which $\phii$ is the undisturbed (assuming no body in the domain) wave potential of the incident wave, $\phid$ is the diffraction potential of the 
incident wave about the stationary body, and $\phir$ is the radiation potential due to an oscillatory motion of the body in still water. 
If the motion of an oscillating body such as a wave energy converter is not affected by nonlinearities in the system (like those 
arising from nonlinear power take off units), frequency-domain models are generally used to obtain the solution of motion $\V \X$
due to a monochromatic harmonic wave excitation of angular frequency $\omega$
\begin{align}
\left(\V M+ \V{A}(\omega) \right)\ddot{\X} + \V{B}(\omega)\dot{\X} + \V{K}_{\text{hydro}} \X = \F_{\text{e}}(\omega).
\end{align}  
Here $\V{M}$ is the mass matrix of the buoy, $\V{A}(\omega)$ and $\V{B}(\omega)$ are the frequency-dependent added mass and 
damping matrix of the buoy, respectively,  $\V{K}_{\text{hydro}}$  is the linear hydrostatic stiffness arising from the buoyancy force for a \emph{floating} buoy, and 
$\F_{\text{e}}(\omega)$ is the wave excitation force (Froude-Krylov and diffraction). The displacement, velocity, and acceleration of the body are denoted by   
$\V{X}$,  $\dot{\X}$, and $\ddot{\X}$, respectively. The dimensions of the matrices and vectors depend on the degrees of freedom~\cite{Sirigu2019numerical}.  
The frequency-domain models are typically resolved using the boundary element 
method (BEM) and this approach has been widely adopted in commercial codes like ANSYS AQWA~\cite{Ansys2014} or WAMIT~\cite{Lee1995}.  

When  nonlinear  effects  such  as  viscous  forces  or  nonlinear PTO interactions  with the buoy are considered,  the  linearity  
assumption of frequency-domain models is  no longer  valid.  A common approach to overcome this limitation for many seakeeping 
applications is to use the time-domain model based on the Cummins equation~\cite{Cummins1962,Ogilvie1964}, in which the nonlinearities 
are modeled as time-varying coefficients of a system of ordinary differential equations~\cite{Giorgi2018l,Giorgi2018e}. Cummins equation is a vector integro-differential
equation which involves a convolution of radiation impulse response function (RIRF) with the velocity of the body, and reads as

\begin{align}
\left(\V M+ \V{A}_\infty \right) \ddot{\X}(t) + \int_{0}^{t} \V{K}_{\text{r}}(t-\tau)\dot{\X}(\tau)\; \text{d}\tau +  \V{K}_{\text{hydro}} \X(t)  = \F_{\text{e}}(t) + \F_{\text{ext}}( \X(t),\dot{\X}(t), t ),  \label{eqn_cummins}
\end{align}  
in which $\V{A}_\infty$ is the added mass at infinite frequency, given by 
\begin{equation} 
\V{A}_\infty = \Lim{\omega \rightarrow \infty}  \V{A}(\omega),
\end{equation}
and $\V{K}_{\text{r}}$ is the radiation impulse response function (RIRF). The radiation convolution function is also referred to 
as memory function because it represents a fluid memory effect due to the radiation forces emanated by the oscillating body 
in the past. In the Cummins equation, all nonlinear effects are lumped into $\F_{\text{ext}}$ term, 
which represents external forces applied to the system. The nonlinear external 
forces could arise, for example, due to viscous drag or PTO/mooring forces.  One of the computational challenges to the solution of the 
time-domain Cummins equation is the  convolution integral involving the memory function. The time-varying  
RIRF  can  be  evaluated by  a  number  of  numerical methods, see for example works of  Yu and Falnes~\cite{Yu95}, Jefferys~\cite{Jefferys84}, Damaren~\cite{Damaren00}, McCabe et al.~\cite{Mccabe05}, and Cl\'ement~\cite{Clement95}.  In this work we follow the state-space representation approach of Fossen and Perez~\cite{Perez2008joint,Perez2008time} to approximate the radiation convolution integral by

\begin{align}
\F_{\text{r}}(t)=\int_{0}^{t} \V{K}_{\text{r}}(t-\tau)\dot{\X}(\tau)\; \text{d}\tau \simeq
\begin{cases} 
\dot{\V{\zeta}}_{\text{r}} (t)= \V{\mathcal{A}}_{\text{r}} \V{\zeta}_{\text{r}}(t) +  \V{\mathcal{B}}_{\text{r}}\dot{\X}(t), \\
\F_{\text{r}}(t) = \V{\mathcal{C}}_{\text{r}} \V{\zeta}_{\text{r}}(t),
\end{cases} 
\end{align}   
in which $\V{\mathcal{A}}_{\text{r}}$, $\V{\mathcal{B}}_{\text{r}}$, and $\V{\mathcal{C}}_{\text{r}}$ are the state-space matrices 
for carrying out time-domain analysis. It is also possible to evaluate the RIRF in frequency domain and subsequently transform it back to the time domain~\cite{Taghipour2008,Roessling2015,Jefferys84}.

\begin{figure}
  \centering
    \includegraphics[scale = 0.3]{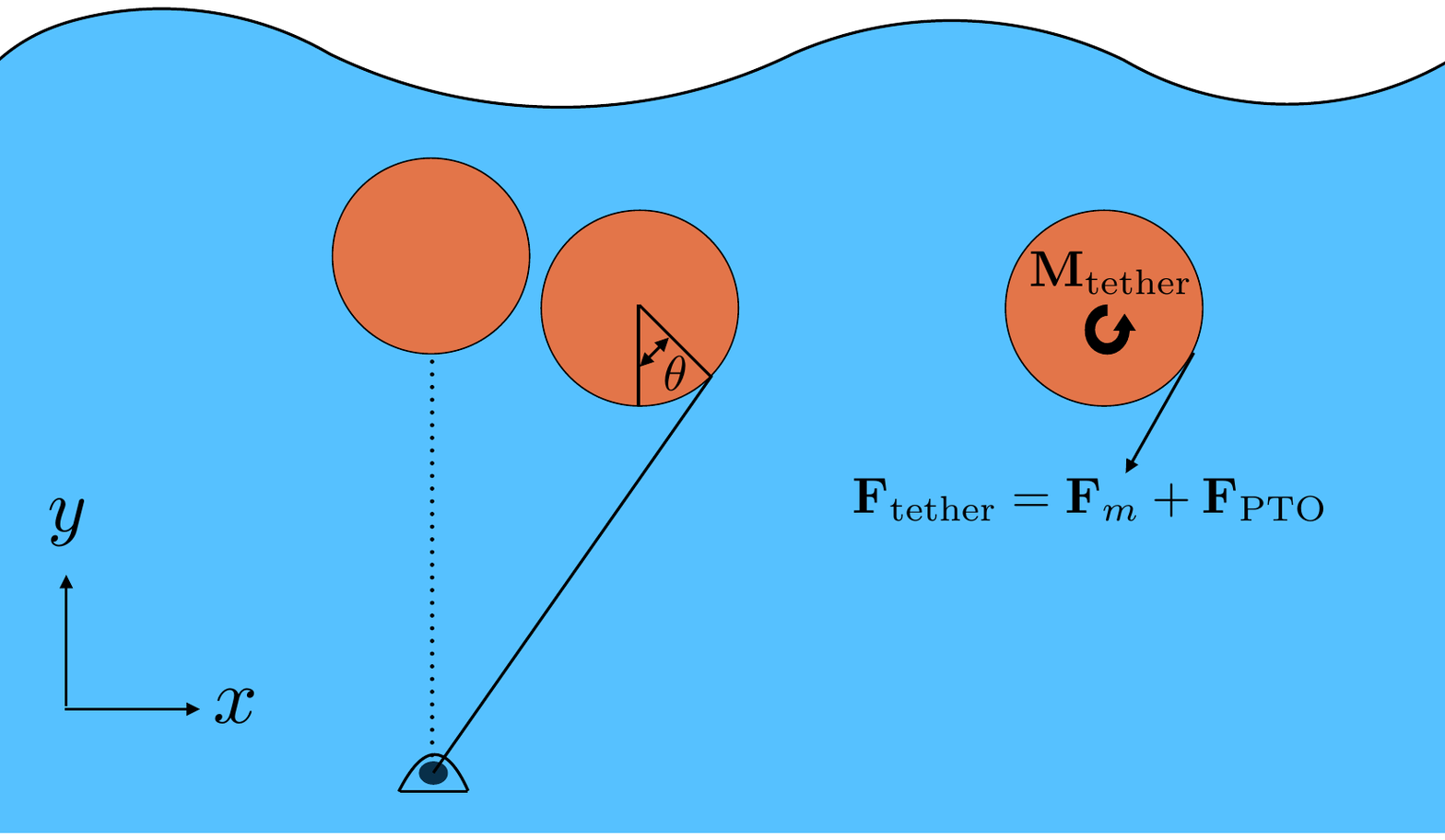}
  \caption{Schematic representation of a submerged point absorber tethered to the sea floor by the PTO system.
}
\label{fig_schematicPA} 
\end{figure}

Fig.~\ref{fig_schematicPA} shows the schematic representation of a submerged point absorber tethered to the sea floor by the PTO system, 
as modeled in this work. In our model, the PTO provides both damping and stiffness loads on the submerged buoy. 
For the PA system considered here, the external forces arise from the nonlinear viscous drag $\F_{\text{drag}}$, and 
the PTO stiffness and damping loads, which are denoted by $\F_{\text{m}}$ and $\F_{\text{PTO}}$, respectively. Since the point absorber is taken to 
be completely submerged under water, the hydrostatic stiffness arising from the buoyancy force is $\V{K}_{\text{hydro}} = \V{0}$. Instead, 
the buoy experiences a permanent hydrostatic force in the upward direction. Accounting for all the external forces acting on the buoy, the Cummins equation for a submerged two-dimensional buoy of density $\rho_{\text{s}}$, diameter $D$, and volume $V_{\text{buoy}} = \pi D^2/4$ reads as
\begin{align}
\left(\V M+ \V{A}_\infty \right) \ddot{\X}(t) + \F_{\text{r}}   &= \F_{\text{e}} + \F_{\text{drag}} +  \F_{\text{hydrostatic}} + \F_{\text{m}} + \F_{\text{PTO}},  \label{eqn_cummins_detailed} \\
\F_{\text{hydrostatic}} &= (\rho_{\text{w}} - \rho_{\text{s}}) \, g\, V_{\text{buoy}} \; \hat{\V{y}}, \label{eqn_fhydro} \\
\F_{\text{m}} &= - k_{\text{PTO}} \left(  \Delta l +  \Delta l_0 \right) \; \V{\hat{l}},  \label{eqn_fmooring} \\
\F_{\text{PTO}} &= -b_{\text{PTO}} \d{ \Delta l}{t} \; \hat{\V l}, \label{eqn_fdamper}
\end{align}
in which $k_{\text{PTO}}$ and $b_{\text{PTO}} $ are the stiffness and damping constants of the PTO,  $ \Delta l$ is 
the elongation of the PTO from a reference length,  $ \Delta l_0 =  |\F_{\text{hydrostatic}}|/k_{\text{PTO}}$ is the permanent extension 
of the PTO to balance the hydrostatic force $\F_{\text{hydrostatic}}$, and  $\hat{\V l}$ is a unit vector along PTO in the current configuration.  
The nonlinear viscous drag on the two-dimensional cylindrical buoy is modeled following the resistive drag model of 
Ding et al.~\cite{Ding2016}
 
 \begin{align}
\F_{\text{drag,x}} &= -\frac{1}{2} \rho_{\text{w}} C_{x} S_{x} |u|u  \; \hat{\V{x}}, \\
\F_{\text{drag,y}} &= -\frac{1}{2} \rho_{\text{w}} C_{y} S_{y}|v|v \; \hat{\V{y}}, \\
\M_{\text{drag},\theta}& = - \frac{1}{2} \rho_{\text{w}} C_{\theta}D^5|\dot{\theta}|\dot{\theta} \; \hat{\V{z}},
\end{align} 
in which $C_{x}$, $C_{y}$, and  $C_{\theta}$ are the drag coefficients in the surge ($\hat{\V{x}}$), heave ($\hat{\V{y}}$) and pitch ($\hat{\V{z}}$)
directions, respectively,  and $S_x = S_y = D$ is the planar cross-section area of the disk.

\subsection{Calculation of 2D coefficients} \label{sec_2dcoefs}
\begin{figure}[]
  \centering
  \subfigure[Surge force]{
    \includegraphics[scale = 0.32]{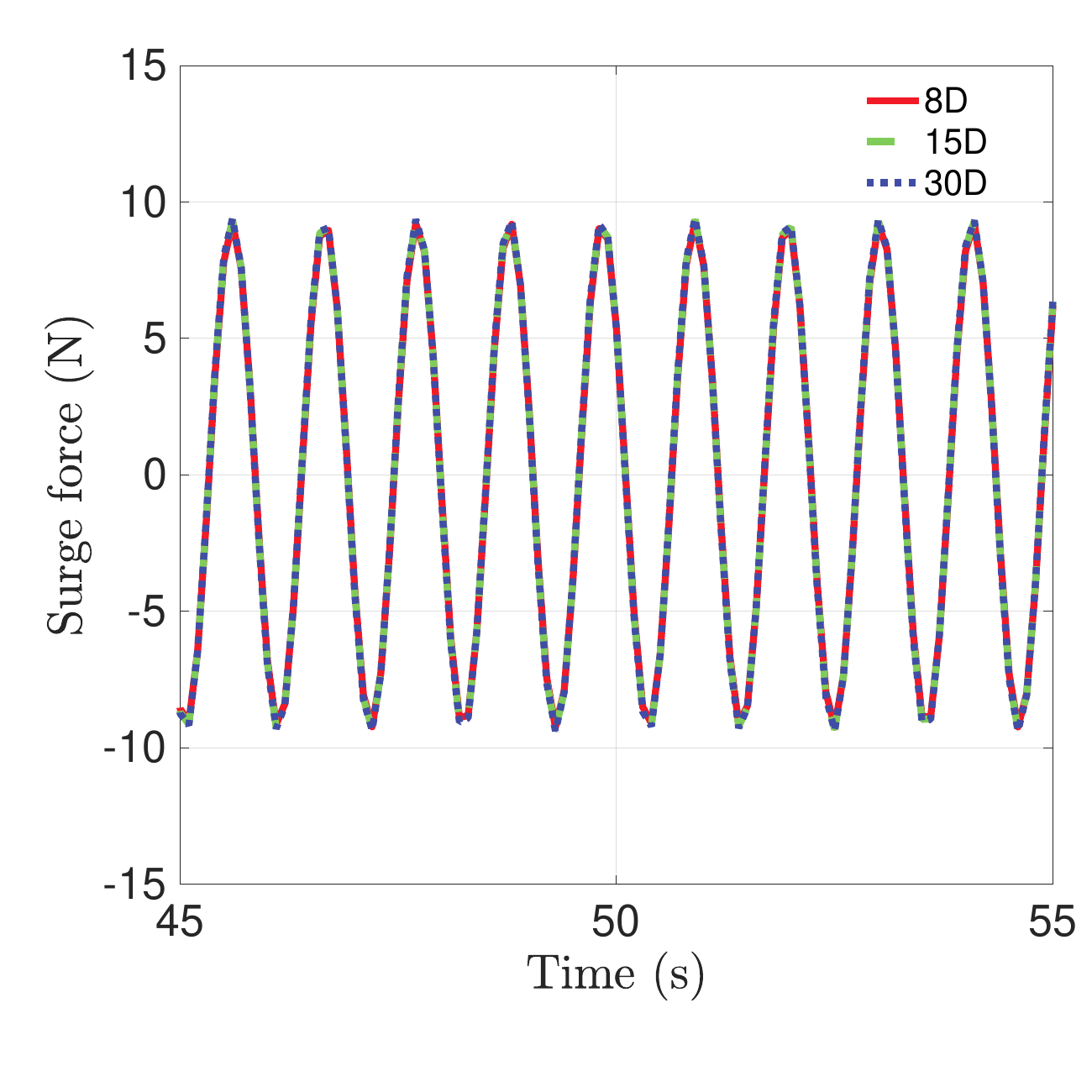}. 
    \label{fig_surge_normalized}
  }
   \subfigure[Heave force]{
    \includegraphics[scale = 0.32]{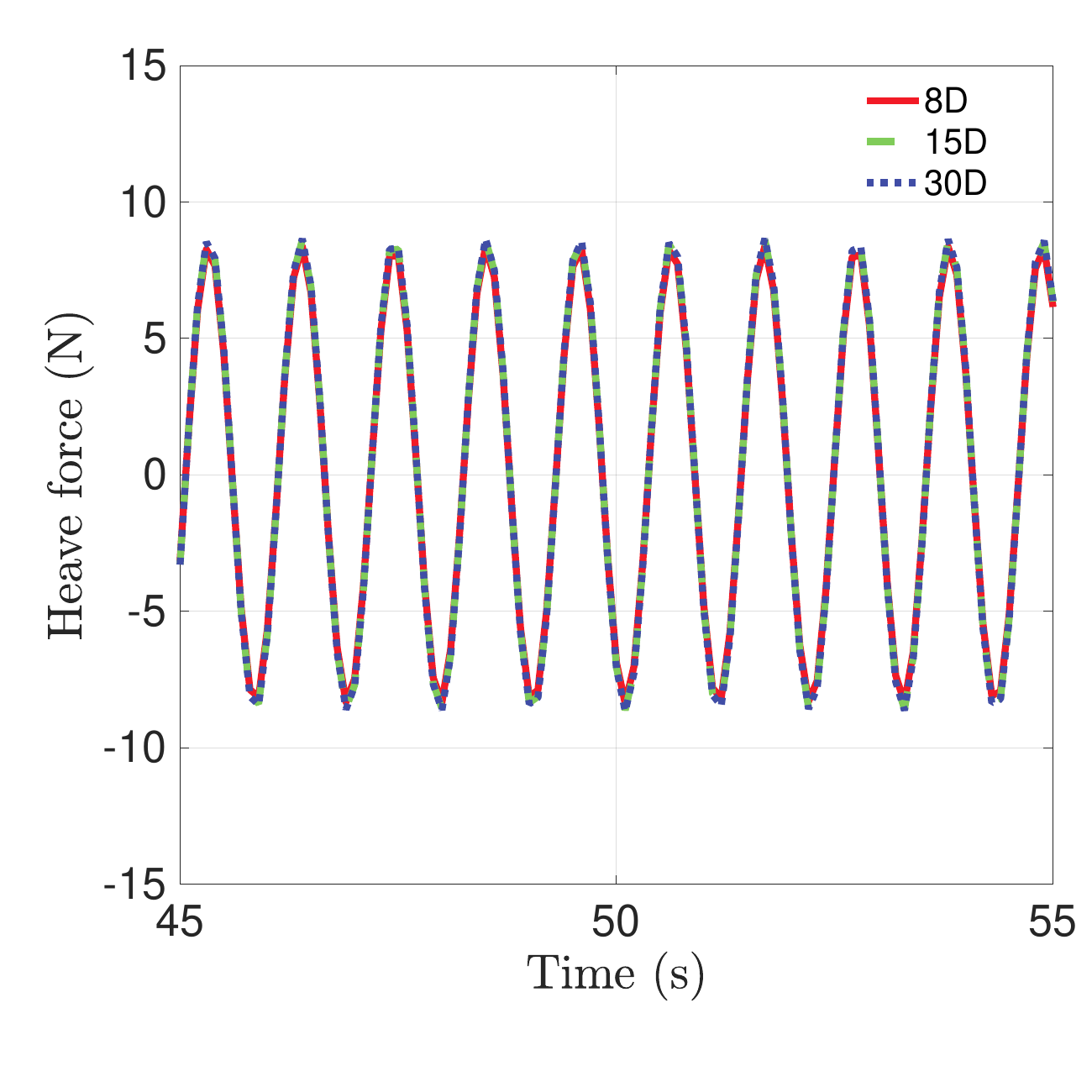}
    \label{fig_heave_normalized}
  }
  \caption{Time history of normalized~\subref{fig_surge_normalized} surge and~\subref{fig_heave_normalized} heave force on a three-dimensional cylinder of varying length $L = 8D$, $15D$, and $30D$. Forces correspond to a regular wave of height $\mathcal{H} = 0.01$ m and time period $\mathcal{T}=0.8838$ s.
  }
  \label{fig_force_normalization}
\end{figure}

To solve the time-domain Cummins Eq.~\eqref{eqn_cummins_detailed} for a  cylindrical shaped submerged point absorber, 
we use ANSYS AQWA to extract the frequency-dependent quantities such as:  1) added mass matrix
$\V{A}(\omega)$; 2) wave excitation force $\F_{\text{e}}(\omega)$; 3) and radiation damping matrix $\B(\omega)$. Further post-processing 
is done to transform the  AQWA data to time-domain amenable quantities like $\A_\infty$, 
$\V{\mathcal{A}}_{\text{r}}$, $\V{\mathcal{B}}_{\text{r}}$, and $\V{\mathcal{C}}_{\text{r}}$.
The approach described in Fossen and Perez~\cite{Perez2008joint,Perez2008time} is implemented using custom MATLAB scripts 
for this purpose. Finally, the time-domain Eqs.~\eqref{eqn_cummins_detailed}-\eqref{eqn_fdamper} are solved using an in-house 
Simulink-based code.

We remark that ANSYS AQWA provides frequency-dependent data for three-dimensional geometries. To obtain the two-dimensional data
for a submerged disk, we simulate a three-dimensional cylinder of length $L$ and diameter $D$, 
and normalize the extracted quantities by $L$. In order to ensure that the extracted quantities are not affected by finite length truncation effects, 
we perform three BEM simulations in AQWA by taking $L = 8D$, $15D$, and $30D$. Fig.~\ref{fig_force_normalization} shows the 
time history of normalized surge and heave force on a three-dimensional cylinder with different lengths.  
Since the length-normalized forces are almost identical for the three chosen lengths, the finite length truncation effects are negligible 
for a cylinder of length $8D$ and beyond. Similar trends are also obtained for normalized added mass $\V{A}(\omega)$ and 
radiation damping  $\B(\omega)$ matrices in Sec.~\ref{sec_pto_coefs}.

%%%%%%%%%%%%%%%%%%%%%%%%%%%%%%
\section{Numerical model based on the incompressible Navier-Stokes equations} \label{sec_wsi_eqs}
We use fictitious domain Brinkman penalization (FD/BP) method to perform fully-resolved wave-structure interaction simulations. 
The FD/BP method is a fully-Eulerian approach to FSI modeling. Contrary to the body-conforming mesh techniques, FD/BP method 
extends the fluid domain equations \emph{into} the solid domain and a common set of equations are written for the two domains.
This modeling approach makes the fictitious domain methods computationally more efficient for moving bodies compared to the body-conforming mesh techniques.  

The WSI framework is implemented within IBAMR~\cite{BhallaBP2019}, which is an open-source C++
library providing support for immersed boundary methods with
adaptive mesh refinement~\cite{IBAMR-web-page}.
IBAMR relies on SAMRAI \cite{HornungKohn02, samrai-web-page} for Cartesian grid 
management and the AMR framework and on PETSc~\cite{petsc-efficient, petsc-user-ref, petsc-web-page} 
for linear solver support.  

We begin by first describing the continuous equations of motion for the multiphase FD/BP method and thereafter detail its spatiotemporal discretization. 

\subsection{Continuous equations of motion} \label{sec_cont_eqs}

We state the governing equations for a coupled multiphase fluid-structure system in a fixed region of space  $\Omega \subset \mathbb{R}^d$, 
for $d = 2$ spatial dimensions. A fixed Eulerian coordinate system $\x = (x, y) \in \Omega$ is used to describe the momentum and 
incompressibility of fluid and solid domains. The spatial location of the immersed body $\Omega_{\rm b}(t) \subset \Omega$ is tracked using an indicator function $\chi(\x,t)$ which is nonzero in the solid domain and zero in the fluid domain $\Omega_{\rm f}(t) \subset \Omega$. The time-varying 
solid and fluid domains are non-overlapping and their union occupies the entire domain  $\Omega = \Omega_{\rm f}(t) \cup \Omega_{\rm b}(t)$. 
We employ spatially and temporally varying density $\rho(\x,t)$ and viscosity $\mu(\x,t)$ fields to describe the coupled three phase system. The equations of motion for the multiphase fluid-structure interaction system read as

\begin{align}
  \D{\rho \u(\x,t)}{t} + \div \rho\u(\x,t)\u(\x,t) &= -\grad p(\x,t) + \div \left[\mu \left(\grad \u(\x,t) + \grad \u(\x,t)^T\right) \right]+ \rho\g + \fc(\x,t), \label{eqn_momentum}\\
  \div \u(\x,t) &= 0. \label{eqn_continuity} 
\end{align}
Eq.~\eqref{eqn_momentum} is the conservative form of momentum equation, whereas Eq.~\eqref{eqn_continuity} describes the incompressibility 
of the system. The quantity $\u(\x,t)$ expresses the velocity, $p(\x,t)$ the mechanical pressure, and $\fc(\x,t)$ represents the 
Eulerian constraint force density which is nonzero only in the solid domain. The acceleration due to gravity is taken to be   
in negative $y$-direction $\g = (0, -g)$. In the FD/BP method, the rigidity-enforcing constraint force $\fc(\x,t)$ is 
defined as a penalization force that enforces a rigid body velocity $\ub(\x,t)$ in $\Omega_b(t)$. 
By treating the immersed structure as a porous region of vanishing permeability $\kappa \ll 1$, the Brinkman penalized constraint 
force is formulated as 
\begin{align}  
\fc(\x,t)  &= \frac{\chi(\x,t)}{\kappa}\left(\ub(\x,t) - \u(\x,t)\right). \label{eqn_brinkman_force}
\end{align}
Sec.~\ref{sec_fsi_coupling} describes the fluid-structure coupling algorithm and the rigid body velocity $\ub(\x,t)$ calculation.

\subsection{Phase tracking}

We use scalar level set function $\phi(\x,t)$ to identify liquid and gas regions, $\Omegal \subset \Omega$ and $\Omegag \subset \Omega$, respectively, in the computational domain. \REVIEW{The combined liquid and gas region denotes the fluid region described in the previous section, i.e., $\Omegal \cup \Omegag = \Omega_{\rm f}$.} The zero-contour of $\phi$ function implicitly defines the liquid-gas interface 
$\Gamma(t) = \Omegal \cap \Omegag$. Similarly, the Eulerian indicator function $\chi(\x,t)$ of the immersed body is 
expressed in terms of the level set function $\psi(\x,t)$~\footnote{\REVIEW{$\chi(\x,t) = 1 - H^{\rm body}$, in which $H^{\rm body}$ is the body Heaviside function defined in Eq.~\ref{eqn_Hbody} using $\psi(\x,t)$. }}, whose zero-contour implicitly defines the surface of the 
body $\Sb(t) = \partial \Vb(t)$; see Fig.~\ref{fig_cfd_domains}. \REVIEW{Without loss of generality,  $\phi$ level set (signed distance) values
are taken to be negative in the liquid phase and positive in the air phase. Similarly, $\psi$ level set values are taken 
to be negative inside the solid body, whereas they are taken to be positive outside the solid region}.  Using the signed distance property of the level set functions, the material properties like 
density and viscosity can be conveniently expressed as a function of  $\phi(\x,t)$ and $\psi(\x,t)$ fields    
\begin{align}
\rho (\x,t) &= \rho(\phi(\x,t), \psi(\x,t)), \label{eq_rho_ls}\\
\mu (\x,t) &= \mu(\phi(\x,t), \psi(\x,t)) \label{eq_mu_ls}.
\end{align}

As the simulation advances in time, the phase transport is described by the advection of level set fields by the local fluid velocity, which in 
conservative form reads as
\begin{align}
\D{\phi}{t} + \div (\phi \u) &= 0, \label{eq_ls_fluid_advection} \\
\D{\psi}{t} + \div (\psi \u) &= 0. \label{eq_ls_solid_advection}
\end{align}

The signed distance property of $\phi$ and $\psi$ is generally disrupted under linear advection,
Eqs.~\eqref{eq_ls_fluid_advection} and~\eqref{eq_ls_solid_advection}. Let $\widetilde{\phi}^{n+1}$
denote the flow level set function following an advective transport after
time stepping through the interval $\left[t^{n}, t^{n+1}\right]$. The flow level set is \emph{reinitialized}
to obtain a signed distance field $\phi^{n+1}$ by computing a steady-state solution to the Hamilton-Jacobi
equation

\begin{align}
&\D{\phi}{\tau} + \sgn\left(\widetilde{\phi}^{n+1}\right)\left(\|\grad \phi \| - 1\right) = 0, \label{eq_eikonal} \\
& \phi(\x, \tau = 0) = \widetilde{\phi}^{n+1}(\x), \label{eq_eikonal_init}
\end{align}
which yields a solution to the Eikonal equation $\|\grad \phi \|  = 1$ at the end of each time step. 
We refer the readers to~\cite{Nangia2019MF} for more details on the specific discretization of
Eqs.~\eqref{eq_eikonal} and~\eqref{eq_eikonal_init}, which employs second-order
ENO finite differences combined with a subcell-fix method described by Min~\cite{Min2010},
and an immobile interface condition described by Son~\cite{Son2005}. Both subcell-fix method and
the immobile interface condition have been shown to be effective in conserving mass of the flowing phases
in the literature.

Since we consider a simple geometry (cylinder) in this work, the solid level set $\psi^{n+1}$ is analytically calculated 
and reinitialized by using the new location of center of mass at  $t^{n+1}$. For more complex structures,
computational geometry techniques can be employed to compute the signed distance 
function~\footnote{In our code we use surface triangulation of complex geometries to compute the signed distance function.}.

\subsection{Spatial discretization} 

\begin{figure}[]
  \centering
  \subfigure[Continuous domain]{
    \includegraphics[scale = 0.2]{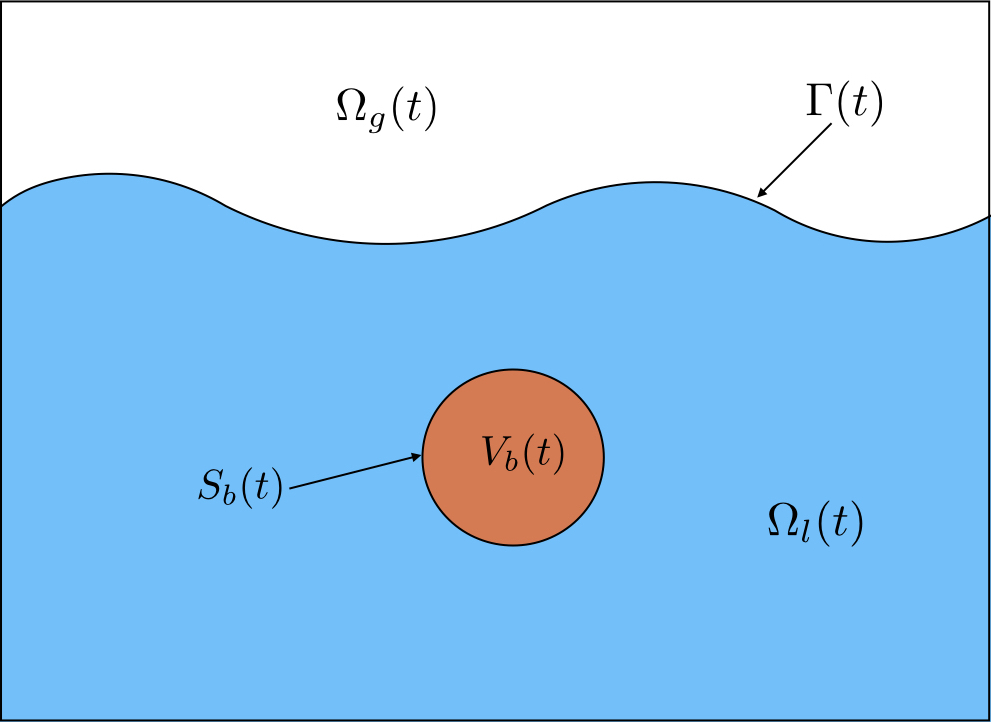} 
    \label{fig_cont_domain}
  }
   \subfigure[FD/BP discretized domain]{
    \includegraphics[scale = 0.2]{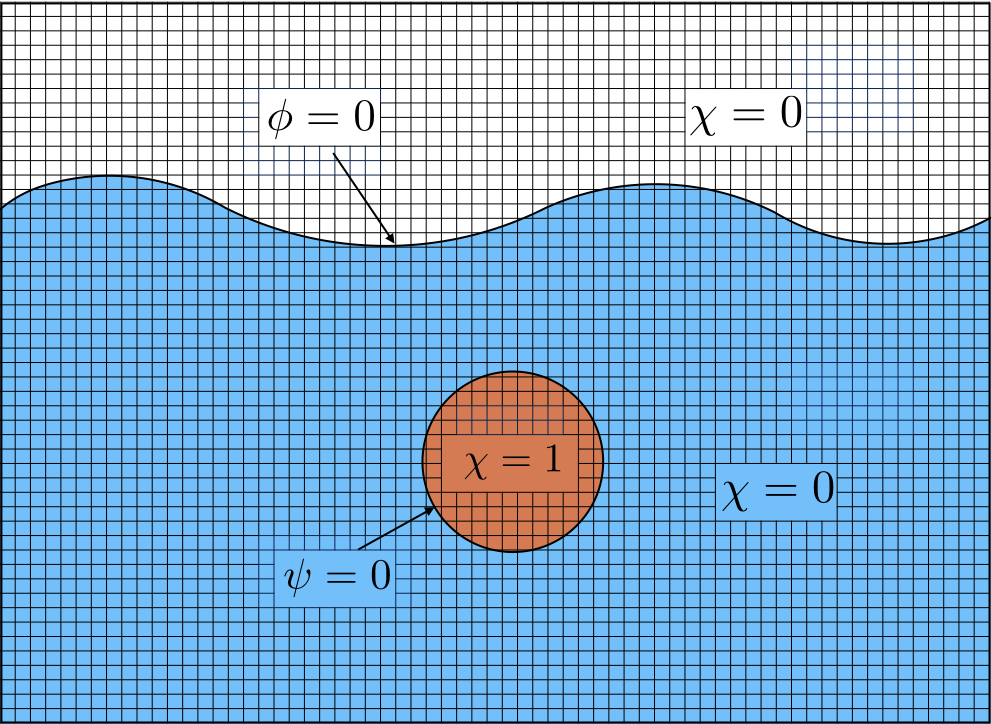}
    \label{fig_discrete_domain}
  }
  \caption{\subref{fig_cont_domain} Sketch of the immersed structure interacting with gas and liquid phases in a rectangular 
  domain $\Omega$.~\subref{fig_discrete_domain} Discretization of the domain $\Omega$ into Eulerian grid cells and the indicator 
  function $\chi$ used in the FD/BP method to differentiate between the fluid and solid regions; $\chi= 1$ inside the structure domain 
  and $\chi = 0$ in liquid and gas domains. $\phi = 0$ contour represents the liquid-gas interface and $\psi = 0$ defines the solid-fluid interface.
  }
  \label{fig_cfd_domains}
\end{figure}

The continuous equations of motion Eqs.~\eqref{eqn_momentum}-\eqref{eqn_continuity} are discretized on a staggered 
Cartesian grid as shown in Fig.~\ref{fig_cfd_domains}. The discretized domain is made up of $\Nx \times \Ny$ rectangular cells that cover the physical 
domain $\Omega$. The mesh spacing in the two directions is denoted by $\dx$ and $\dy$. Taking the lower left corner of the rectangular 
domain to be the origin $(0,0)$ of the coordinate system, each cell center of the grid has
position $\x_{i,j} = \left((i + \half)\dx,(j + \half)\dy\right)$ for $i = 0, \ldots, \Nx - 1$ and $j = 0, \ldots, \Ny - 1$.
The physical location of the \emph{vertical} cell face is half a grid space away from $\x_{i,j}$ in the $x$-direction and 
is given by $\x_{i-\half,j} = \left(i\dx,(j + \half)\dy\right)$. \REVIEW{Similarly, $\x_{i,j-\half} =\left((i + \half)\dx,j\dy\right)$} is the physical 
location of the \emph{horizontal} cell face that is half a grid cell away from  $\x_{i,j}$ in the $y$-direction. The level set fields,
pressure degrees of freedom, and the material properties are all approximated at cell centers and are denoted by 
$\phi_{i,j}^{n} \approx \phi\left(\x_{i,j}, t^n\right)$, $\psi_{i,j}^{n} \approx \psi\left(\x_{i,j}, t^n\right)$,  
$p_{i,j}^{n} \approx p\left(\x_{i,j},t^{n}\right)$, $\rho_{i,j}^{n} \approx \rho\left(\x_{i,j},t^{n}\right)$
and  $\mu_{i,j}^{n} \approx \mu\left(\x_{i,j},t^{n}\right)$, respectively; some of these quantities are interpolated onto the required 
degrees of freedom as needed (see~\cite{Nangia2019MF} for further details). Here, $t^n$ denotes the time at time step $n$.
The velocity degrees of freedom are defined on the cell faces and are denoted by $u_{i-\half,j}^{n} \approx u\left(\x_{i-\half,j}, t^{n}\right)$, and
$v_{i,j-\half}^{n} \approx v\left(\x_{i,j-\half}, t^{n}\right)$. Additional body forces on the right-hand 
side of the momentum equation are approximated on the cell faces of the staggered grid.  

We use second-order finite differences to discretize all spatial derivative
operators. The discretized version of the \REVIEW{spatial operator is denoted with a $h$ subscript}; i.e., $\grad \approx \grad_h$. Further details 
on the spatial discretization can be obtained in prior studies~\cite{Nangia2019MF,Cai2014,Griffith2009,Bhalla13}.

\subsection{Solution methodology} \label{sec_sol_method}

\subsubsection{Material property specification}
\begin{figure}[]
  \centering
  \subfigure[Material characteristics for fluids]{
    \includegraphics[scale = 0.2]{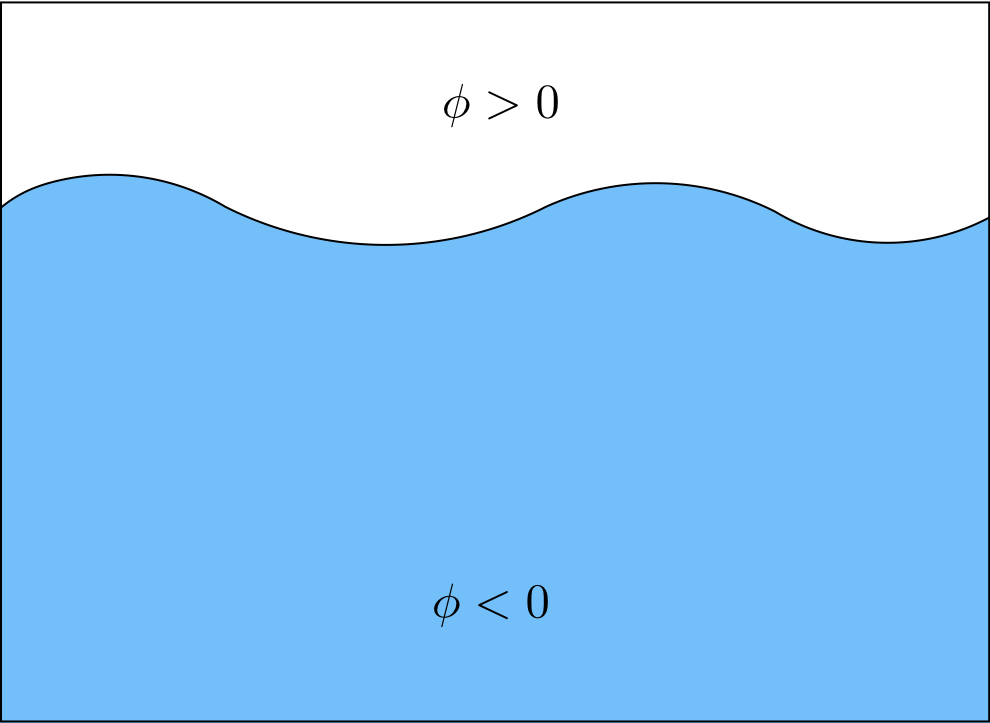}
    \label{ls_flow_diagram}
  }
   \subfigure[Material characteristics in the entire domain]{
    \includegraphics[scale = 0.2]{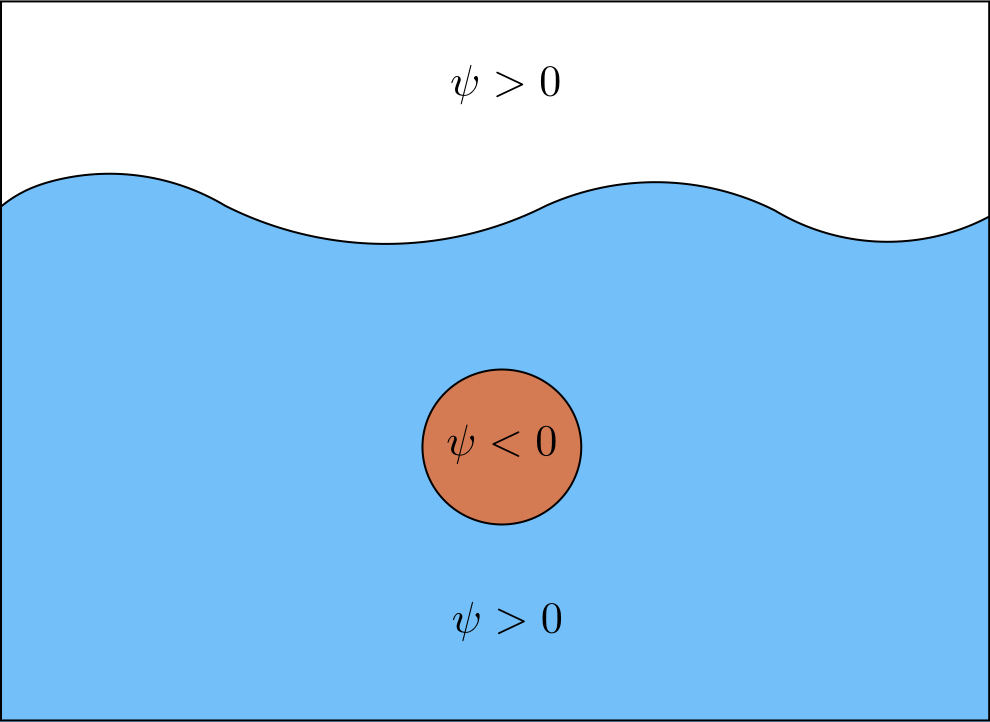}
    \label{ls_solid_diagram}
  }
  \caption{Sketch of the two-stage process for setting the density and viscosity in the computational domain.
  \subref{ls_flow_diagram} Material properties are first prescribed in the ``flowing" phase based on the liquid-gas level set
  function $\phi$ (---, black) and ignoring the structure level set function $\psi$ (\texttt{---}, orange).
  \subref{ls_solid_diagram} Material properties are then corrected in the phase occupied by the immersed body.}
  \label{fig_ls_diagram}
\end{figure}

Discretely, we use  smoothed Heaviside functions to transition from one material phase to the other. 
The transition zone occurs within $\ncells$ grid cells on either side of water-air interface $\Gamma$ or fluid-solid 
interface $\Sb$. Correspondingly, two numerical Heaviside functions are defined:

\begin{align}
\widetilde{H}^{\text{flow}}_{i,j} &= 
\begin{cases} 
       0,  & \phi_{i,j} < -\ncells \dx,\\
        \frac{1}{2}\left(1 + \frac{1}{\ncells \dx} \phi_{i,j} + \frac{1}{\pi} \sin\left(\frac{\pi}{ \ncells \dx} \phi_{i,j}\right)\right) ,  & |\phi_{i,j}| \le \ncells \dx,\\
        1,  & \textrm{otherwise},
\end{cases}       \label{eqn_Hflow} \\
\widetilde{H}^{\text{body}}_{i,j} &= 
\begin{cases} 
       0,  & \psi_{i,j} < -\ncells \dx,\\
        \frac{1}{2}\left(1 + \frac{1}{\ncells \dx} \psi_{i,j} + \frac{1}{\pi} \sin\left(\frac{\pi}{ \ncells \dx} \psi_{i,j}\right)\right) ,  & |\psi_{i,j}| \le \ncells \dx,\\
        1,  & \textrm{otherwise},  \label{eqn_Hbody}
\end{cases}
\end{align}
The number of transition cells across $\Gamma$ or $\Sb$ interface is assumed to be the same. This is not a strict requirement 
of our numerical method, but is holds true for all the WSI cases simulated in this work. A two-step process (see Fig.~\ref{fig_ls_diagram}) 
is used to prescribe a given material property $\Im$ (such as $\rho$ or $\mu$) in the whole domain $\Omega$:

\begin{itemize}
 \item First, the material property in the “flowing” phase is set via the liquid-gas level set function

\begin{equation}
\label{eqn_ls_flow}
\Im^{\text{flow}}_{i,j} = \Im_\text{l} + (\Im_\text{g} - \Im_\text{l}) \widetilde{H}^{\text{flow}}_{i,j}. 
\end{equation}

\item Next, $\Im^{\text{flow}}$ is corrected by accounting for the structural material property to obtain $\Im_{i,j}^{\text{full}}$ throughout 
the computational domain

\begin{equation}
\label{eqn_ls_solid}
\Im_{i,j}^{\text{full}} = \Im_\text{s} + (\Im^{\text{flow}}_{i,j} - \Im_\text{s}) \widetilde{H}^{\text{body}}_{i,j}. 
\end{equation}

\end{itemize}
%Consistent with Eqs.~\eqref{eqn_Hflow}-\eqref{eqn_ls_solid} and without loss of generality,  $\phi$ signed distance values
%are taken to be negative in the liquid phase and positive in the air phase. Similarly, $\psi$ signed distance values are taken 
%to be negative inside the solid body, whereas they are taken to be positive outside the solid region.    

\subsection{Time stepping scheme}

We employ a fixed-point iteration time stepping scheme with $\ncycles = 2$ cycles per time step to advance quantities of
interest at time level $t^n$ to time level $t^{n+1} = t^n + \Delta t$. If $k$ superscript denotes the cycle number of the fixed-point iteration, 
then at the beginning of each time step we set  $k = 0$,  $\u^{n+1,0} = \u^{n}$, $p^{n+\half,0} = p^{n-\half}$, $\phi^{n+1,0} = \phi^{n}$, and
$\psi^{n+1,0} = \psi^{n}$. For $n = 0$ initial time level, all of the physical quantities have a prescribed initial condition. 
 
\subsubsection{Level set advection}
The level set functions are time-marched using an explicit advection scheme
\begin{align}
\frac{\phi^{n+1,k+1} - \phi^{n}}{\dt} + Q\left(\u^{n+\half,k}, \phi^{n+\half,k}\right) &= 0, \\
\frac{\psi^{n+1,k+1} - \psi^{n}}{\dt} + Q\left(\u^{n+\half,k}, \psi^{n+\half,k}\right) &= 0,
\end{align}
in which $Q(\cdot,\cdot)$ represents an explicit piecewise parabolic method (xsPPM7-limited) approximation to the 
linear advection terms on cell centers~\cite{Griffith2009,Rider2007}.

\subsubsection{Incompressible Navier-Stokes solution}

The spatiotemporal discretization of the conservative form of incompressible Navier-Stokes Eqs.~\eqref{eqn_momentum}-\eqref{eqn_continuity} 
reads as
\begin{align}
	\frac{\breve{\V \rho}^{n+1,k+1} \u^{n+1,k+1} - { \V \rho}^{n} \u^n}{\dt} + \C^{n+1,k} &= -\grad_h p^{n+\half, k+1}
	+ \left(\L_{\mu} \u\right)^{n+\half, k+1}
	+  \V \wp^{n+1,k+1}\g +  \fc^{n+1,k+1}, \label{eqn_c_discrete_momentum}\\
	 \grad_h \cdot \u^{n+1,k+1} &= 0, \label{eqn_c_discrete_continuity}
\end{align}
in which the newest approximation to density $\breve{\V \rho}^{n+1,k+1}$ and the discretization of the 
convective term $\C^{n+1,k}$ are computed such that they satisfy consistent mass/momentum transport, 
which is required to maintain numerical stability for high value of air-water density ratio. We refer
the reader to Nangia et al.~\cite{Nangia2019MF,BhallaBP2019} for more details on consistent mass/momentum 
transport scheme. Note that $\left(\L_{\mu} \u\right)^{n+\half, k+1} =  \half\left[\left(\L_{\mu} \u\right)^{n+1,k+1} + \left(\L_{\mu} \u\right)^n\right]$
is a semi-implicit approximation to the viscous strain rate with
$\left(\L_{\mu}\right)^{n+1} = \grad_h \cdot \left[\mu^{n+1} \left(\grad_h \u + \grad_h \u^T\right)^{n+1}\right]$.
The newest approximation to viscosity $\mu^{n+1,k+1}$ is obtained via the two-stage process
described in Eqs.~\eqref{eqn_ls_flow} and~\eqref{eqn_ls_solid}. 

\REVIEW{In Eq.~\eqref{eqn_c_discrete_momentum}, ${ \V \rho}^{n}$ is the face-centered value of the density field $\rho^{\rm full}$ (refer Eq.~\eqref{eqn_ls_solid} for the definition of $\rho^{\rm full}$), and the density field $\breve{\V \rho}^{n+1,k+1}$ is obtained by advecting ${ \V \rho}^{n}$ discretely.  The specific value of the density field $\V \wp$  
used to compute the gravitational body force $\V \wp \g$ is explained next in the context
of FD/BP fluid-structure coupling algorithm.} 

\subsubsection{Fluid-structure coupling} \label{sec_fsi_coupling}

The Brinkman penalization term that imposes the rigidity constraint in the solid region is treated implicitly in 
Eq.~\eqref{eqn_c_discrete_momentum} and is expressed as
\begin{align}
\fc^{n+1,k+1} = \frac{\widetilde{\chi}}{\kappa}\left(\ub^{n+1,k+1} - \u^{n+1,k+1}\right),  \label{eqn_bp_discrete}
\end{align}
in which $\widetilde{\chi} = 1 - \widetilde{H}^{\text{body}}$,  $ \widetilde{H}^{\text{body}}$ represents the regularized structure Heaviside 
function (Eq.~\eqref{eqn_Hbody}) and $\kappa \sim \cO(10^{-8})$; this permeability value has been found sufficiently small to enforce the 
rigidity constraint effectively in the prior studies~\cite{Gazzola2011,BhallaBP2019}. In Eq.~\eqref{eqn_bp_discrete}, $\ub$ is the 
solid body velocity which can be expressed as a sum of translational $\Ur$ and rotational $\Wr$ velocities
\begin{equation}
\ub^{n+1,k+1} = \Ur^{n+1,k+1} + \Wr^{n+1,k+1} \times \left(\x - \Xcom^{n+1,k+1}\right).
\end{equation}
\REVIEW{The center of mass point $\Xcom^{n+1,k+1}$ is updated using rigid body translational velocity as 
\begin{equation}
\Xcom^{n+1,k+1} = \Xcom^n + \dt \,  \Ur^{n+1,k+1}.
\end{equation}
}
The rigid body velocity is computed by integrating Newton's second law of motion
\begin{align}
		\Mb \frac{\Ur^{n+1,k+1} - \Ur^n}{\dt} &=  \cF^{n+1,k} + \Mb \g + \F_{\text{m}}^{n+1,k} + \F_{\text{PTO}}^{n+1,k},  \label{eqn_newton_u} \\
		\Ib \frac{\Wr^{n+1,k+1} - \Wr^n}{\dt} &=  \cM^{n+1,k} + \M_{\text{m}}^{n+1,k} + \M_{\text{PTO}}^{n+1,k},  \label{eqn_newton_w}
\end{align}
in which $\Mb$ is the mass, $\Ib$ is the moment of inertia matrix,  $\cF$  and $\cM$ are the net hydrodynamic force and torque, respectively, $\Mb \g$ is the net gravitational force, $\F_{\text{m}}$ and $\F_{\text{PTO}}$ are the PTO stiffness and damping force, respectively (see Eqs.~\eqref{eqn_fmooring}-\eqref{eqn_fdamper}), and  $\M_{\text{m}}$ and $\M_{\text{PTO}}$  are the corresponding PTO stiffness and damping torque about the center of mass point $\Xcom$, respectively.  The net hydrodynamic force $\cF$ and torque $\cM$ on the immersed body is computed by summing the contributions from pressure 
and viscous forces acting on the body 
\REVIEW{
\begin{align}
\cF^{n+1,k}  &= \sum_{\rm face}    \left(-p^{n+1,k} \n_{\rm face} + \mu_{\rm f} \left(\grad_h \u^{n+1,k} +  \left(\grad_h \u^{n+1,k}\right)^T \right)\cdot \n_{\rm face} \right) \Delta A_{\rm face},   \label{eqn_int_fh} \\
\cM^{n+1,k}  &= \sum_{\rm face}  \left(\x - \Xcom^{n+1,k}\right) \times  \left(-p^{n+1,k} \n_{\rm face} + \mu_{\rm f} \left(\grad_h \u^{n+1,k} +  \left(\grad_h \u^{n+1,k}\right)^T \right)\cdot \n_{\rm face} \right) \Delta A_{\rm face}.  \label{eqn_int_mh}
\end{align}
The hydrodynamic traction in these two equations is evaluated on Cartesian grid faces  that define a stair-step representation of the body on the Eulerian mesh~\cite{BhallaBP2019}. In Eqs.~\eqref{eqn_int_fh} and~\eqref{eqn_int_mh}, $\n_{\rm face}$ denotes the unit normal to the face (pointing away from the solid and into the fluid), and $\Delta A_{\rm face}$ is the area of the face.
}

Since the net gravitational force on the body is already included in Eq.~\eqref{eqn_newton_u}, we exclude it from the body region 
in the INS momentum Eq.~\eqref{eqn_c_discrete_momentum}. Therefore we set  $\V \wp \g = \bm{\rho}^{\text{flow}} \g$ in Eq.~\eqref{eqn_c_discrete_momentum}, which also avoids spurious currents due to large density variation near the fluid-solid interface~\cite{Nangia2019WSI}. \REVIEW{More specifically, $\bm{\rho}^{\text{flow}}$ is obtained using Eq.~\eqref{eqn_ls_flow} and \emph{not} through Eq.~\eqref{eqn_ls_solid}.}

\subsection{FD/BP validation} \label{sec_validation}

To ensure that the fully-resolved CFD model performs reliably for the submerged point absorber problem 
and remains stable under high density contrast between different phases, two validation cases are considered. 
For the first validation case, damped oscillations of a dense and a light cylinder in various damping regimes are considered and the simulated results are compared against analytical solutions~\cite{Rao2011}. For the second validation case, we perform grid convergence tests for a submerged point absorber that oscillates in the heave direction under wave excitation loads. For validation of other aspects of FD/BP method  and numerical wave tank 
implementation, we refer readers to our prior works~\cite{BhallaBP2019,Nangia2019WSI}.   

For all the cases considered in this work, water and air densities are taken to be $1025$ kg/m$^3$ and $1.2$ kg/m$^3$, respectively, and their respective viscosities are taken to be $10^{-3}$ Pa$\cdot$s and $1.8 \times 10^{-5}$ Pa$\cdot$s. Surface tensions effects are neglected as they do not affect the wave and converter dynamics at the scale of the problems considered in this work.

\subsubsection{Damped oscillations of a cylinder}

\begin{figure}[]
  \centering
  \subfigure[Schematic]{
    \includegraphics[scale = 0.28]{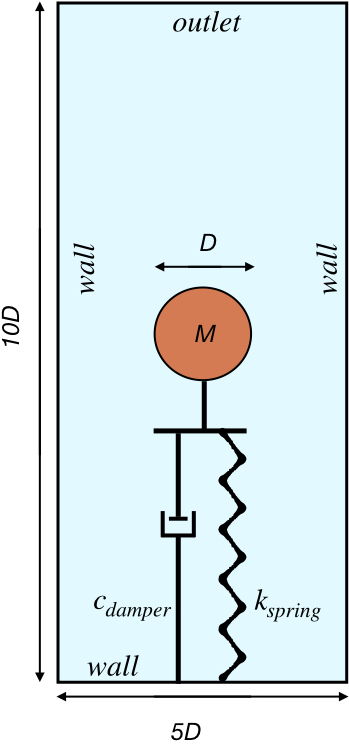}
    \label{fig_fdo_schematic}
  }
   \subfigure[$m^* = 100$]{
    \includegraphics[scale = 0.32]{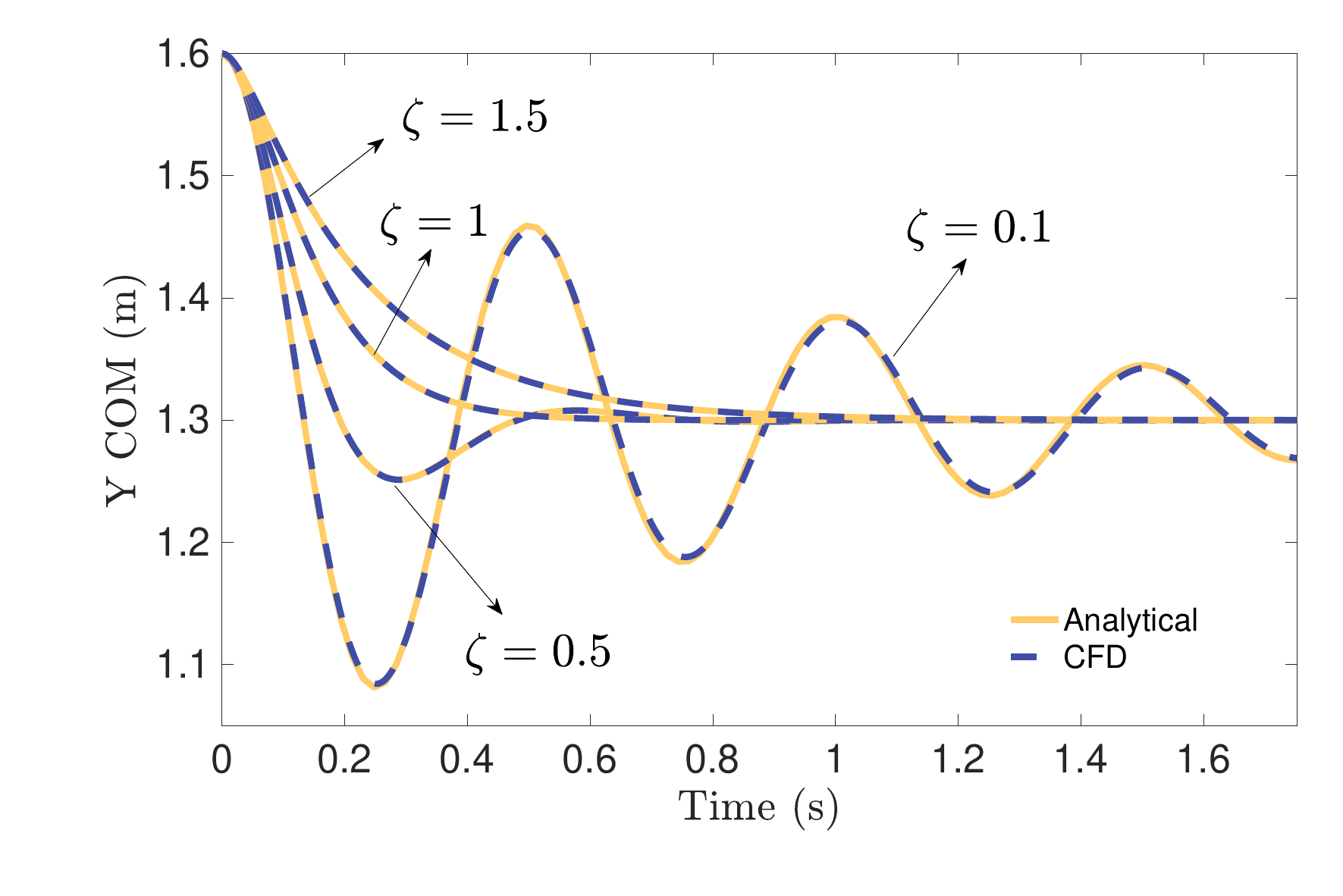}
    \label{fig_decay_results}
  }
  \caption{ \subref{fig_fdo_schematic} Schematic of the damped-oscillatory system. \subref{fig_decay_results} Center of mass vertical position as a function of time for various values of damping ratio $\zeta$. (----, solid yellow line) Analytical;  (\texttt{---}, dashed blue line) CFD model.
  }
\end{figure}

\begin{figure}[]
  \centering
  \subfigure[]{
    \includegraphics[scale = 0.25]{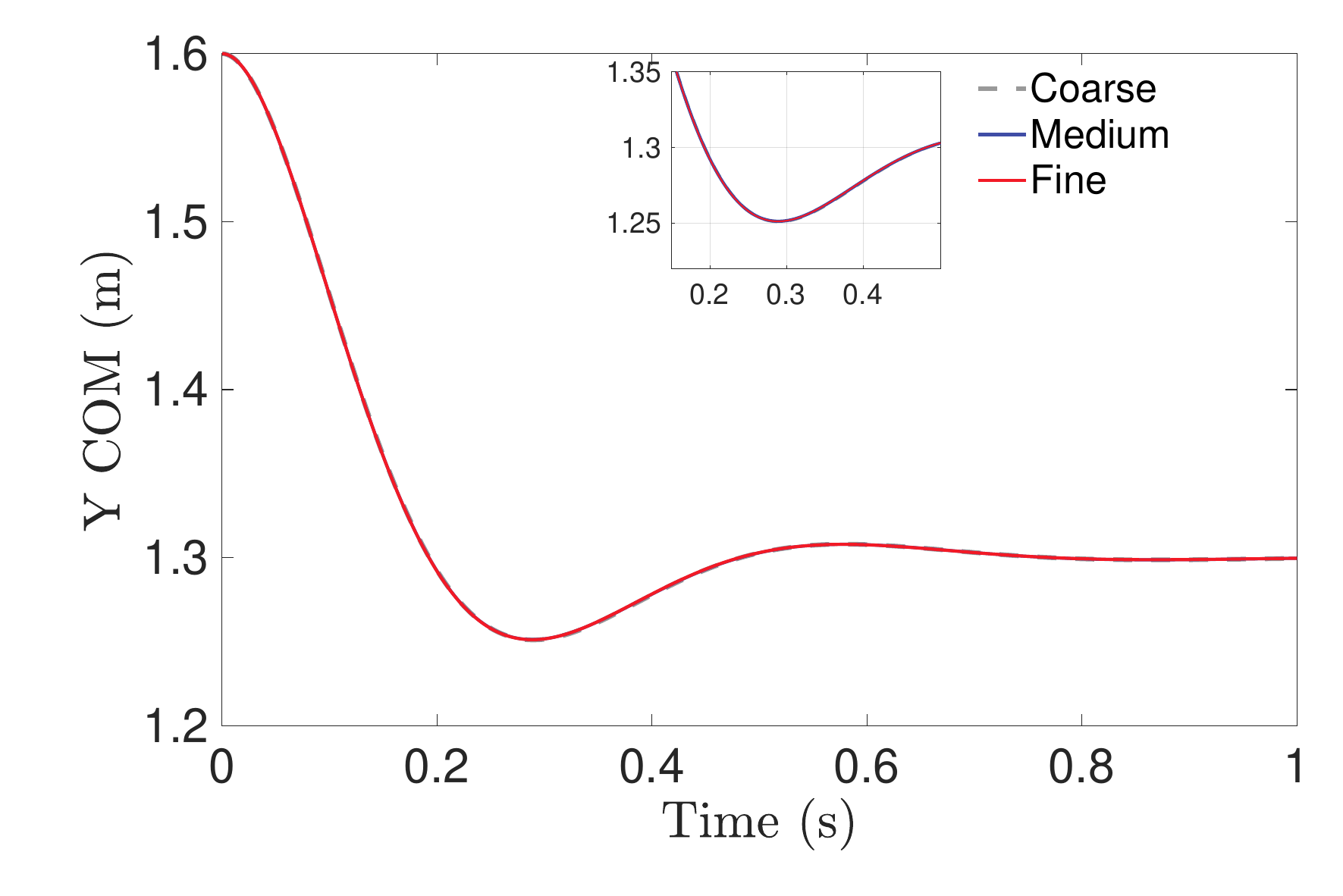}
    \label{fig_fdo_yres}
  }
   \subfigure[]{
    \includegraphics[scale = 0.25]{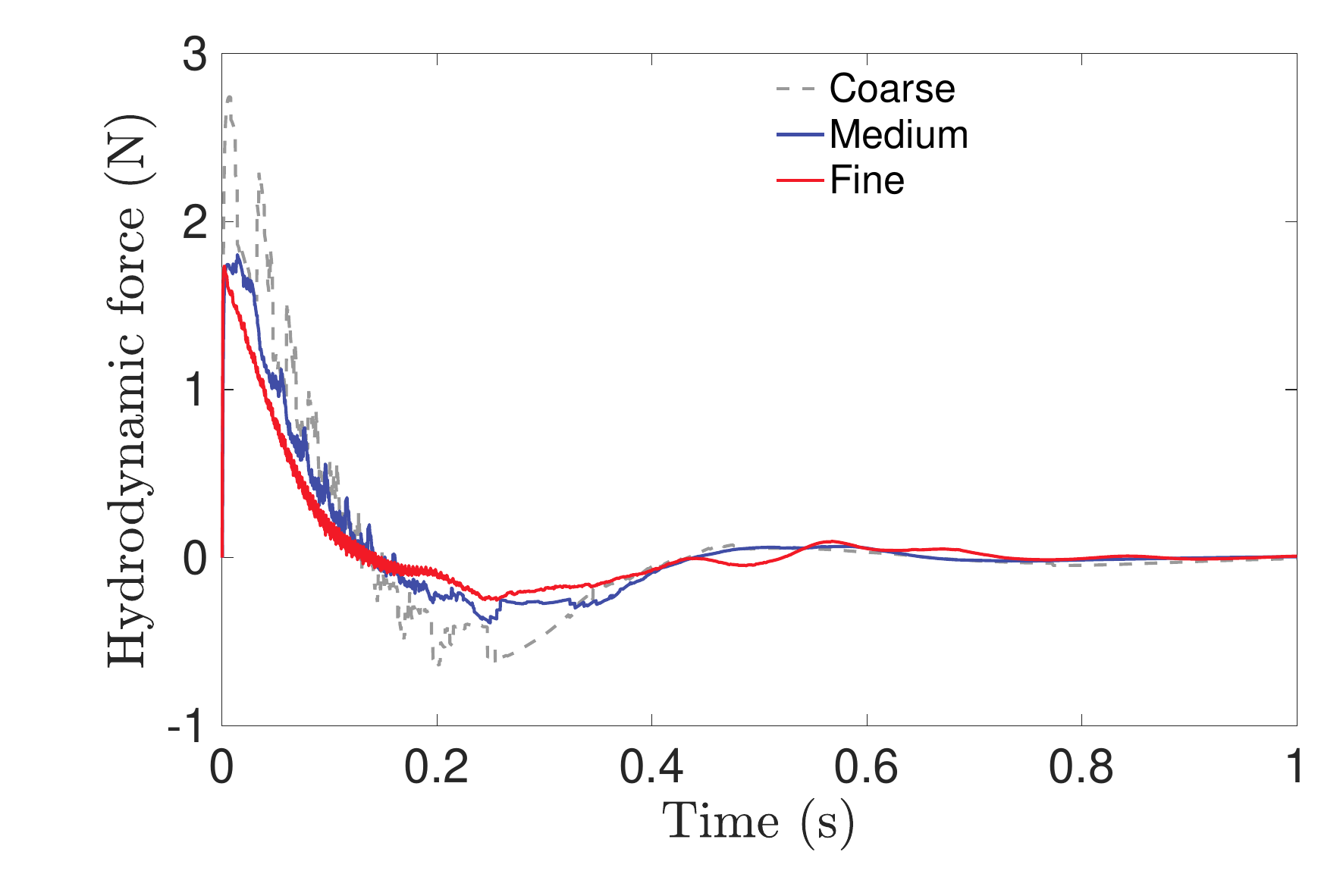}
    \label{fig_fdo_fyres}
  }
  \caption{ Grid convergence of~\subref{fig_fdo_yres} vertical displacement of the center of mass, and~\subref{fig_fdo_fyres} vertical component of the hydrodynamic (pressure and viscous) force acting on the cylinder. Here,  $m^* = 100$ and $\zeta = 0.5$ case is considered. 
  }
     \label{fig_fdo_grid_resolution}
\end{figure}

\begin{figure}[]
  \centering
    \subfigure[]{
    	\includegraphics[scale = 0.24]{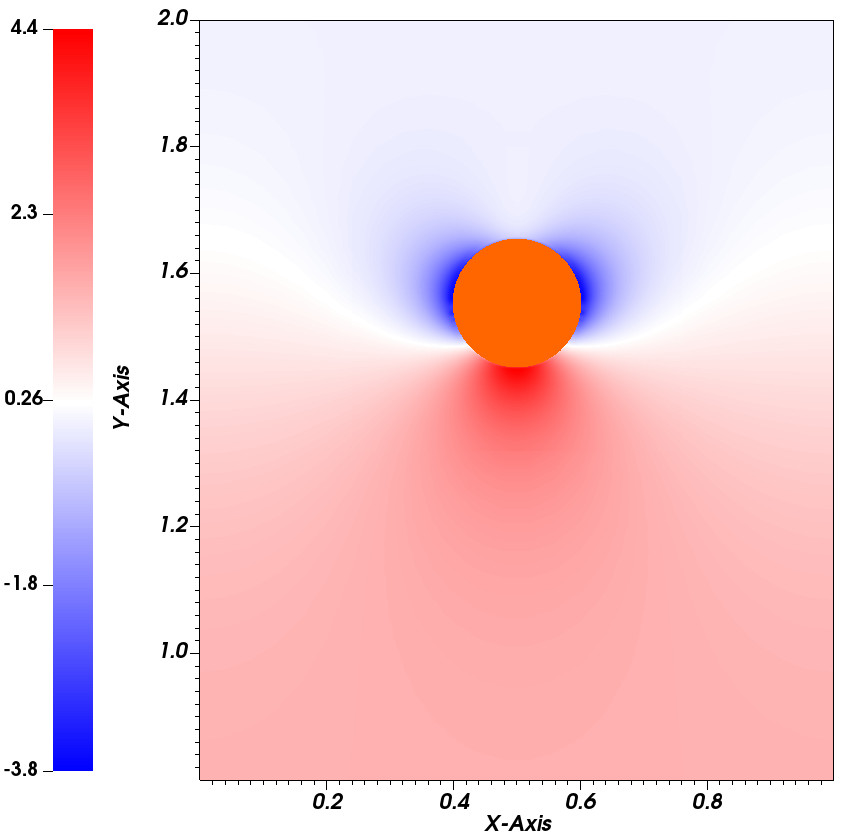}
	 \label{fig_fdo_pressure}
    }
     \subfigure[]{
    	\includegraphics[scale = 0.24]{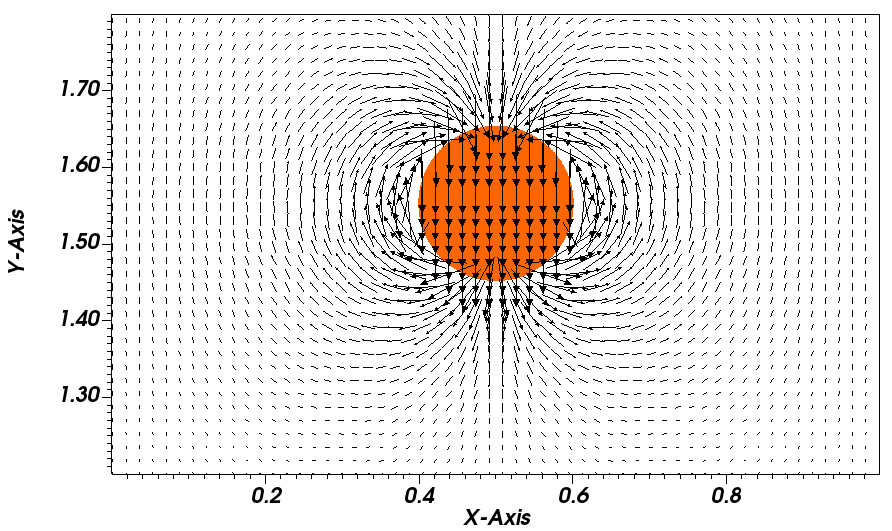}
	 \label{fig_fdo_vel}
    }   
  \caption{Distribution of~\subref{fig_fdo_pressure} pressure, and~\subref{fig_fdo_vel} velocity vectors around the cylinder during its downward motion in an under-damped regime ($\zeta = 0.5$)  at time $t = 0.05$ s. A medium grid resolution is used for the case shown here. 
}
  \label{fig_fdo_pressure_vorticity}
\end{figure}

We begin by considering a simplified version of the point absorber problem. A planar disk of 
diameter $D$, mass density $\rho_{\text{s}}$ (or mass $M = \rho_{\text{s}} \pi D^2/4$), is attached to 
a spring of stiffness value $k_{\text{spring}}$, and a mechanical damper of damping 
coefficient $c_{\text{damper}}/c_{\text{critical}} = \zeta$. Here, $c_{\text{critical}} = 2 \sqrt{k_{\text{spring}} M}$ is the 
critical damping coefficient. The value of $\zeta$ determines the behavior of the system:
$\zeta < 1$ leads to under-damping, $\zeta > 1$ leads to over-damping, and  $\zeta = 1$ results in a critically damped system. 

The computational domain is taken to be $\Omega$: $[0,5D] \times [0,10D]$ and the cylinder's initial center of mass  
point is released from $(X_0,Y_0) = (2.5D, 8D)$. The rest length of the spring is taken as $6.5D$, which gives the initial 
extension of the spring to be $1.5D$. The spring and the damper are connected to the bottom of the domain as shown 
in the schematic~\ref{fig_fdo_schematic}. The cylinder is surrounded by air phase. The density ratio between the solid and 
the fluid phase is denoted by $m^*$.

Next, we consider the damped oscillation of the cylinder at different density ratios $m^*$ = 100, 2, and 0.8. We fix $D = 0.2$ m, and 
$k_{\rm spring} = 500$ N/m in these simulations. Four test cases corresponding to $\zeta = $ 0.1, 0.5, 1.0, and 1.5 are simulated 
with each density ratio.  

\subparagraph{High density ratio $\V{m^* =  100}$:}

We discretize the domain by a $100 \times 200$ uniform grid and use a constant time step size of $\Delta t = 2.5 \times 10^{-4}$ s to simulate the four damping ratio cases. Fig.~\ref{fig_decay_results} compares the center of mass vertical position as a function of time with the analytical solutions in different damping regimes.  The analytical solutions are derived by neglecting the fluid forces on the mass. With a large density difference between the solid and air phase, the hydrodynamic forces are small compared to the inertia of the system, and the fluid-structure interaction solution matches the analytical solution quite well as observed in Fig.~\ref{fig_decay_results}. Moreover, the results also show that the FD/BP methodology remains stable for large density ratios.

\begin{table}[]
    \caption{Grid resolutions to simulate the forced damped oscillation of a cylinder.}
    \centering
    \scalebox{1.1}
    {
        \begin{tabular}{|c|c|c|c|}
         \hline
         Resolution & Coarse & Medium & Fine \\
         \hline
         $N_x$ & 50& 100 &  400 \\
          \hline
          $N_y$ & 100 & 200 &  800 \\
          \hline
          $\Delta t$ (s) & $2.5 \times 10^{-4}$  & $2.5 \times 10^{-4}$  & $10^{-4}$ \\
          \hline
          \end{tabular} 
      }
      \label{tab_fdo_grid_resolution}
\end{table}

To ensure that the aforementioned spatiotemporal resolution produces a converged FSI solution, we perform a convergence study  with coarse, medium and fine grids for $\zeta = 0.5$ case. Table~\ref{tab_fdo_grid_resolution} tabulates the spatiotemporal resolutions used for these grids. Fig.~\ref{fig_fdo_grid_resolution} shows the grid convergence plots for the cylinder displacement and the hydrodynamic force acting in the vertical direction. As observed in the figure, a medium grid resolution of $100 \times 200$ is sufficient to resolve the FSI dynamics. Moreover, the spurious oscillations (as a function of time) in the hydrodynamic forces are significantly reduced at higher (medium and fine in this case) spatiotemporal resolutions, which is an expected trend for fictitious domain/immersed techniques~\cite{lee2011sources}. Fig.~\ref{fig_fdo_pressure_vorticity} shows the pressure and flow distribution around the cylinder at $t = 0.05$ s.

\subparagraph{Low density ratios $\V{m^*}$ = 2 and $\V{m^*}$ = 0.8:}

\begin{figure}[]
  \centering
 \subfigure[$m^* = 2$]{
    	\includegraphics[scale = 0.25]{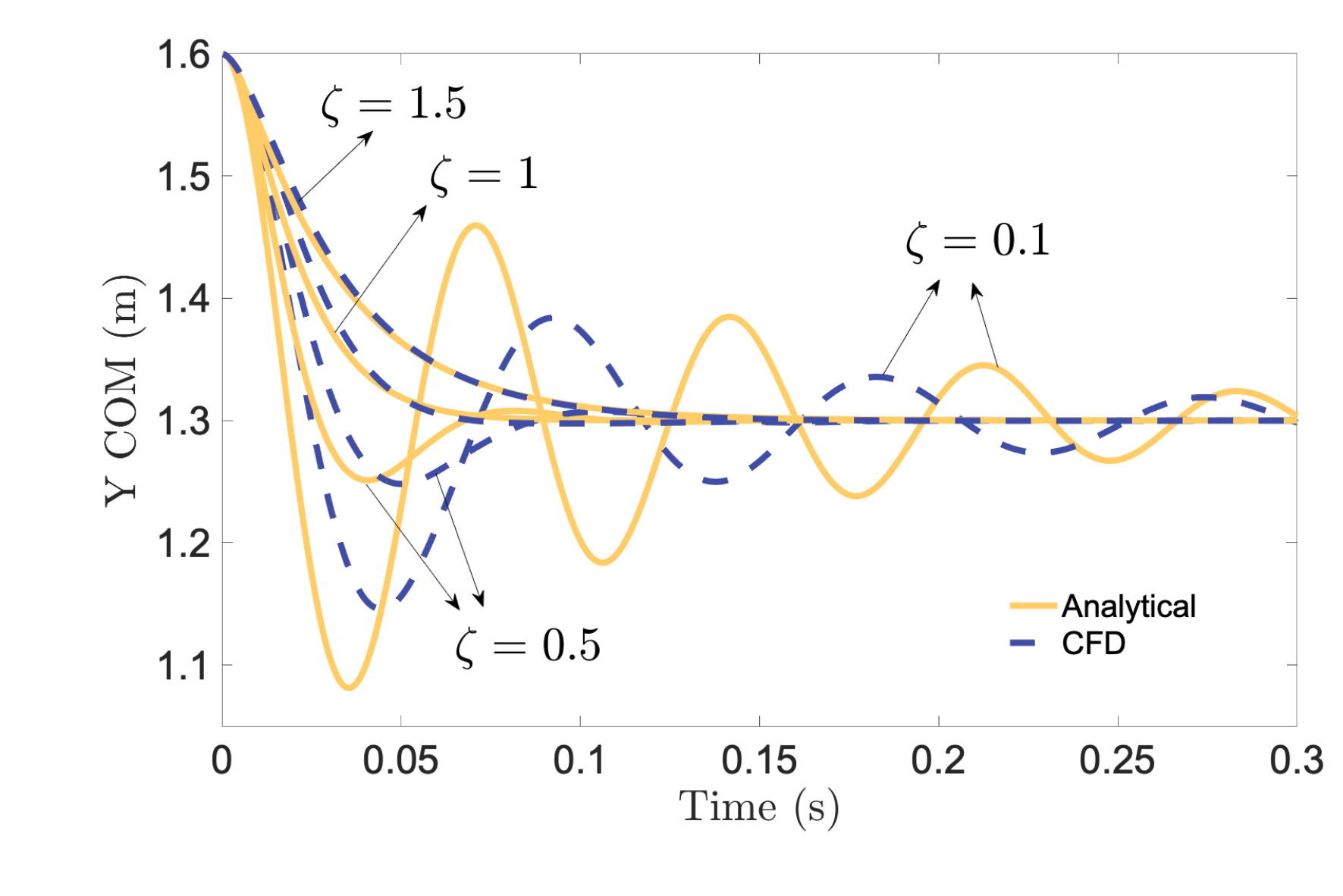}
	\label{fig_decay_results_R2}
    }
 \subfigure[$m^* = 0.8$]{
    	\includegraphics[scale = 0.25]{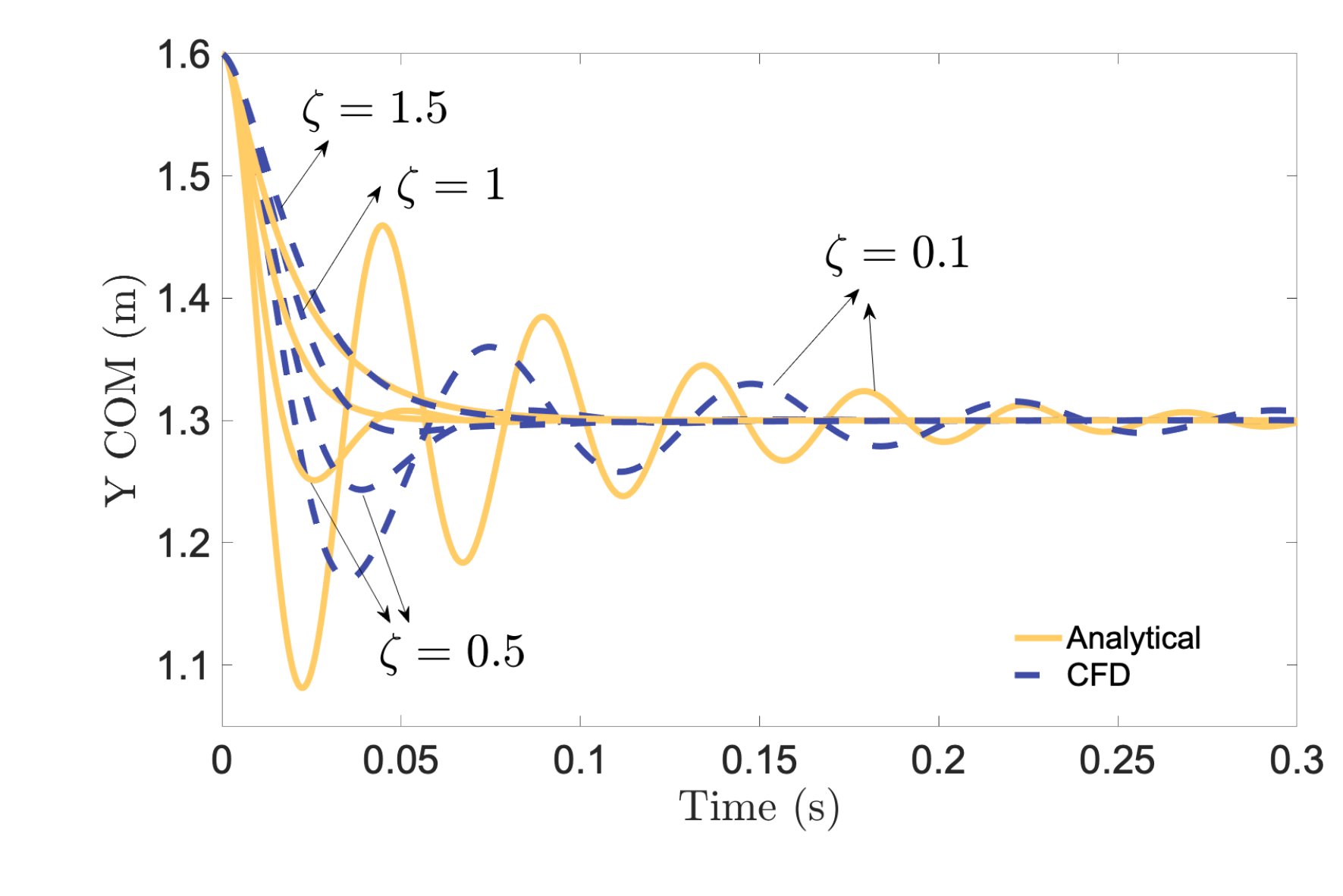}
	\label{fig_decay_results_R0pt8}
    }    
  \caption{ Center of mass vertical position as a function of time for various values of damping ratio $\zeta$ with~\subref{fig_decay_results_R2}  $m^* = 2$ and~\subref{fig_decay_results_R0pt8} $m^* = 0.8$.  A medium grid resolution of $100 \times 200$ is used for these cases. (----, solid yellow line) Analytical;  (\texttt{---}, dashed blue line) CFD model. 
        \label{fig_decay_results_R20pt8}
  }
\end{figure}

If the density of the solid is comparable to the surrounding fluid, the fluid forces are not negligible and they affect the motion of the 
solid, particularly at low damping ratios. We simulate the previous case with low density ratios of $m^*$ = 2 and 0.8. Fig.~\ref{fig_decay_results_R20pt8} compares the cylinder displacement with the analytical solution. As observed in Figs.~\ref{fig_decay_results_R2} and \ref{fig_decay_results_R0pt8}, the CFD solution deviates significantly from the analytical solution for $\zeta = $ 0.1 and 0.5 with $m^* = 2$, and for $\zeta  = $ 0.1, 0.5, and 1 for $m^* = 0.8$, respectively. This deviation is expected physically for these cases, however. \\

The results in this section show that the fully-resolved FD/BP model can accurately capture the rigid body dynamics of a system in 
which an external forcing is provided by mechanical springs and dampers. The method remains stable for large density ratio between different phases and it does not suffer from the \emph{added mass} effects that become prominent when $m^* \lessapprox 1$~\cite{Nobile99,Causin05,Burman09,Banks17,Serino19}. \REVIEW{ We also refer the readers to our prior work on FD/BP method~\cite{BhallaBP2019} that considers additional low-density ratio cases and validates them against  numerical and experimental results from the literature.}

\subsubsection{Motion under wave excitation} \label{sec_heave_gridconverge}

As a next validation case of our fully-resolved WSI model, we consider the heave motion of a submerged cylindrical buoy induced by 
the incoming water waves. This case is close to the actual problem of interest and serves to provide 
an estimate of mesh resolution needed to resolve the WSI dynamics adequately. First we describe the problem setup.    

\begin{figure}[]
  \centering
    \includegraphics[scale = 0.28]{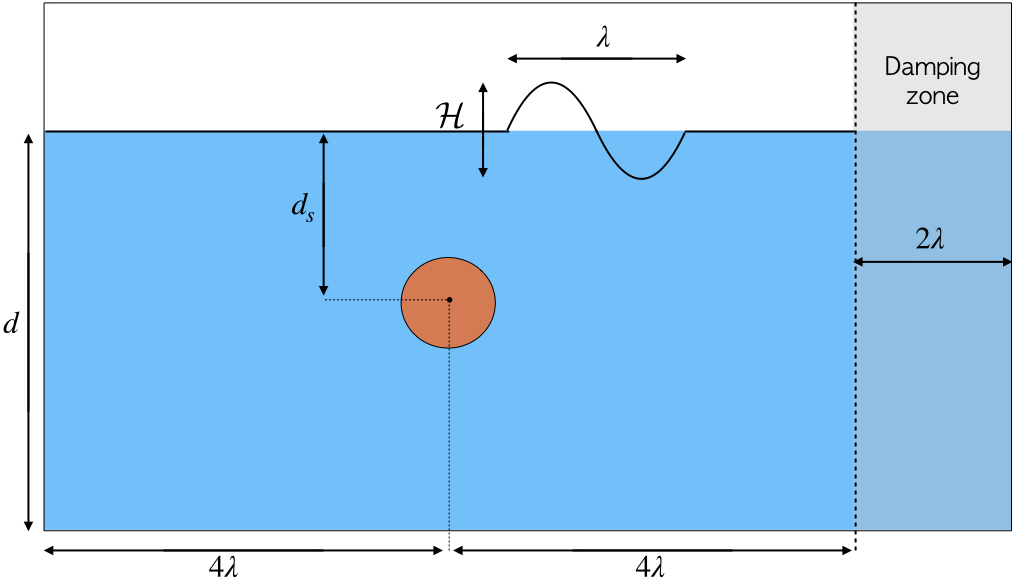}
  \caption{Schematic representation of a submerged point absorber in a numerical wave tank. Blue shade represents the water phase, whereas
   the air phase atop the water surface is represented by white shade. 
}
  \label{fig_nwt}
\end{figure}

\begin{figure}[]
  \centering
    \includegraphics[scale = 0.32]{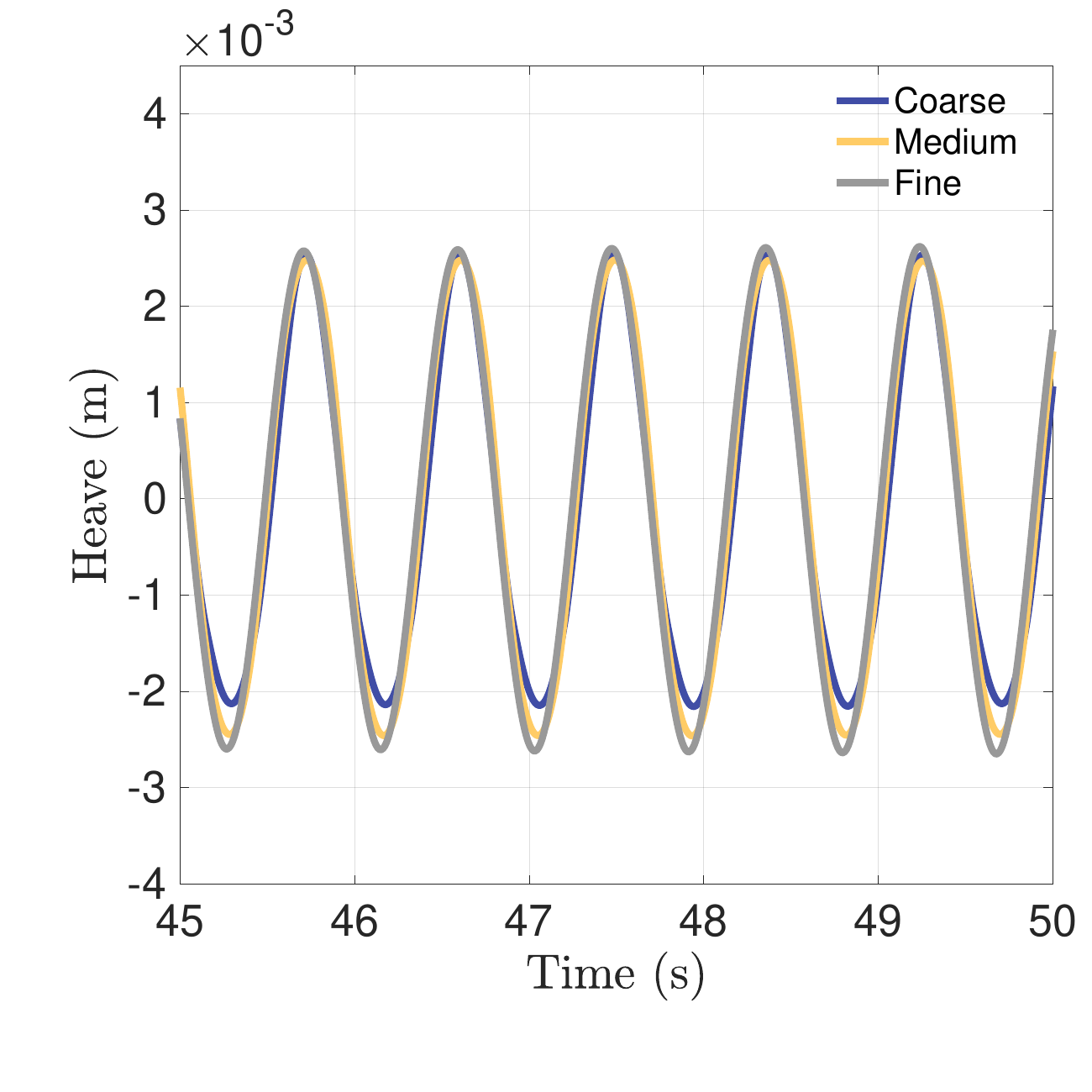}
  \caption{Heave dynamics of a submerged cylindrical buoy of $D = 0.16$ m using coarse (blue), medium (yellow), and fine (grey) grid resolutions with regular wave 
  of $\mathcal{H} = 0.01$ m, $\lambda = 1.216$ m,  $\mathcal{T} = 0.8838$ s, $d_{\text{s}} = 0.25$ m, and $d = 0.65$ m. 
}
  \label{fig_heave_resolution}
\end{figure}

Fig.~\ref{fig_nwt} shows the schematic of the problem and the numerical wave tank (NWT) layout. Water waves of height $\mathcal{H}$, 
wavelength $\lambda$, and time period $\mathcal{T}$ are generated at the left end of the domain using fifth-order Stokes wave theory~\cite{Fenton1985}. The waves propagate a distance of $8 \lambda$ in the positive $x$-direction before getting attenuated 
in the damping zone of length $2 \lambda$ located towards the right end of the domain. The submerged point absorber is placed midway 
at a distance of $4 \lambda$ in the working zone of NWT. The depth of submergence of the point absorber is $d_{\text{s}}$, whereas the mean 
depth of water in the NWT is $d$. Regular water waves of $\mathcal{H} = 0.01$ m, 
$\lambda = 1.216$ m,  $\mathcal{T} = 0.8838$ s and $d = 0.65$ m are generated in the NWT. The simulated waves satisfy the 
dispersion relationship 
\begin{equation}
\lambda = \frac{g}{2\pi} \mathcal{T}^2 \tanh \left( \frac{2\pi d}{\lambda} \right), \label{eqn_dispersion}
\end{equation} 
and are in deep water regime $d/\lambda > 0.5$. 

The diameter of the buoy with relative density $\rho_{\text{s}}/\rho_{\text{w}} = 0.9$ is taken to be $D = 0.16$ m.
The PTO stiffness and damping coefficients are taken as $k_{\text{PTO}} = 1995.2$ N/m and 
$b_{\text{PTO}} = 80.64$ N$\cdot$s/m, respectively. The initial submergence depth of the disk is $d_{\text{s}} = 0.25$ m. 
Three grid resolutions corresponding to coarse, medium, and fine grids are used to simulate the heave dynamics 
of the buoy. The  spatiotemporal resolutions for the three simulations are tabulated in Table~\ref{tab_grid_resolution}. 
Fig.~\ref{fig_heave_resolution} compares the vertical center of mass displacement (heave motion) as a function of time for the 
three grid resolutions. The results suggest that the medium grid resolution \REVIEW{with approximately 20 cells per diameter of the buoy, 150 cells per wavelength, and 5 cells per wave height} is sufficient to resolve the WSI dynamics adequately. \REVIEW{We use this resolution to simulate the rest of the CFD cases considered in this paper.}

\begin{table}[]
    \caption{Grid resolution to simulate the heave dynamics of a submerged point absorber.}
    \centering
    \scalebox{1.1}
    {
        \begin{tabular}{|c|c|c|c|}
         \hline
         Resolution & Coarse & Medium & Fine \\
         \hline
         $N_y$ & 125& 250 &  400 \\
          \hline
          $N_x$ & 750 & 1500 &  2400 \\
          \hline
          $\Delta t$ (s) & $10^{-3}$  & $5 \times 10^{-4}$  & $2.5 \times 10^{-4}$ \\
          \hline
          \end{tabular} 
      }
      \label{tab_grid_resolution}
\end{table}

%%%%%%%%%%%%%%%%%%%%%%%%%%%%%%
\section{Comparison between Cummins and CFD models} \label{sec_model_compare}

In this section we compare the simulated dynamics of a submerged cylindrical point absorber using Cummins equation-based Simulink model
and fully-resolved CFD model implemented in IBAMR. The dynamics of the buoy and the generated PTO power at steady-state 
are compared using the two models. We take the same buoy setup, \REVIEW{PTO coefficients, and wave characteristics} of the previous section~\ref{sec_heave_gridconverge}
to compare the dynamics. Since Cummins equation is based on linear wave theory, the wave parameters should correspond to
first-order Stokes/Airy wave theory. Using the wave classification phase space described by Le M\'ehaut\'e~\cite{LeMehaute2013},
it can be verified that wave parameters of Sec.~\ref{sec_heave_gridconverge} closely correspond to the linear wave theory. Next, we compare 
the dynamic response of the buoy in various degrees of freedom using the two models to assess the performance of Cummins equation-based models
for practical wave energy conversion applications.

\subsection{One degree of freedom}

\begin{figure}[]
  \centering
  \subfigure[Heave dynamics of a 1-DOF buoy]{
    \includegraphics[scale = 0.32]{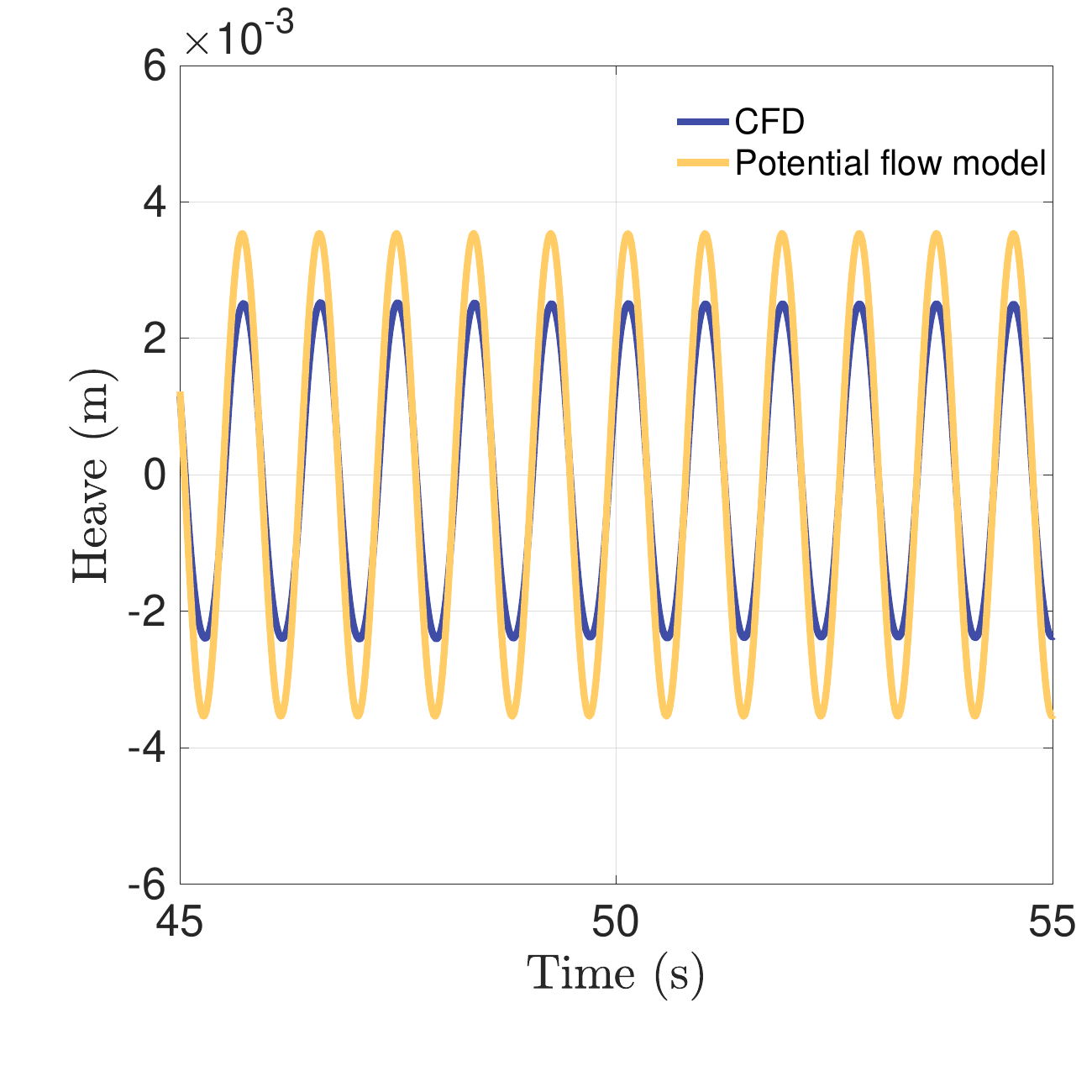}. 
    \label{fig_heave_1dof}
  }
   \subfigure[Generated PTO power of a 1-DOF buoy] {
    \includegraphics[scale = 0.32]{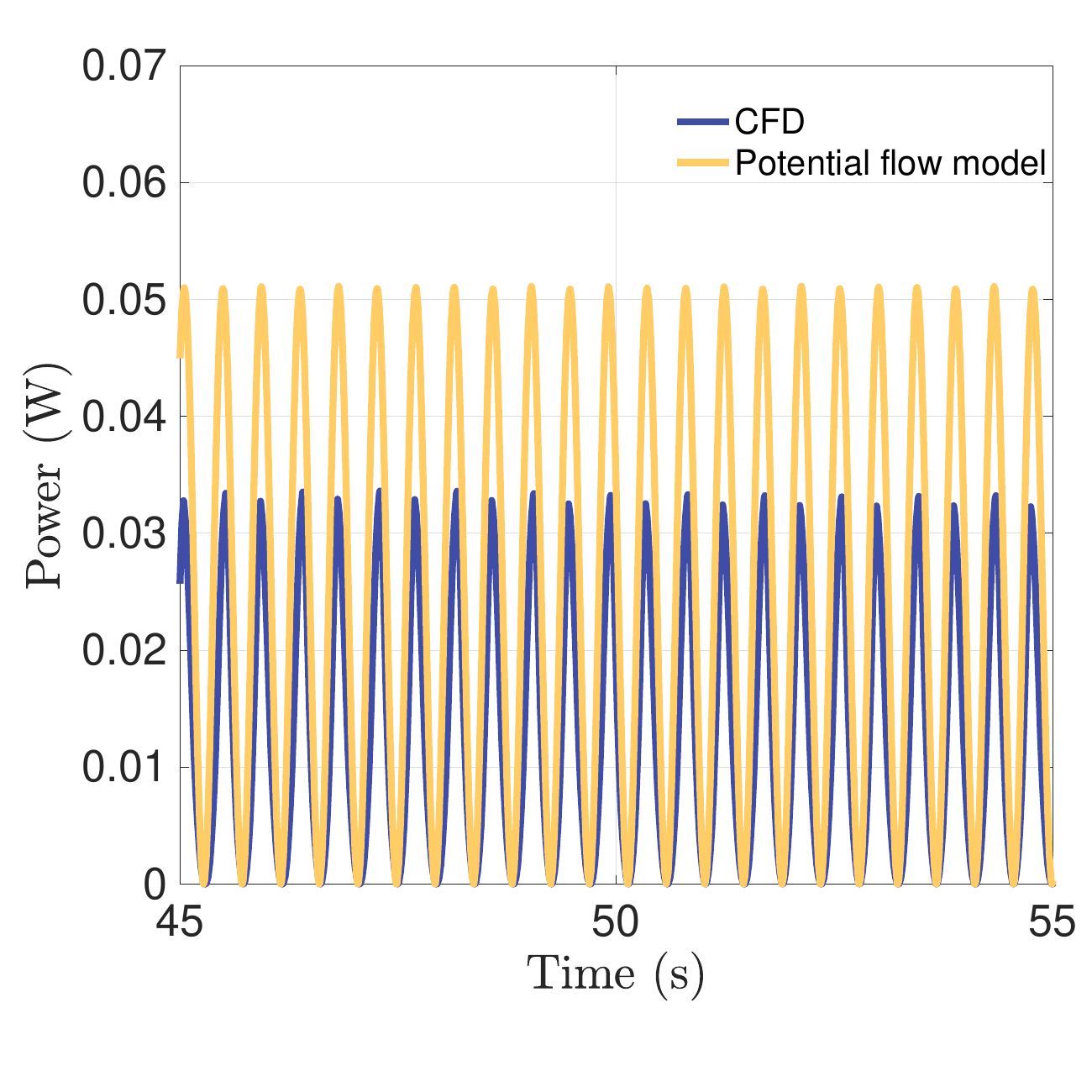}
    \label{fig_power_1dof}
  }
  \caption{Comparison of rigid body dynamics of a 1-DOF buoy simulated using potential flow and CFD models.~\subref{fig_heave_1dof} Heave dynamics.~\subref{fig_power_1dof} Generated PTO power.  
  }
  \label{fig_1dof_comparison}
\end{figure}

We first simulate one degree of freedom (1-DOF) buoy using the two models. 
Fig.~\ref{fig_1dof_comparison} compares the heave dynamics of 
the submerged buoy and the generated PTO power during steady-state.  As observed in Fig.~\ref{fig_heave_1dof}, 
Cummins model over-predicts the heave amplitude due to the fact that linear potential theory over-estimates the 
Froude-Krylov forces or the wave excitation loads on the submerged buoy~\cite{Penalba2017,Yu2013,Anbarsooz2014}. The generated 
PTO power is also higher using the Cummins model as observed in Fig.~\ref{fig_power_1dof}. \REVIEW{Nevertheless, 
the  dynamics obtained using the two models are qualitatively the same.} 
%and the Cummins equation-based model can be reliably used to 
%simulate heaving dynamics of the 1-DOF buoy.

\subsection{Two degrees of freedom}
\begin{figure}[]
  \centering
  \subfigure[Heave dynamics of a 2-DOF buoy]{
    \includegraphics[scale = 0.32]{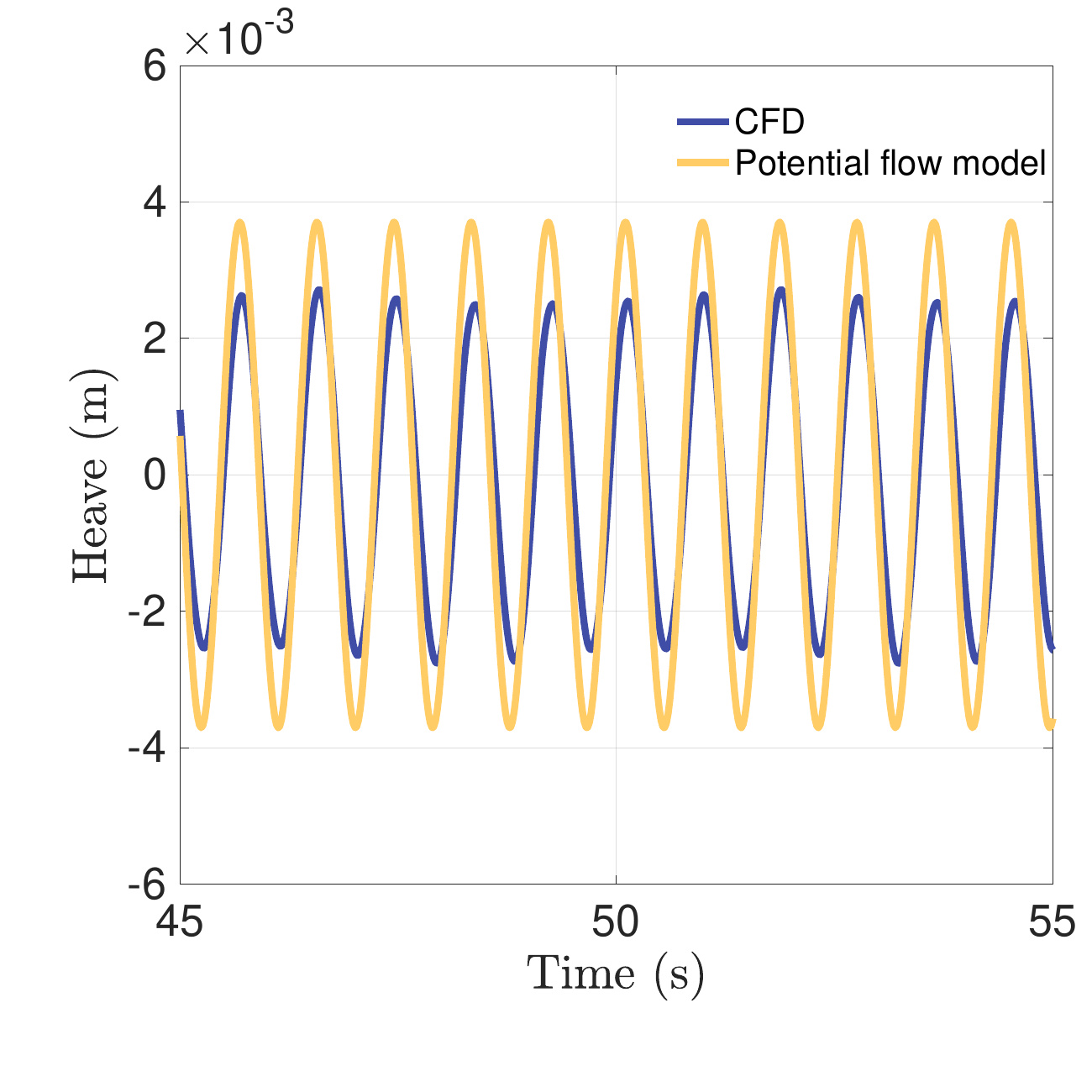} 
    \label{fig_heave_2dof}
  }
   \subfigure[Generated PTO power of a 2-DOF buoy]{
    \includegraphics[scale = 0.32]{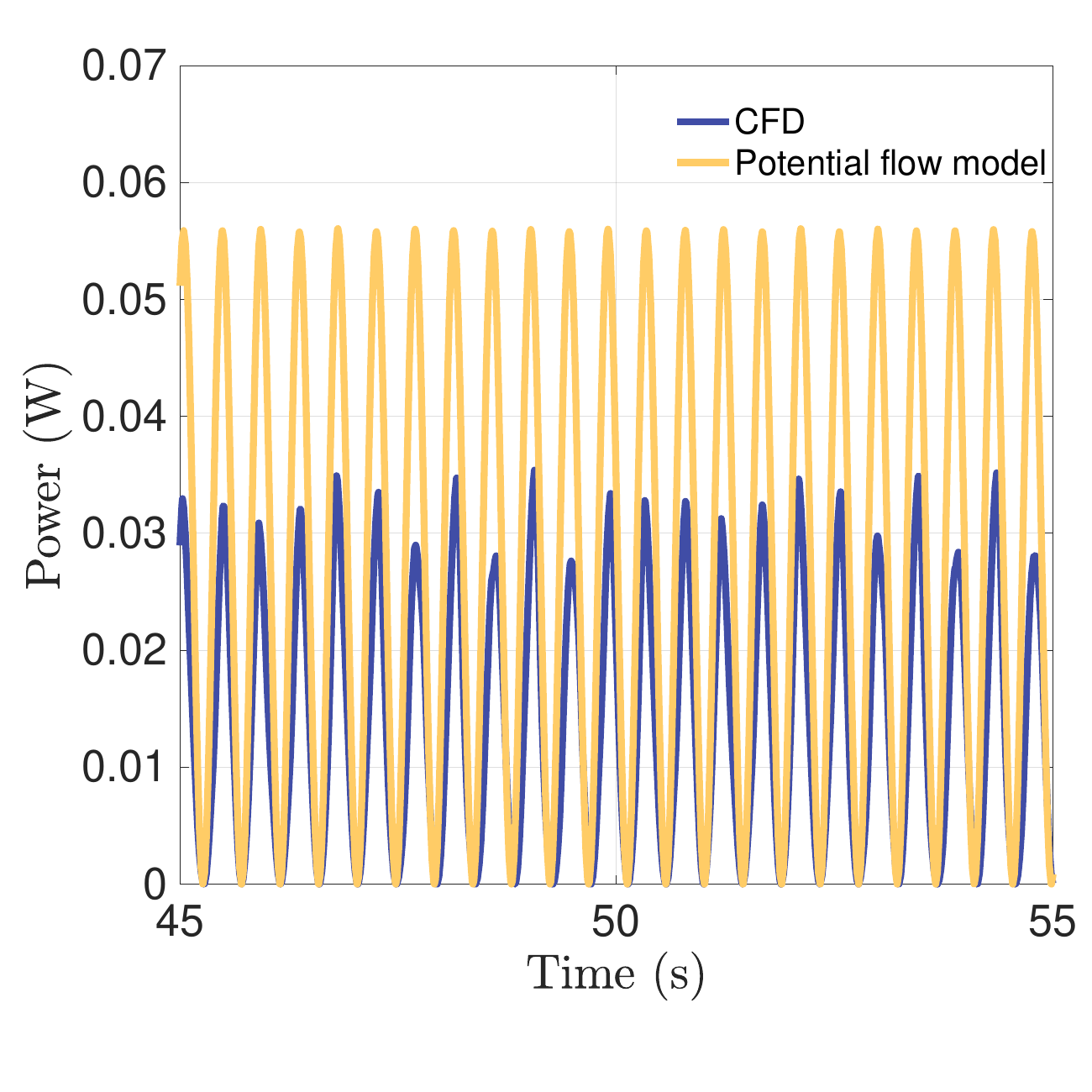}
    \label{fig_power_2dof}
  }
  \subfigure[Surge dynamics of a 2-DOF buoy]{
    \includegraphics[scale = 0.32]{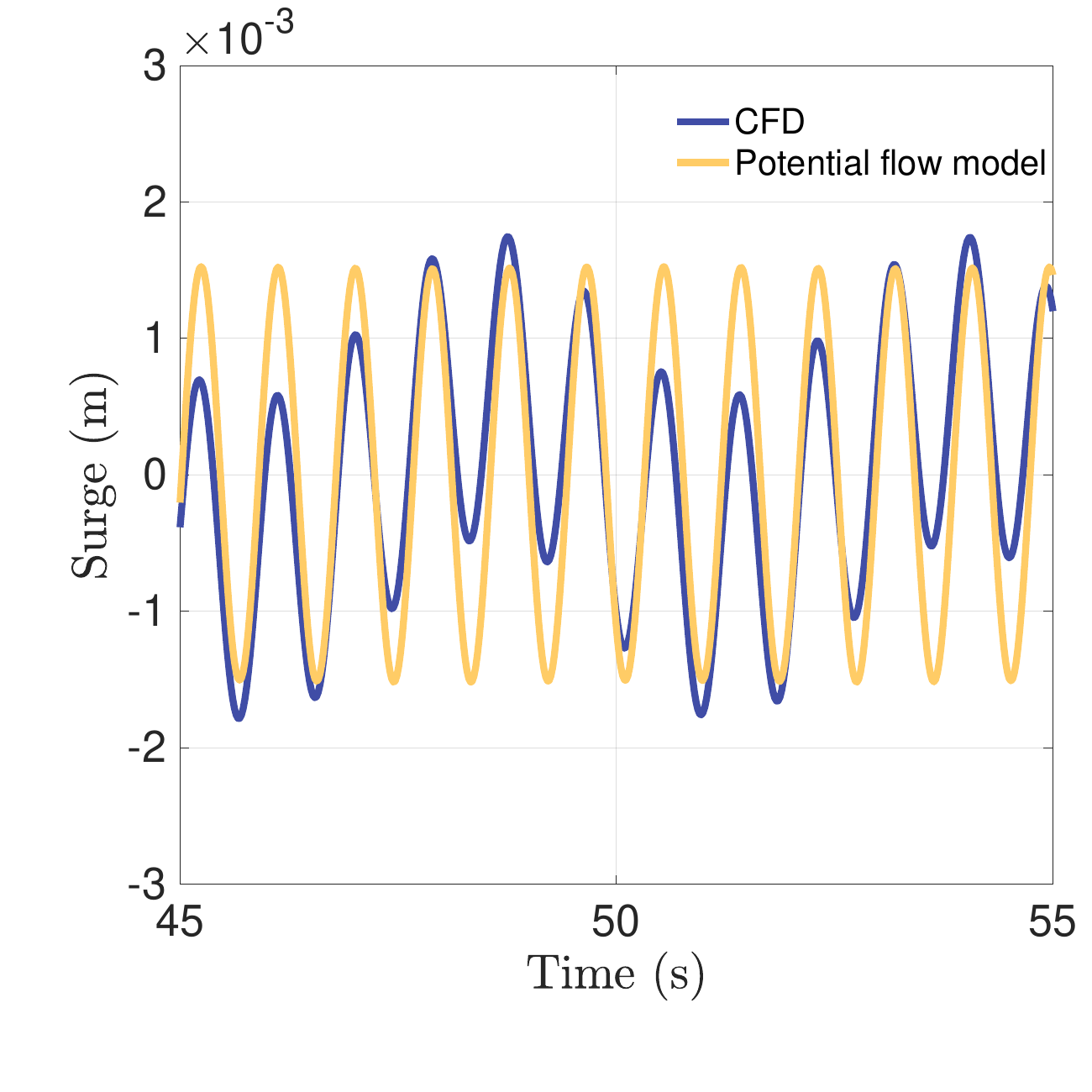} 
    \label{fig_surge_2dof}
  }
   \subfigure[Surge dynamics of a 2-DOF without drift phenomenon]{
    \includegraphics[scale = 0.32]{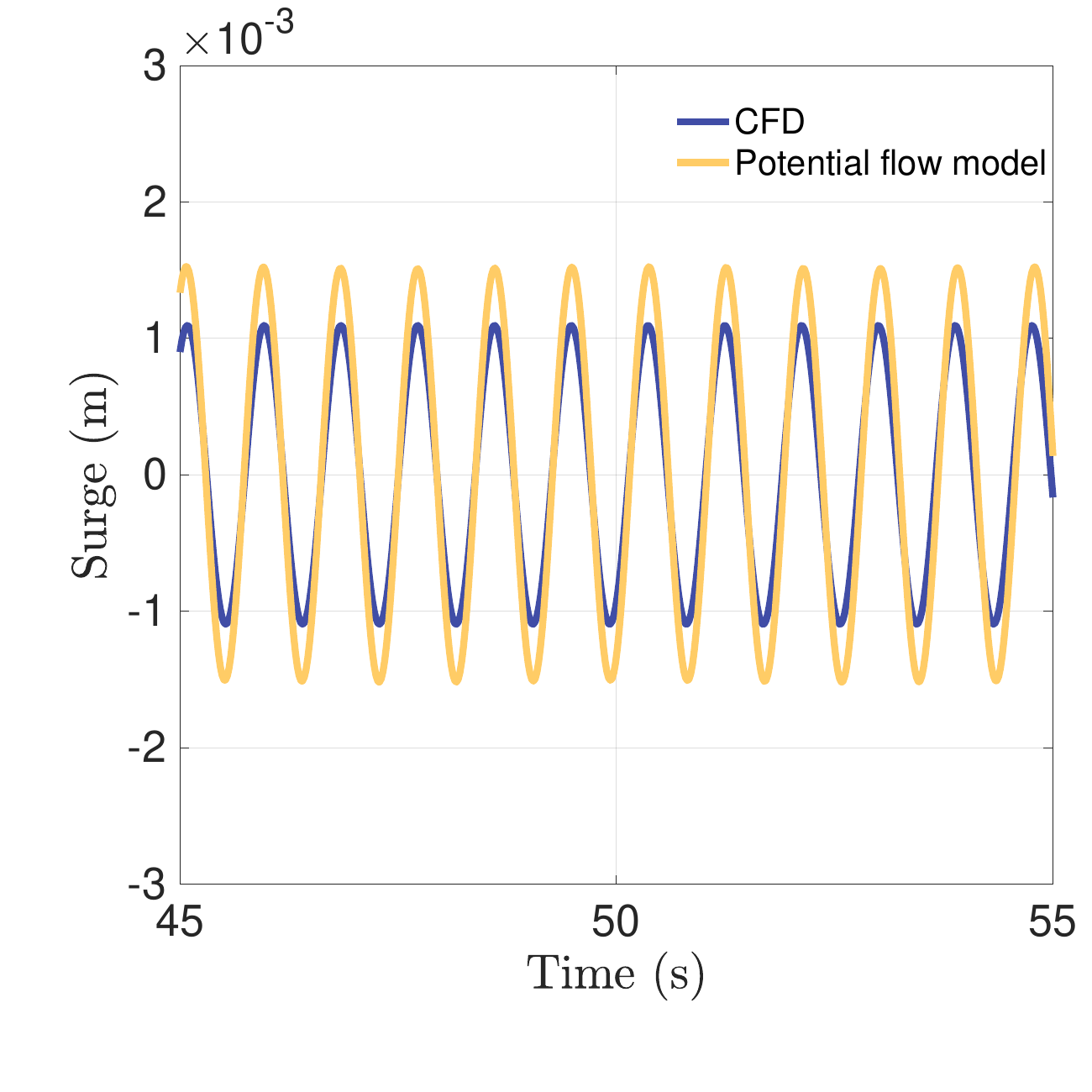}
    \label{fig_surge_2dof_nodrift}
  }
  \caption{Comparison of rigid body dynamics of a 2-DOF buoy simulated using potential flow and CFD models.~\subref{fig_heave_2dof} Heave dynamics.~\subref{fig_power_2dof} Generated PTO power.~\subref{fig_surge_2dof} Surge dynamics.~\subref{fig_surge_2dof_nodrift} 
  Surge dynamics without the slow drift.   
  }
  \label{fig_2dof_comparison}
\end{figure}

\REVIEW{Next, we simulate a 2-DOF buoy that oscillates in heave and surge directions using the two models.} Fig.~\ref{fig_2dof_comparison} compares the 
heave and surge dynamics of the submerged buoy and the generated PTO power during steady-state. The amplitude of 
the heave motion remains almost the same as that of 1-DOF buoy. \REVIEW{The amplitude of surge is lower than heave, 
which can be observed in Fig.~\ref{fig_surge_2dof_nodrift}}. However, both models estimate slightly higher PTO power \REVIEW{(7 - 10 \% more for this case)}  compared 
to the 1-DOF buoy as shown in Fig.~\ref{fig_power_2dof}.  This is in agreement with Falnes~\cite{Falnes2002}, who also predicts 
an increase in power absorption efficiency of a point absorber with more than one degree of freedom using theoretical analyses.

Notice that for the 2-DOF buoy, the fully-resolved WSI model is able to capture the slow drift 
phenomenon induced by the background flow. On the contrary, the potential flow model is unable to capture the slow drift of the buoy, as 
evidenced in Fig.~\ref{fig_surge_2dof}. To compare the amplitude of the high-frequency surge motion, 
Fig.~\ref{fig_surge_2dof_nodrift} is plotted by eliminating the slow drift frequency from the surge dynamics 
obtained from CFD using a high-pass filter. Similar to the heave dynamics, the potential flow model over-predicts the 
 surge amplitude. The inability of the Cummins model to capture the slow drift phenomenon can be 
explained as follows. According to Chakrabarti~\cite{Chakrabarti1984}, the drift forces estimated by the potential flow theory tend to zero 
for  $ k a  < 0.5$, in which  $k = 2 \pi / \lambda$ is the wave number and  $a = \mathcal{H}/2$ is the wave 
amplitude. In the present simulation, $k a   = 0.0258$, and consequently, the Cummins equation-based model under-predicts the drift force.
We remark that estimating drift forces is an important criteria for designing a robust mooring system for a 
practical WEC~\cite{Faizal2014}. Therefore, fully-resolved simulations are more reliable for such calculations. 
 
\subsection{Three degrees of freedom}
\begin{figure}[]
  \centering
  \subfigure[Heave dynamics of a 3-DOF buoy]{
    \includegraphics[scale = 0.32]{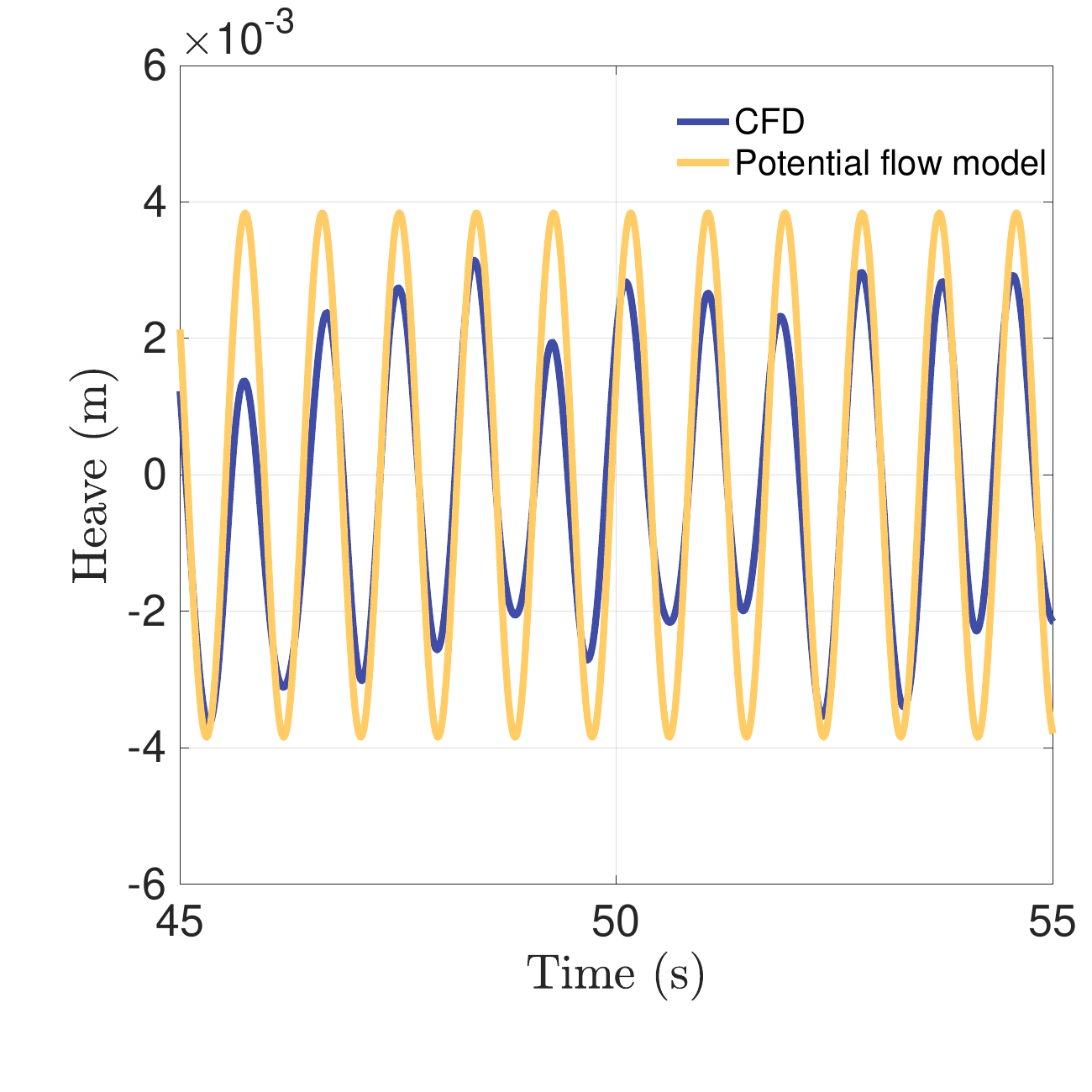} 
    \label{fig_heave_3dof}
    }
   \subfigure[Surge dynamics of a 3-DOF buoy]{
    \includegraphics[scale = 0.32]{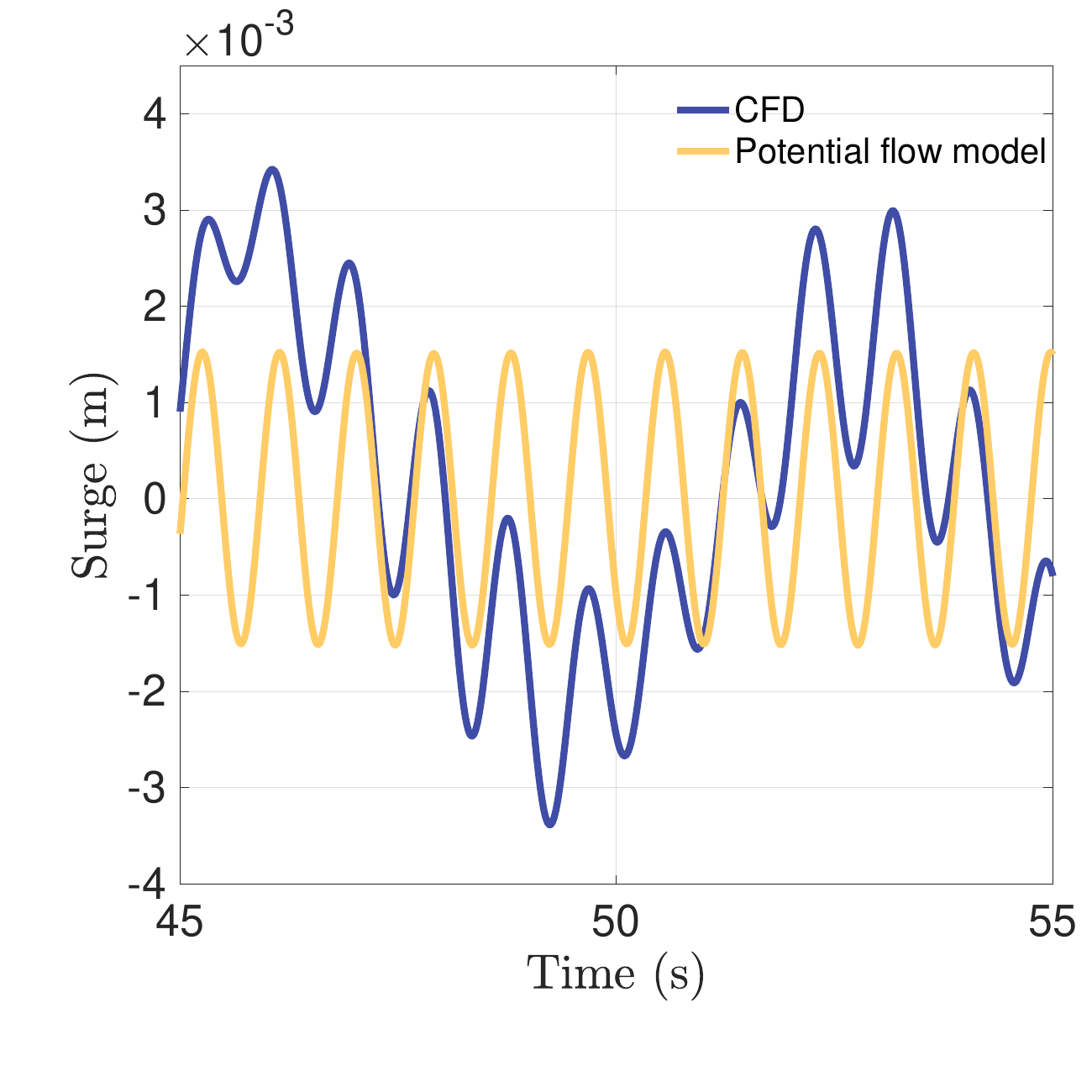}
    \label{fig_surge_3dof}
  }
    \subfigure[Pitch dynamics of a 3-DOF buoy]{
    \includegraphics[scale = 0.32]{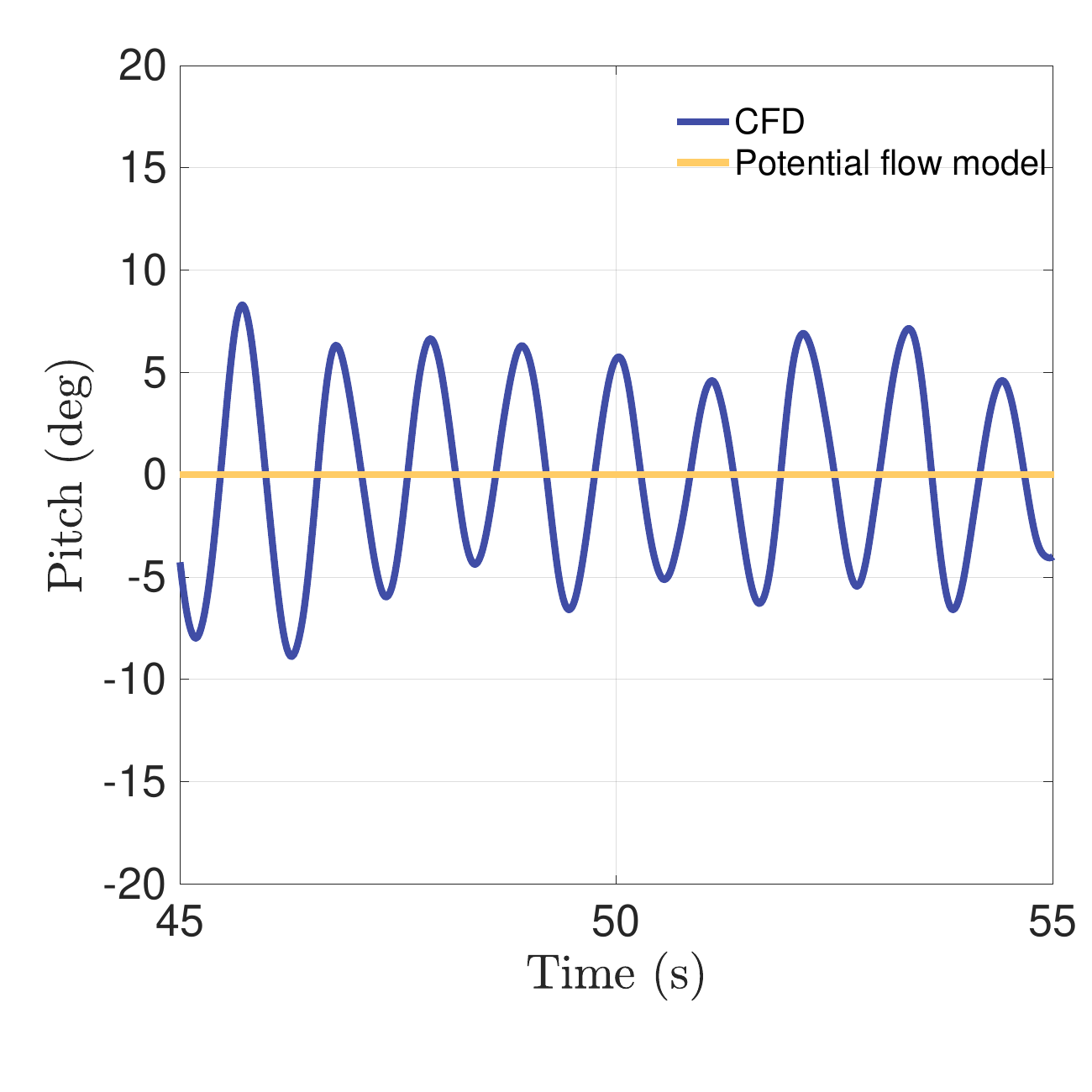} 
    \label{fig_pitch_3dof}
  }
    \caption{Comparison of rigid body dynamics of a 3-DOF buoy simulated using potential flow and CFD models.~\subref{fig_heave_3dof} Heave dynamics.~\subref{fig_surge_3dof} Surge dynamics.~\subref{fig_pitch_3dof} Pitch dynamics.  
  }
 \label{fig_3dof_comparison} 
\end{figure}

\begin{figure}[]
  \centering
  \subfigure[Complete trajectory]{
    \includegraphics[scale = 0.32]{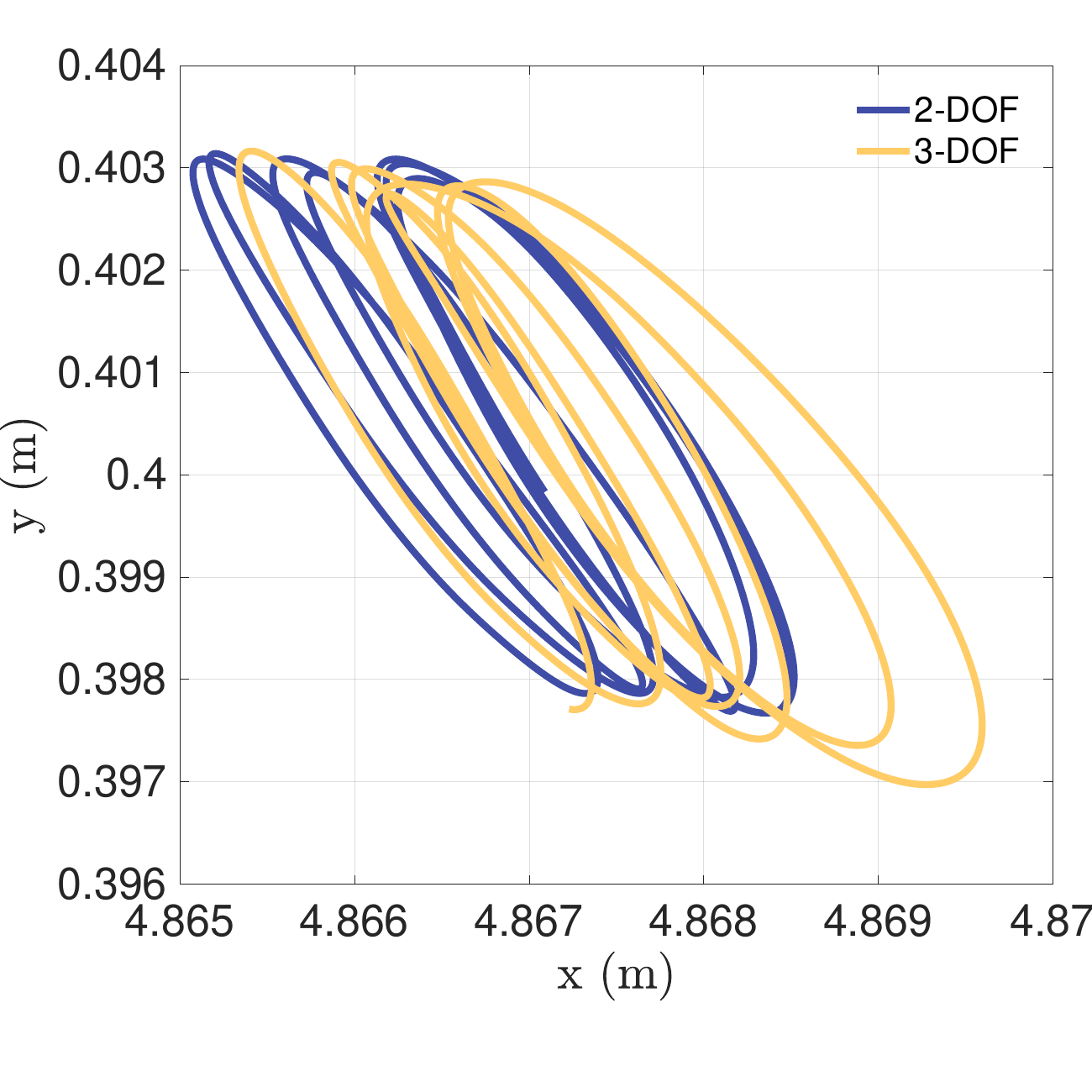}. 
    \label{fig_com_traj}
  }
   \subfigure[Trajectory without slow drift]{
    \includegraphics[scale = 0.32]{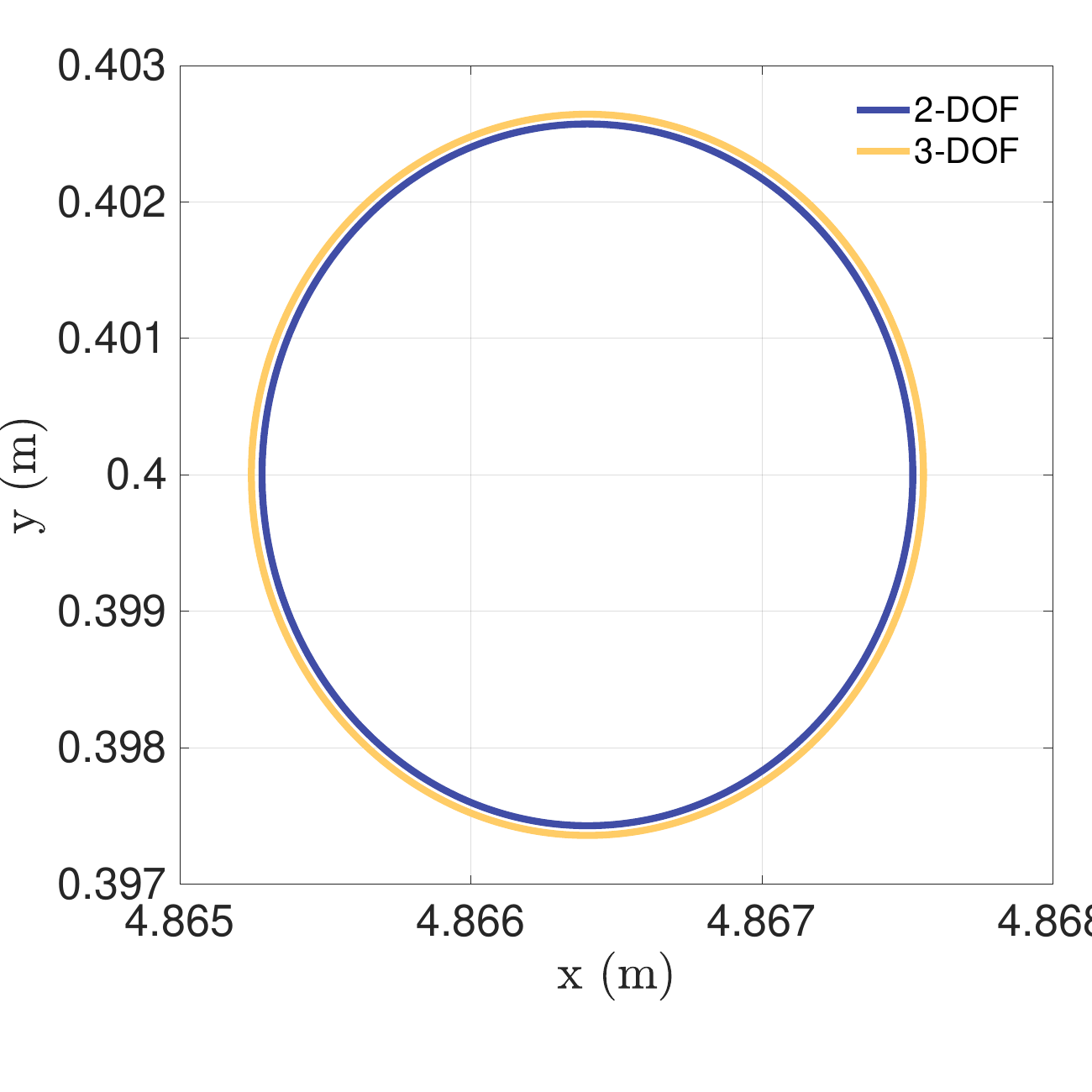}
    \label{fig_com_traj_high_freq}
  }
  \caption{\REVIEW{Center of mass point trajectory of 2- and 3-DOF buoys traced between the time interval $64 - 70$ s.~\subref{fig_com_traj} Complete trajectory.~\subref{fig_com_traj_high_freq} Trajectory without slow drift.}   
  }
  \label{fig_com_2_3dof}
\end{figure}

Lastly, we analyze the 3-DOF buoy that can undergo heave, surge, and pitching motions. Fig.~\ref{fig_3dof_comparison} shows the heave, surge, and pitch dynamics of the buoy. Compared to the 2-DOF buoy dynamics, the heave 
and surge dynamics (Fig.~\ref{fig_heave_3dof} and Fig.~\ref{fig_surge_3dof}, respectively) show a larger oscillatory behavior 
when pitch is included in the WSI model. This can be further confirmed from Fig.~\ref{fig_com_2_3dof} which shows the trajectory traced by the
center of mass points of 2- and 3-DOF buoys. Fig.~\ref{fig_com_traj} shows the complete trajectory of the center of mass point, including the slow drift,  
traced during few steady-state periods of motion ($64 - 70$ s). After eliminating the slow drift from the complete trajectory using a high-pass filter, 
Fig.~\ref{fig_com_traj_high_freq} shows a larger trajectory for the 3-DOF buoy. The buoy 
undergoes an average rotation of $5^{\circ}$ at steady-state, which can be observed
in Fig.~\ref{fig_pitch_3dof}. In contrast, the potential flow model results in an insignificant pitch. This can be explained as follows. 
For a fully-submerged axisymmetric body, the wave excitation forces 
due to the induced fluid pressure do not produce any rotational torque, as the pressure forces pass through the center of mass 
of the buoy. Moreover, the hydrodynamic coupling between various degrees of freedom is also weak when linear potential flow equations are considered~\cite{schubert2020linear}. Consequently, the resulting pitch velocity $\dot{\theta}$, and the viscous drag torque $\M_{\text{drag},\theta}$ are negligible. 
The simulated dynamics of a 3-DOF buoy using the potential flow model is therefore quite similar to the prior 2-DOF buoy case. 

\begin{figure}[]
  \centering 
   \subfigure[]{
    \includegraphics[scale = 0.32]{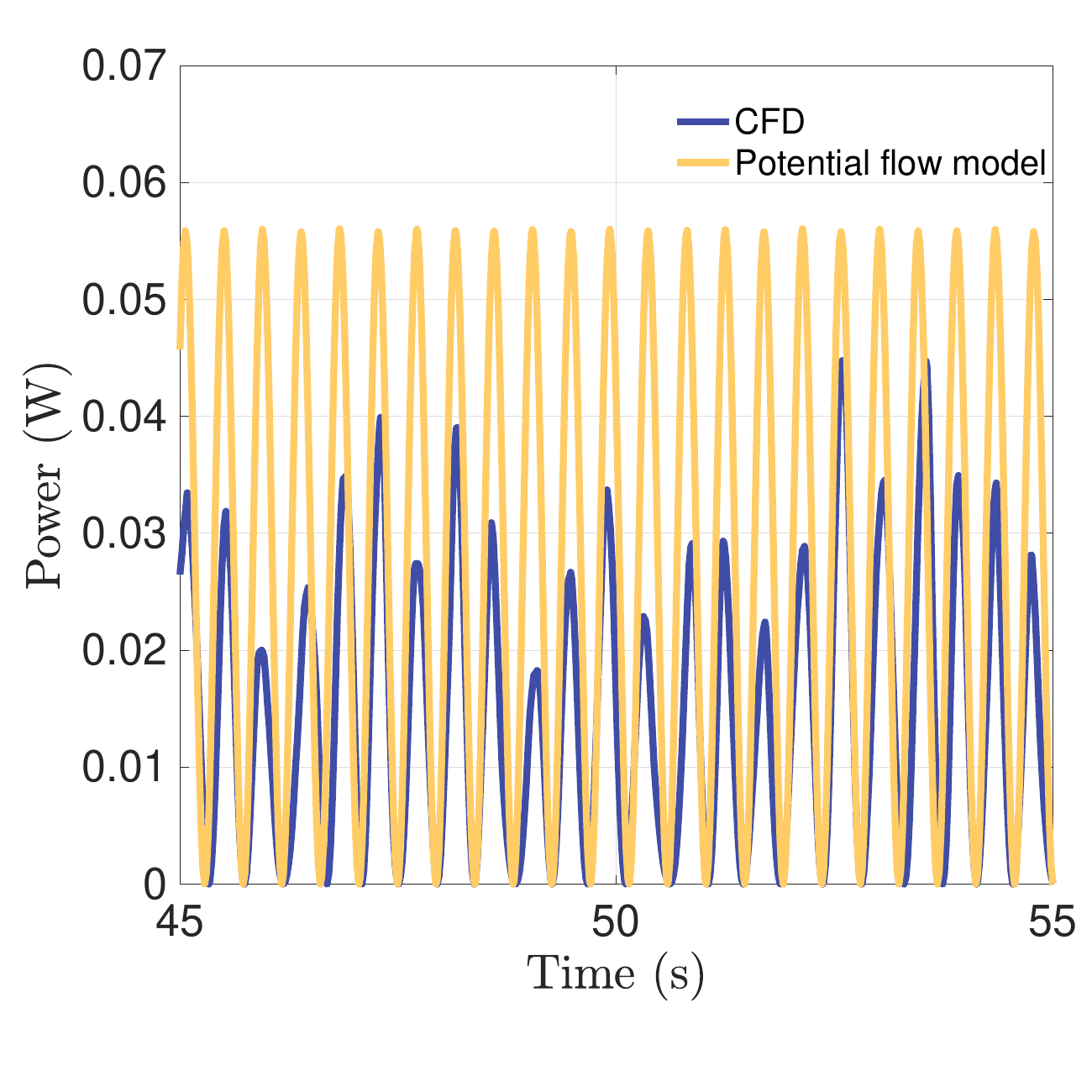}
    \label{fig_power_3dof}
  }
  \subfigure[]{
    \includegraphics[scale = 0.32]{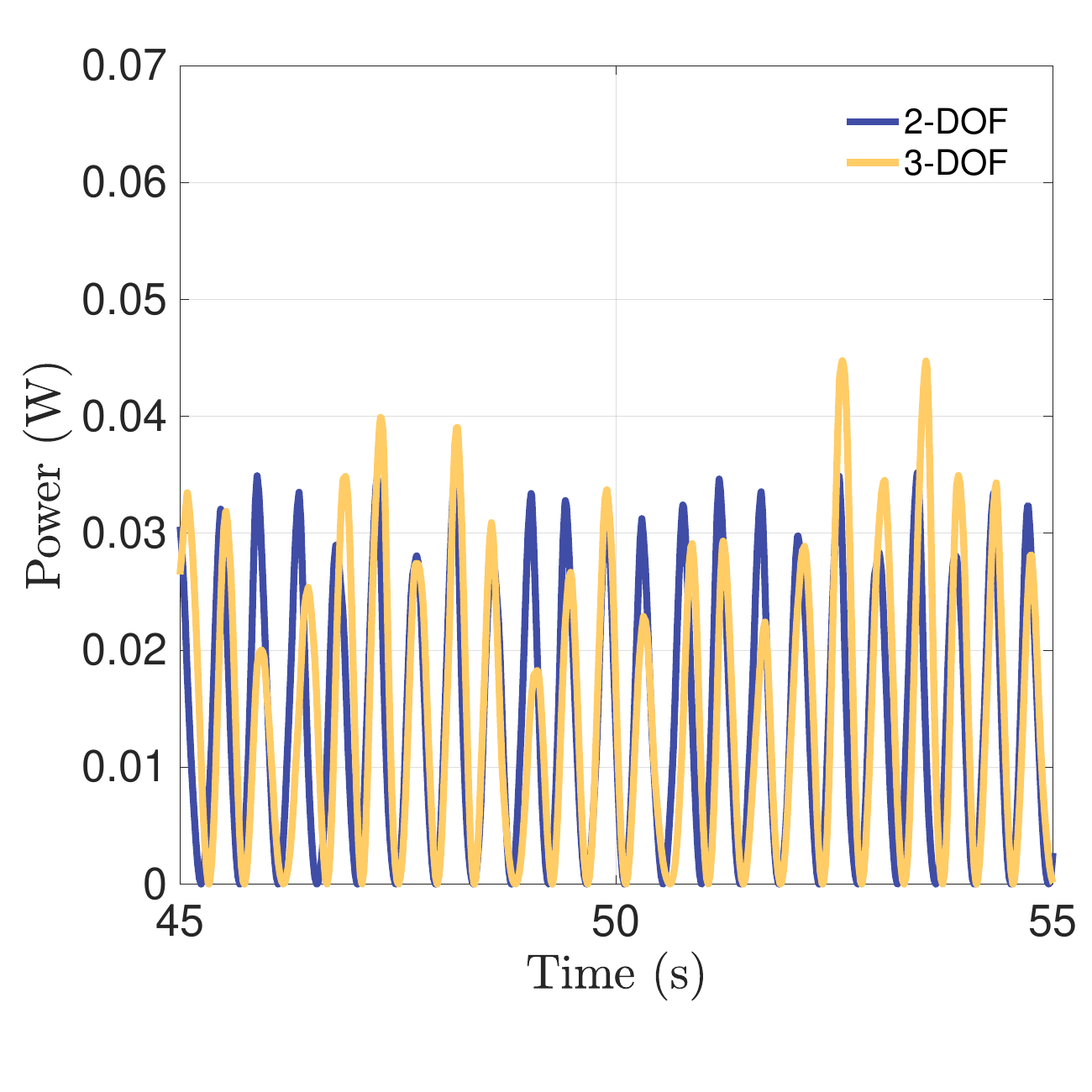}
    \label{fig_power_3_2dof}
    }
  \caption{\REVIEW{\subref{fig_power_3dof} Generated PTO power of a 3-DOF buoy using potential flow and CFD models.~\subref{fig_power_3_2dof} Power comparison with a 2-DOF buoy.}   
  }
  \label{fig_power_3dof} 
\end{figure}

Fig.~\ref{fig_power_3dof} shows the power production of the 3-DOF buoy and 
contrasts it with  2-DOF PTO power generation. As observed in Fig.~\ref{fig_power_3_2dof}, the pitching motion does 
not contribute substantially to the power generation compared to heave and surge motions for the simulated wave parameters. 
However, for higher wave heights and hence more energetic waves, power contribution from pitching motion could in general 
be significant.  Moreover, for point absorbers like ISWEC, the pitch motion of the hull is the primary source of power production. 
Therefore, fully-resolved WSI framework can be reliably used to resolve all modes of motion of a PA.

\subsection{Vortex shedding}

\begin{figure}[]
  \centering 
   \subfigure[1-DOF, $t = 30$ s]{
    \includegraphics[scale = 0.38]{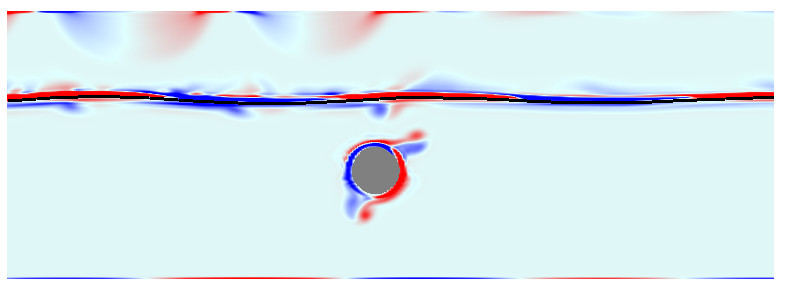}
    \label{fig_vortices_1dof_30s}
  }
 
   \subfigure[1-DOF, $t = 60$ s]{
    \includegraphics[scale = 0.38]{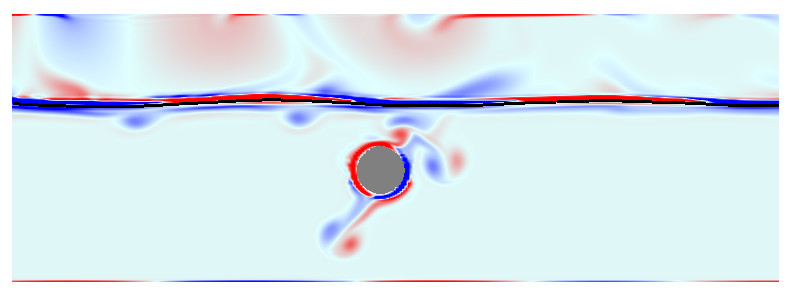}
    \label{fig_vortices_1dof_60s}
  }
  
   \subfigure[2-DOF, $t = 30$ s]{
    \includegraphics[scale = 0.38]{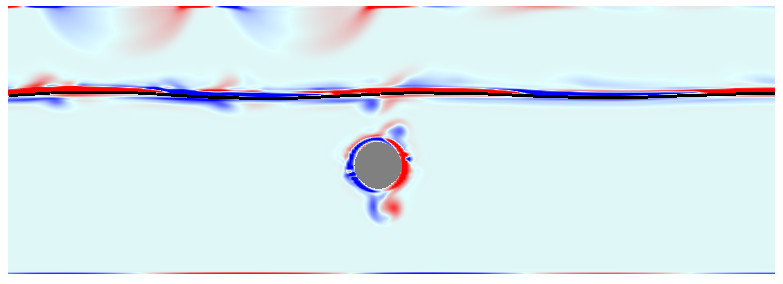}
    \label{fig_vortices_2dof_30s}
    }
  
  \subfigure[2-DOF, $t = 60$ s]{
    \includegraphics[scale = 0.38]{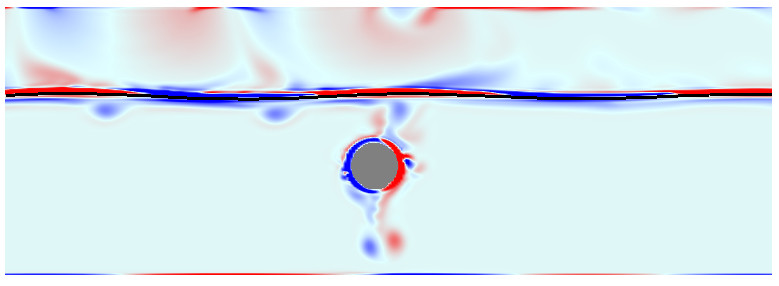}
    \label{fig_vortices_2dof_60s}
    }
  \caption{Vorticity generated due to WSI of 1-DOF (\subref{fig_vortices_1dof_30s} and~\subref{fig_vortices_1dof_60s}) and 2-DOF (\subref{fig_vortices_2dof_30s} and~\subref{fig_vortices_2dof_60s}) buoys at two different time instants. The plotted vorticity is in the range $-1.3$ to $1.3$ s$^{-1}$. 
  Fully-resolved WSI simulations are performed using $\mathcal{H} = 0.03$ m, $\mathcal{T} = 0.909$ m, $d_{\text{s}} = 0.25$ m, $d = 0.65$ m, $D = 0.16$ m, and $\rho_{\text{s}}/\rho_{\text{w}} = 0.9$.        
  }
  \label{fig_vorticity} 
\end{figure}

The prior sections highlighted the differences in the dynamic response of the buoy in one, two and three degrees of freedom. 
Here, we analyze the differences in vortex shedding pattern for a buoy having different degrees of freedom.  Fig.~\ref{fig_vorticity} shows 
the vorticity production and transport due WSI of 1- and 2-DOF buoys. The vortex production of a 3-DOF buoy is qualitatively 
similar to the 2-DOF buoy and is not shown.  

\begin{figure}[]
  \centering 
\subfigure[]{
  \includegraphics[scale = 0.28]{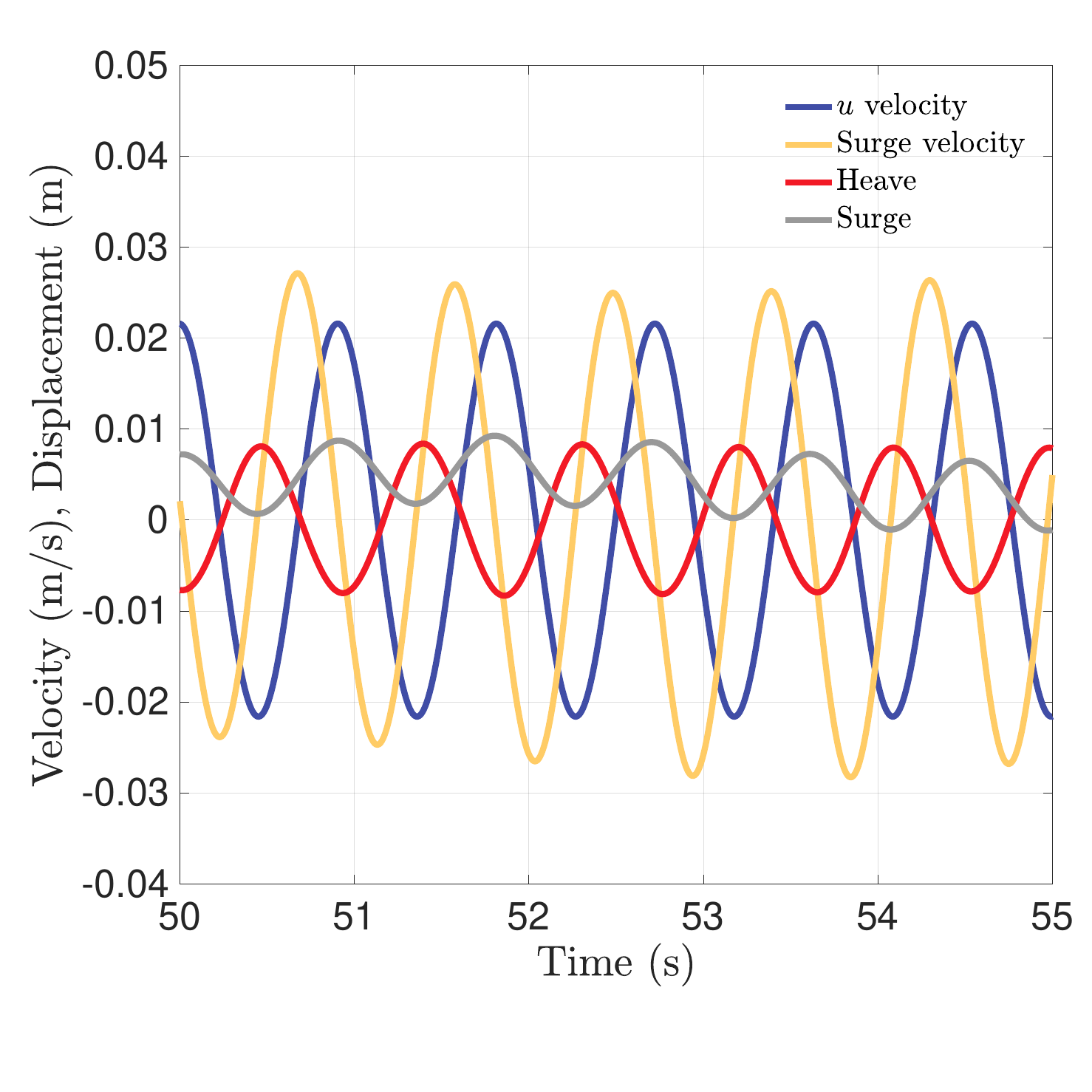}
  \label{fig_vorticity_surge_2dof}
}
\subfigure[]{
  \includegraphics[scale = 0.28]{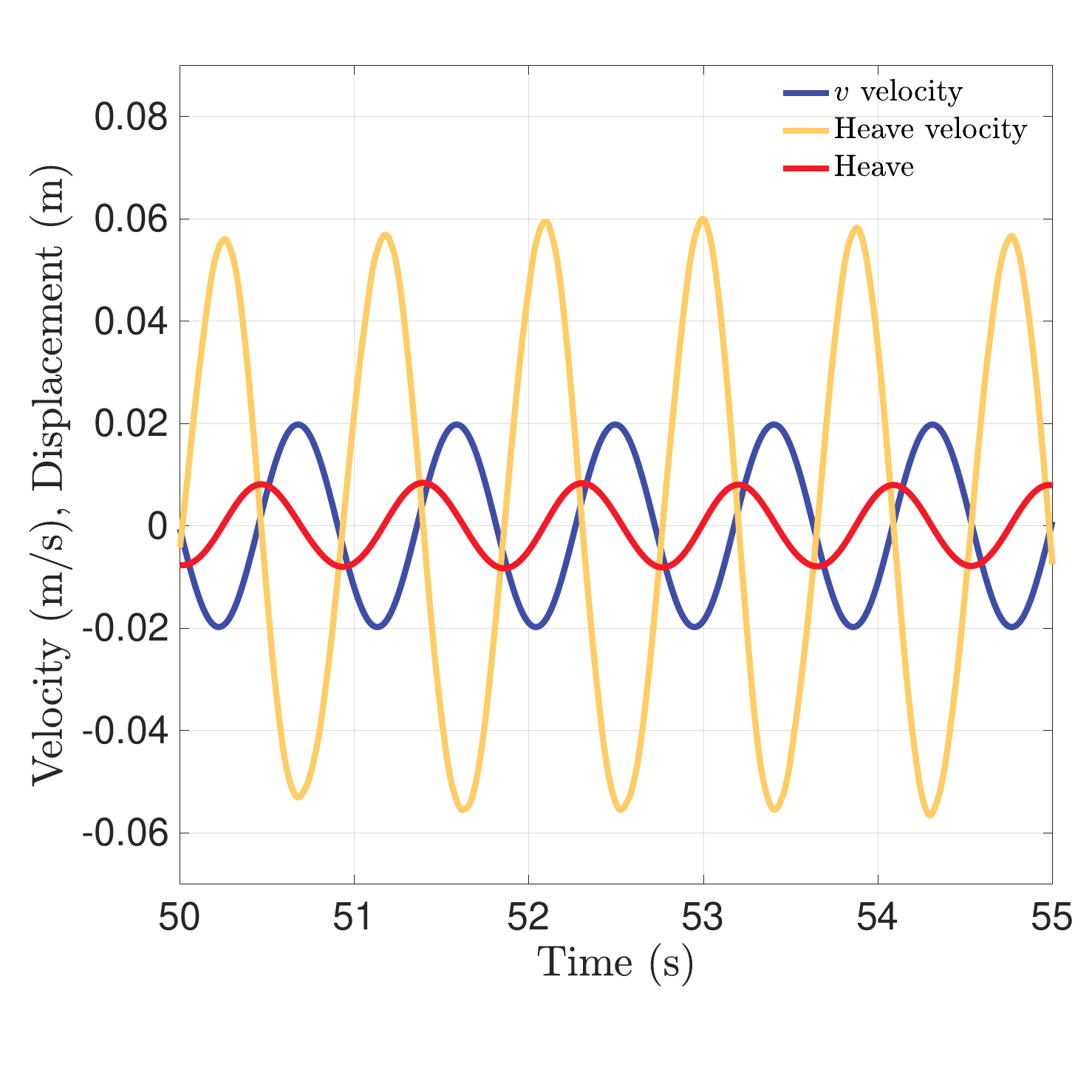}
    \label{fig_vorticity_heave_2dof}
}
  \caption{\REVIEW{\subref{fig_vorticity_surge_2dof} Undisturbed horizontal ($u$) flow velocity at the bottom location of the buoy, along with its heave and surge displacements, and  surge velocity. \subref{fig_vorticity_heave_2dof}  Undisturbed vertical ($v$) flow velocity at the bottom location of the buoy, along with its heave displacement and heave velocity. The bottom location of the buoy is considered at $y = -(d_{\rm s} + D/2)$, with $y = 0$ denoting the undisturbed free-water surface.} }
  \label{fig_vorticity_curves} 
\end{figure}

As observed in Fig.~\ref{fig_vorticity}, the 1-DOF buoy sheds vortices in an inclined direction, whereas the 2-DOF one sheds it in the vertical direction. This observation 
can be explained as follows. For the 1-DOF buoy, vortex structures generated at the surface of the buoy is transported by wave induced horizontal flow velocity $u$, 
along with a vertical flow produced by the heaving buoy. Therefore, the net transport of vorticity is in an inclined direction. \REVIEW{For the 2-DOF buoy, Fig.~\ref{fig_vorticity_surge_2dof} plots the undisturbed horizontal flow velocity $u$ at the bottom most point of the buoy, when the buoy is at its initial equilibrium position.  The heave and surge displacements, as well as the surge velocity of the buoy using the CFD model are also shown in Fig.~\ref{fig_vorticity_surge_2dof}. It can be observed that the surge displacement and the horizontal flow velocity are in phase with each other. Also, the difference in the magnitude of the surge velocity and the $u$ component of the flow velocity is small. These two factors mitigate flow separation in the horizontal direction.  In contrast, the difference in the magnitude of heave velocity of the buoy and $v$ component of the flow field is relatively large, as shown in Fig.~\ref{fig_vorticity_heave_2dof}. It can also be observed that the heave displacement and heave velocity of the buoy are out of phase ($\pi/2$ and $\pi$ radians, respectively) with the $v$ component of the flow field. Therefore, for the 2-DOF buoy, vortex shedding primarily happens due to the heave motion. Moreover, a vortex pair is shed at the end of a heave stroke (at the extremum of the heave curve) when the flow and heave velocities begin to increase from their zero values, but in opposite directions; see Fig.~\ref{fig_vorticity_heave_2dof}.  For the 
2-DOF (and also 3-DOF) buoy, the surge velocity of the buoy is out-of-phase (by approximately $\pi/2$ radians) with the horizontal flow velocity.  At the time of vortex shedding (which will be at some but not all peaks of the heave curve in this case), the surge velocity of the buoy and the horizontal component of the flow velocity are of similar magnitude but have opposite signs (Fig.~\ref{fig_vorticity_surge_2dof}). The difference in the magnitude of surge and $u$ velocities becomes even smaller if $u$ velocity is considered at the lowest $y$ location attained by the buoy. Hence, the main component of the transport velocity is 
in the vertical direction when a vortex pair is released for the 2-DOF buoy.} 

For these simulations, the PTO coefficients are taken to be $k_{\text{PTO}} = 1995.2$ N/m and 
$b_{\text{PTO}} = 80.64$ N$\cdot$s/m. Increasing or decreasing these coefficients by a factor of five changed the phase difference 
between surge and horizontal velocities only slightly, suggesting that the 2- or 3-DOF buoy will shed vortices in the vertical direction irrespective of (physically reasonable) 
PTO coefficients.    

In our simulations we impose zero pressure boundary condition on the top boundary, which is taken rather close to the air-water interface to reduce the overall computational cost. Within this narrow space, the vortex structures in the air phase can reach the top boundary, especially when they are generated by waves of large amplitude. To reduce the boundary artifacts~\footnote{They do not affect the wave and buoy dynamics though.} observed in Fig.~\ref{fig_vorticity}, the top boundary could be modeled further away or a vorticity damping zone can be added near the boundary to dissipate the incoming vortex structures. \REVIEW{At the bottom of the NWT, we use no-slip velocity boundary conditions. This produces a thin boundary layer region at the channel bottom, which can be observed in Fig.~\ref{fig_vorticity}.}

%%%%%%%%%%%%%%%%%%%%%%%%%%%%%

\section{Results and discussion} \label{sec_results}

In this section we use the WSI framework to study the conversion efficiency of a 3-DOF submerged cylindrical buoy. 
More specifically, we analyze the performance of the buoy by  varying

\begin{itemize}

\item PTO stiffness and damping coefficients.

\item Density of the submerged buoy.

\item Wave height.

\end{itemize}

The wave absorption or the conversion efficiency of a wave energy converter is defined as the ratio of the mean absorbed power $\overline{P}_{\text{absorbed}}$ 
to the mean wave power $\overline{P}_{\text{wave}}$. For a regular wave, the wave absorption efficiency can be written as
\begin{align}
\eta & = \frac{ \overline{P}_{\text{absorbed}} }{ \overline{P}_{\text{wave}} } =  
\frac{ \frac{1}{\mathcal{T}} \left[\int_{t}^{t+\mathcal{T}}  P_{\text{absorbed}}(t)\; \text{d}t \right]}{\frac{1}{8}\rho_{\text{w}} g \mathcal{H}^2 c_{\text{g}}}. 
\end{align}
Here, $c_{\text{g}}$ is the wave group velocity and  $P_{\text{absorbed}}(t)$ is the instantaneous power absorbed by the 
PTO damper
\begin{align}
c_{\text{g}} &= \frac{1}{2} \frac{\lambda}{\mathcal{T}} \left[ 1 + \frac{2 k d}{\sinh(2 k d)} \right],  \\
P_{\text{absorbed}}(t) & = b_{\text{PTO}} \left(\d{\Delta l}{t} \right)^2. 
\end{align}

\subsection{PTO coefficients} \label{sec_pto_coefs}

\begin{figure}[]
  \centering
  \subfigure[Added mass coefficient $A_{33}(\omega)$]{
    \includegraphics[scale = 0.32]{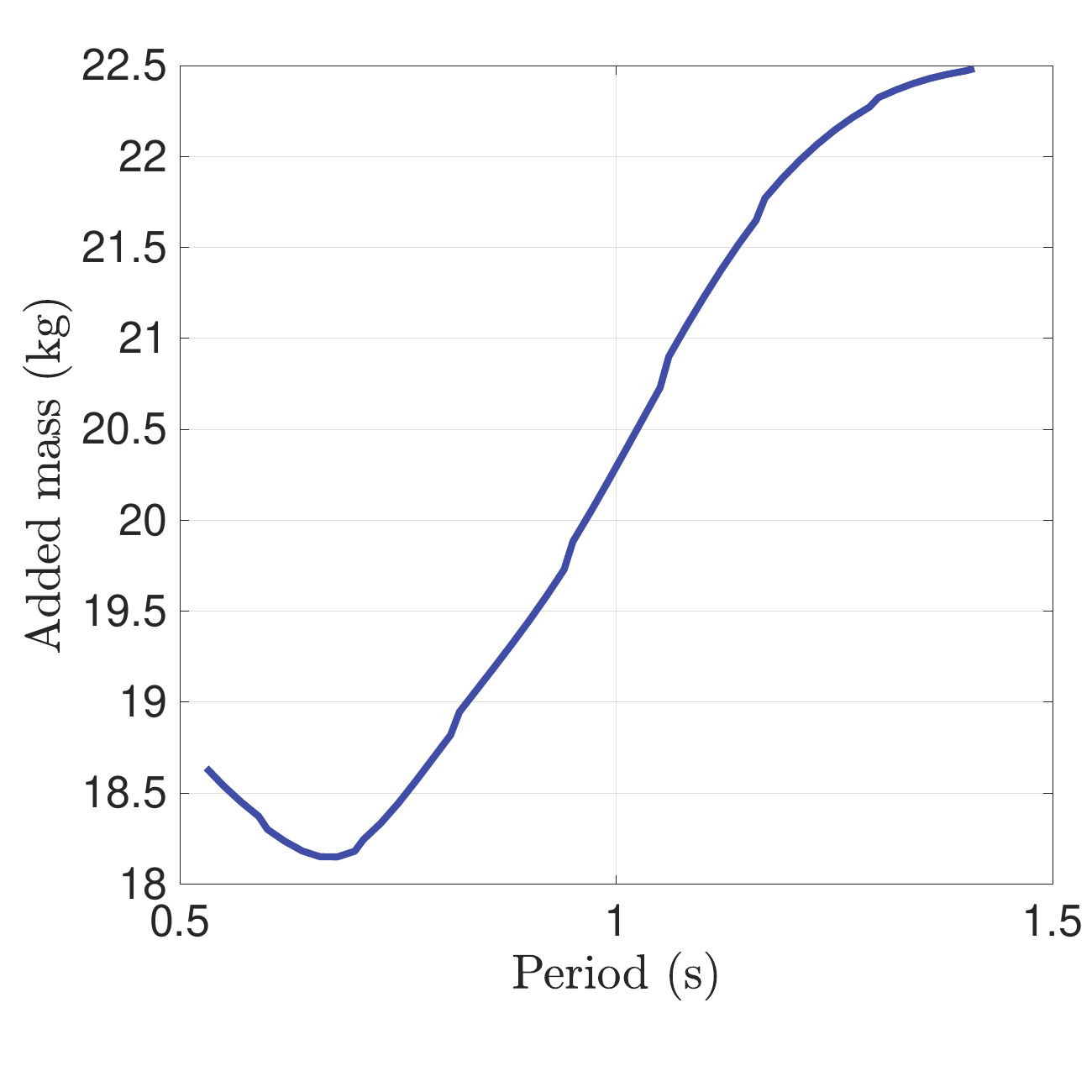} 
    \label{fig_a33}
  }
   \subfigure[Radiation damping coefficient $B_{33}(\omega)$ ]{
    \includegraphics[scale = 0.32]{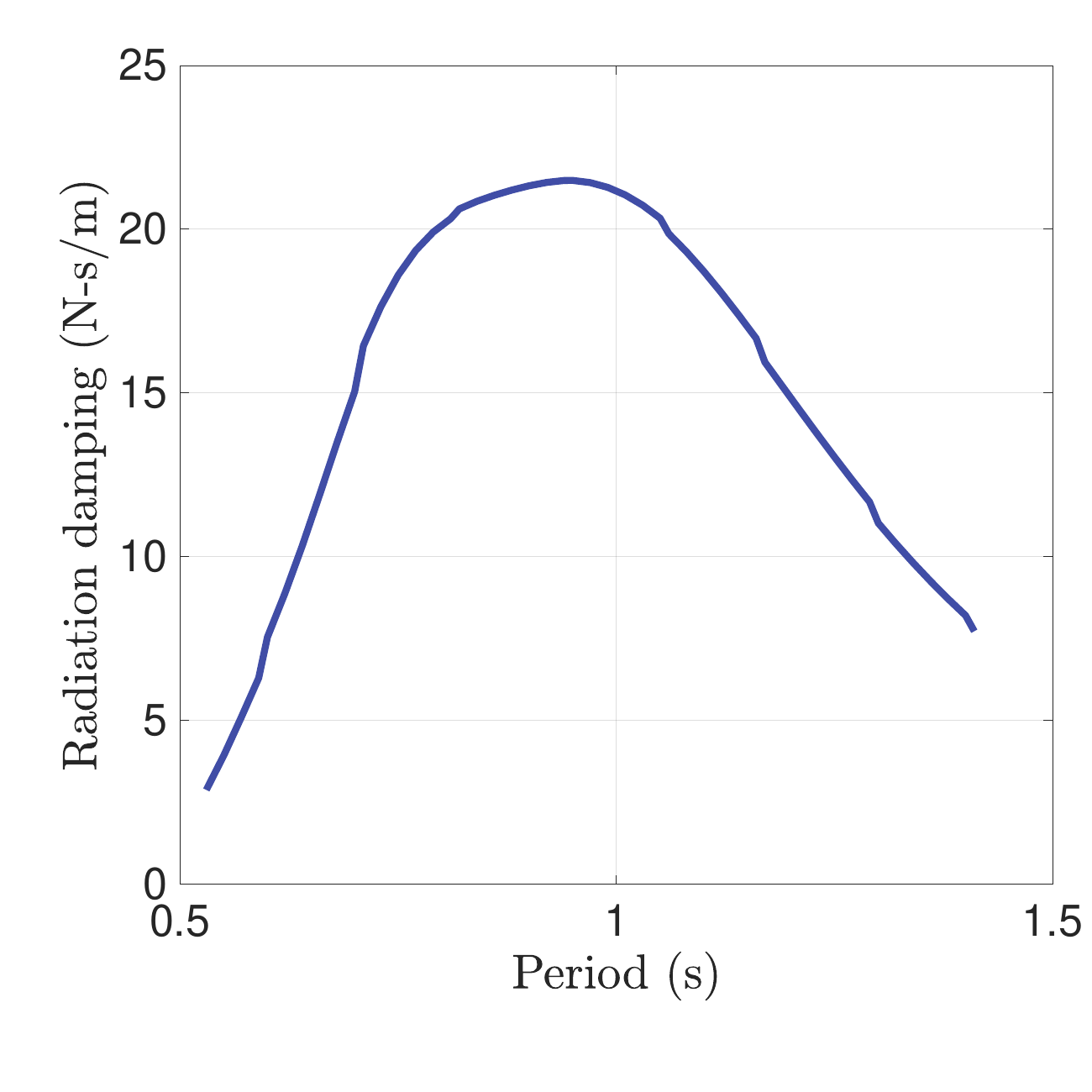}
    \label{fig_b33}
  }
  \caption{\subref{fig_a33} Normalized added mass and~\subref{fig_b33} radiation damping coefficients in the heave direction. 
  }
  \label{fig_a33b33}
\end{figure}

Wave energy absorption efficiency of a converter is maximized if it resonates with the incoming waves~\cite{Falnes2002}. 
In this regard, PTO stiffness and damping coefficients are two tuning parameters that can be effectively 
controlled to synchronize the natural frequency of the mechanical oscillator with the incoming wave frequency.  Accordingly, 
the linear wave theory suggests a reactive control-based strategy to select the PTO stiffness and damping coefficients as
\begin{align}
k_{\text{PTO}} = \omega^2\left(M+A_{33}(\omega)\right)  \quad \text{and}  \quad     b_{\text{PTO}}  =  B_{33}(\omega),  \label{}
\end{align}
in which $M$ is the mass of the converter, and  $A_{33}(\omega)$ and $B_{33}(\omega)$ are the frequency-dependent added mass and 
radiation damping coefficients in the heave direction, respectively. The length-normalized hydrodynamic coefficients of a submerged 
cylinder of length $8D$ for a wave period range of $\mathcal{T} \in [0.625  -1.1]$ s are obtained using AQWA.
Fig.~\ref{fig_a33b33} shows the variation of frequency-dependent hydrodynamic coefficients in the selected $\mathcal{T}$ range. 

\begin{figure}[]
  \centering
    \includegraphics[scale = 0.32]{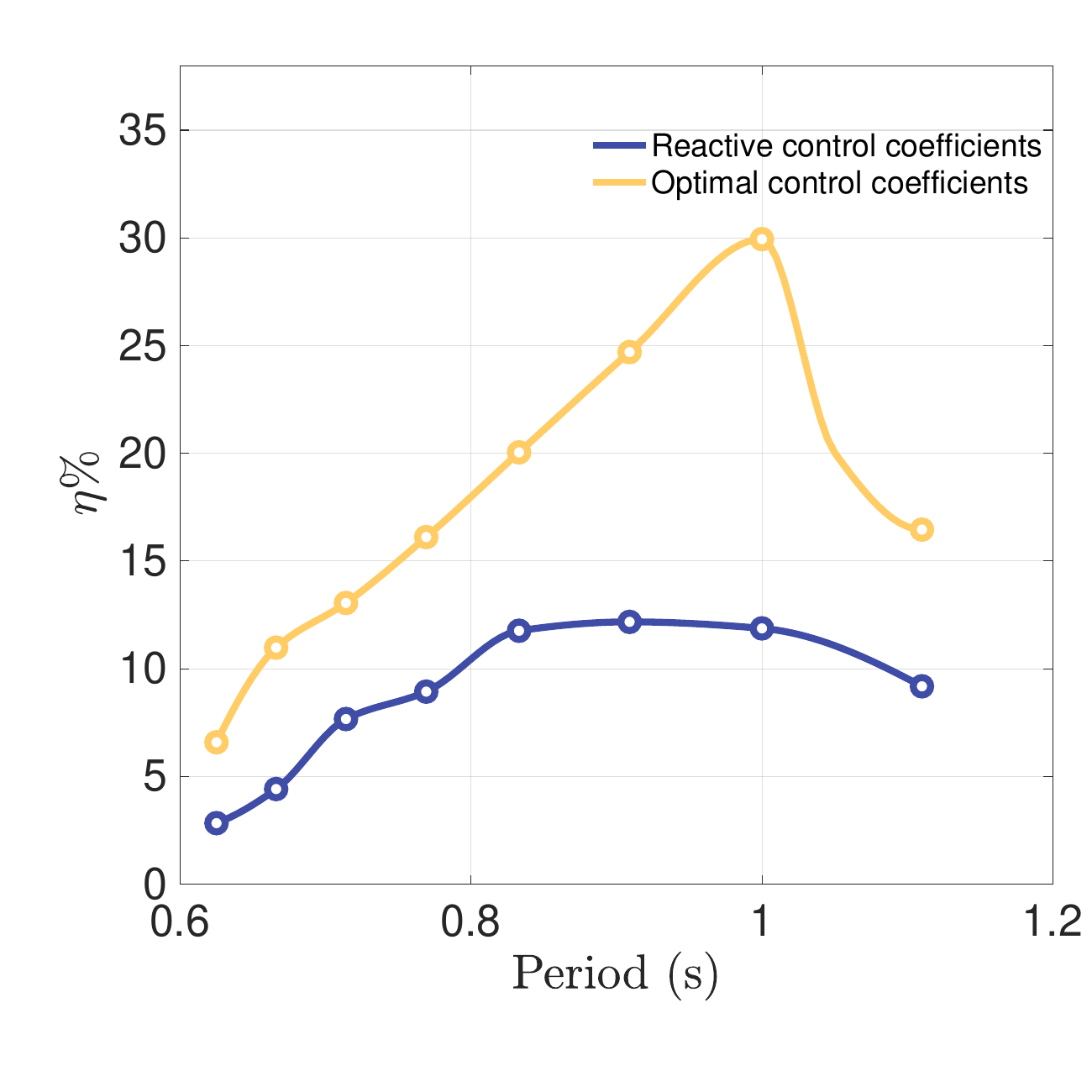}
  \caption{Absorption efficiency of the submerged buoy using reactive control and optimal control PTO coefficients. Fully-resolved WSI simulations are
  performed using $\mathcal{H} = 0.02$ m, $d_{\text{s}} = 0.25$ m, $d = 0.65$ m,  $D = 0.16$ m and $\rho_{\text{s}}/\rho_{\text{w}} = 0.9$.  
}
  \label{fig_reactive_optimal}
\end{figure}

To study the effect of PTO stiffness and damping on the converter efficiency, 
we perform fully-resolved WSI simulations of the submerged buoy for two sets of PTO coefficients: 
\begin{itemize}
\item  Reactive control coefficients: $k_{\text{PTO}} = \omega^2\left(M+A_{33}(\omega)\right)$  and  $b_{\text{PTO}}  =  B_{33}(\omega)$. 
\item  Optimal control coefficients: $k_{\text{PTO}} = \omega^2\left(M+A_{33}(\omega)\right)$  and $b_{\text{PTO}}  =  80.64$  N$\cdot$ m/s.   
\end{itemize}
Fig.~\ref{fig_reactive_optimal} shows the absorption efficiency of the buoy at various wave frequencies using reactive control and 
optimal control PTO coefficients. The curve obtained using reactive control indicates that the converter efficiency saturates 
around $\mathcal{T} = 0.9$ s. Therefore, a higher damping coefficient could be used to narrow down the optimal performance 
period range and to possibly enhance the efficiency of the converter. This is achieved by using the optimal control damping coefficient value of $b_{\text{PTO}}  =  80.64$  N$\cdot$ m/s, which is approximately four times larger than the maximum value of $B_{33}(\omega)$ predicted by 
the linear wave theory for a wave period around $\mathcal{T} = 0.9$ s (see Fig.~\ref{fig_b33}). This optimal value of $b_{\text{PTO}}$ 
increases the absorption efficiency of the buoy for all wave frequencies, which suggests that the reactive control 
damping coefficients predicted by the linear wave theory do not lead to an optimal performance. Further increasing $b_{\text{PTO}}$ value did not enhance the performance of the converter significantly (data not shown). Similar observations were made in Anbarsooz et al.~\cite{Anbarsooz2014}. 
Notice that extremely large values of $b_{\text{PTO}}$ would lead to over-damping of the system, and therefore should also be avoided.
Table~\ref{table_waves} shows the characteristics of the simulated waves including their steepness values. The reactive control curve \REVIEW{of Fig.~\ref{fig_reactive_optimal}} shows a
sub-optimal performance of the buoy for lower wave periods at which the wave steepness is high, i.e.,  the ratio $\mathcal{H}/\lambda > 0.01$. 
Therefore, linear wave theory does not predict optimal PTO coefficients for steeper waves \REVIEW{as it assumes low values of amplitude to wavelength ratio. The sub-optimal values of PTO coefficients predicted by linear wave theory is also reported in~\cite{Anbarsooz2014}.}

\begin{table}[h!]
 \caption{Wave characteristics.}
    \centering
    \scalebox{1.1}{
    \begin{tabular}{|c|c|c|}
     \hline
     Wave period (s) & Wave length (m) & Wave steepness \\
     \hline
     0.625 & 0.6099 &  0.0328 \\
      \hline
      0.666 &  0.6925 &  0.0288 \\
      \hline
      0.714 &  0.7959 &  0.0251 \\ 
      \hline
      0.7692 &  0.9235 & 0.0216 \\
      \hline
      0.833 &  1.0822 & 0.0184 \\
      \hline
      0.909 &  1.2856 & 0.0155 \\
      \hline
          1& 1.5456 & 0.0129 \\
      \hline
          1.11 & 1.875 & 0.0107 \\
      \hline
      \end{tabular} }
      \label{table_waves}
\end{table}

\subsection{Buoy density}

\begin{figure}[]
  \centering
  \subfigure[Efficiency]{
    \includegraphics[scale = 0.32]{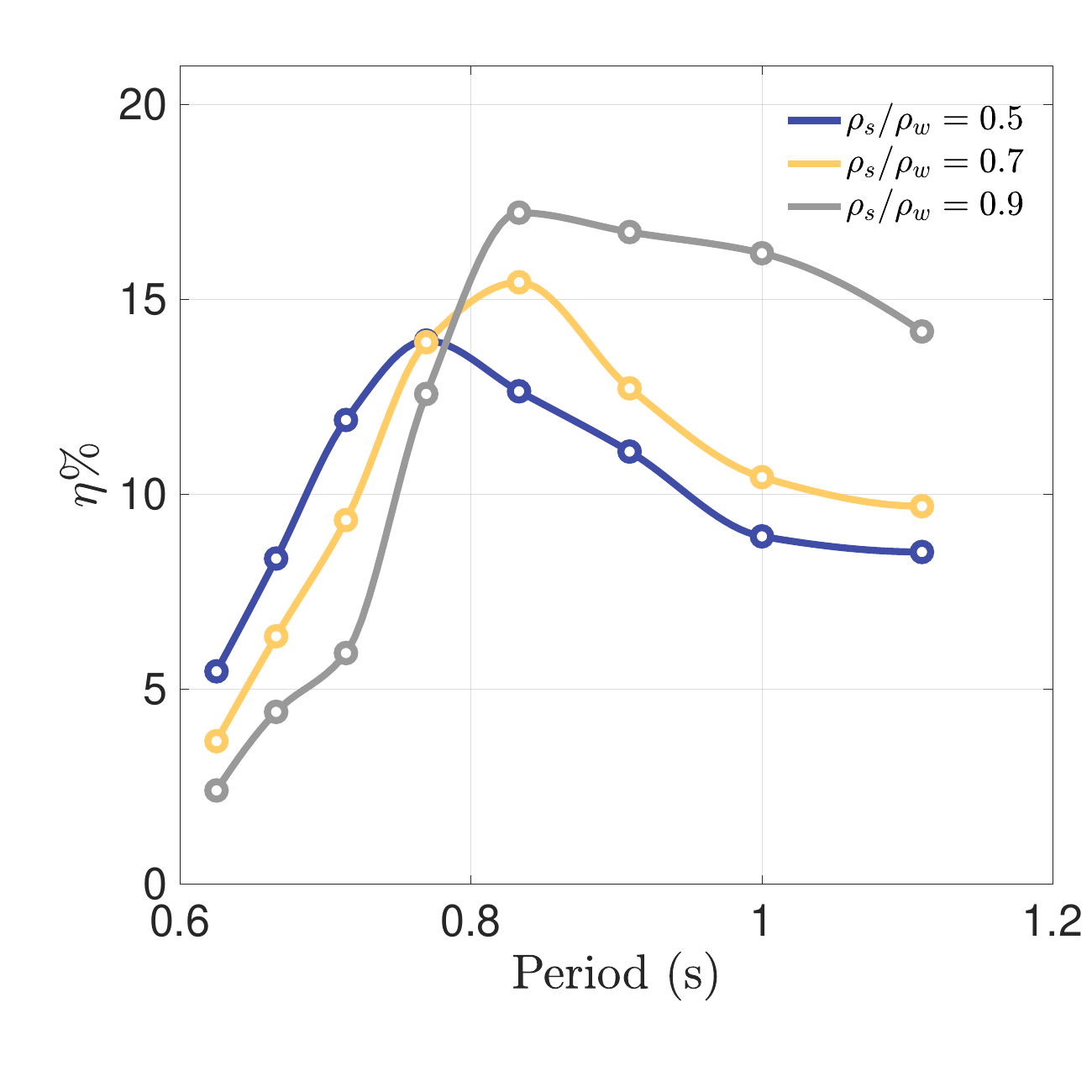} 
    \label{fig_densities}
  }
   \subfigure[Reactive PTO stiffness]{
    \includegraphics[scale = 0.32]{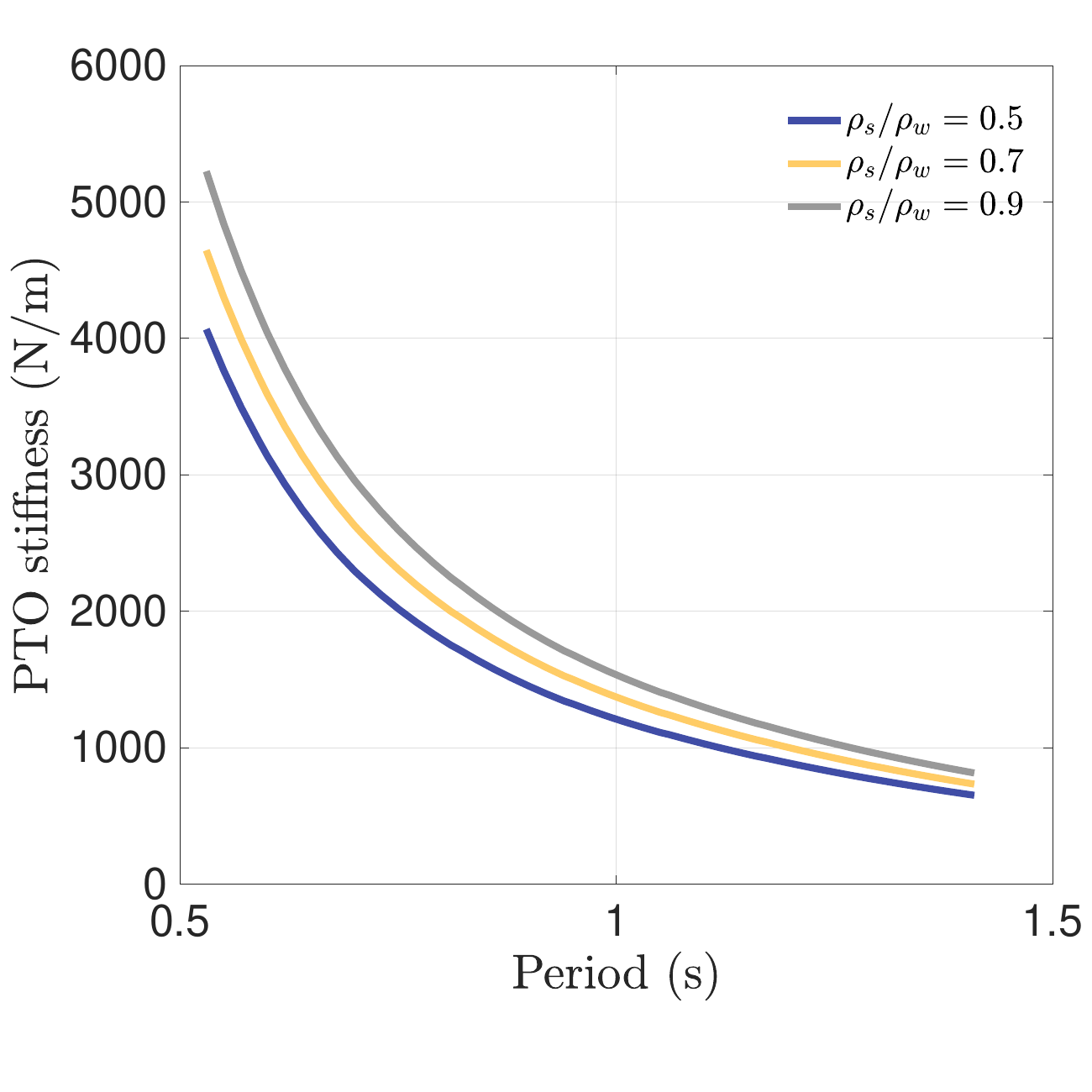}
    \label{fig_density_stiffness}
  }
  \caption{\subref{fig_densities} Absorption efficiency of the submerged buoy with different mass densities. The PTO coefficients are set to 
  $k_{\text{PTO}} = 1995.2$ N/m and  $b_{\text{PTO}}  =  80.64$  N$\cdot$ m/s. Fully-resolved WSI simulations are
  performed using $\mathcal{H} = 0.02$ m, $d_{\text{s}} = 0.25$ m, $d = 0.65$ m,  and $D = 0.16$ m.~\subref{fig_density_stiffness} 
  Reactive PTO stiffness for various densities.   
  }
\end{figure}

Next, we consider the effect of buoy density on its conversion performance. We fix the PTO coefficients to 
$k_{\text{PTO}} = 1995.2$ N/m and  $b_{\text{PTO}}  =  80.64$  N$\cdot$ m/s but vary the mass density for this study. Three relative densities 
of the buoy are considered: $\rho_{\text{s}}/\rho_{\text{w}} = $ 0.5, 0.7, and 0.9.  Fig.~\ref{fig_densities} shows that each density curve results in an optimal performance of the converter in a certain range of wave periods. This optimal range can also be predicted from the reactive control theory
 
\begin{align}
\mathcal{T}_{\text{optimal}} = 2 \pi  \sqrt{\left(M+A_{33}(\omega)\right)/k_{\text{PTO}}} \;.  \label{eqn_natural_freq}
\end{align}
Table~\ref{table_density} shows the optimal performing wave period range calculated using Eq.~\eqref{eqn_natural_freq} for a fixed value of  
$k_{\text{PTO}} = 1995.2$ N/m and by using lowest and highest values of $A_{33}(\omega)$ from Fig.~\ref{fig_a33}. The simulated results and 
the analytically predicted range suggest that the optimal period range increases with increasing mass density of the buoy. Notice that 
since $k_{\text{PTO}} = 1995.2$ N/m is an optimal reactive PTO stiffness for a relative density 0.9, the efficiency curve is higher for 
$\rho_{\text{s}}/\rho_{\text{w}} = 0.9$ compared to other densities. Fig.~\ref{fig_density_stiffness} plots the reactive PTO stiffness for different mass 
densities of the buoy \REVIEW{using Eq.~\eqref{eqn_natural_freq}: $k_{\rm PTO} = 4 \pi^2 \left( M + A_{33}(\omega) \right)/\mathcal{T}^2$.} Lower PTO stiffness coefficients are obtained for lower mass densities \REVIEW{from this relationship}. Therefore, lowering  $k_{\text{PTO}}$ value 
from 1995.2 N/m \REVIEW{(or in other words using a more optimal $k_{\text{PTO}}$ value)}, is expected to increase the conversion efficiency for lighter buoys.

\begin{table}[]
   \caption{Characteristics of the submerged buoy with different relative densities.}
    \centering
    \scalebox{1.1}{
    \begin{tabular}{|c|c|c|c|}
     \hline
    Relative density & Mass (kg) & Optimal wave period range (s) & Permanent tension (N) \\
     \hline
     0.5 & 10.30& [0.74-0.8] &  101.08 \\
      \hline
      0.7 & 14.42 & [0.8-0.85] &  60.65 \\
      \hline
      0.9 & 18.54 & [0.85-0.9] &  20.21 \\ 
      \hline
      \end{tabular} }
      \label{table_density}

\end{table}

We remark that the relative density of the buoy directly influences the permanent tension in its mooring line. For lower mass density values, there is an increased upward hydrostatic force, which is written in the last column of Table~\ref{table_density}. The submerged buoy having $50\%$ water density has five times greater permanent load than a buoy having $90\%$ water density. This factor should be considered while selecting the wave energy converter density. 

\subsection{Wave height}

\begin{figure}[] 
  \centering
    \includegraphics[scale = 0.32]{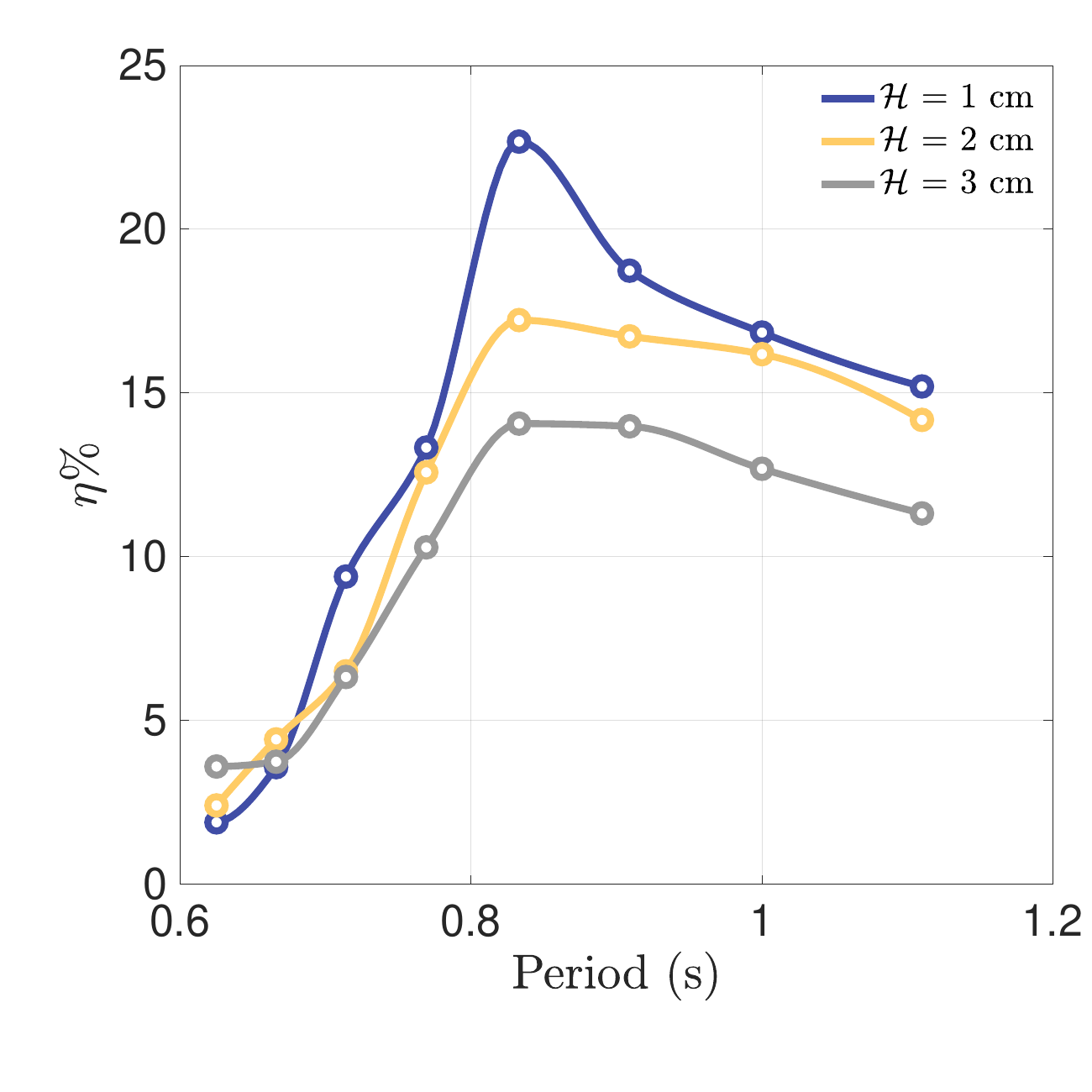}
  \caption{Absorption efficiency of the submerged point absorber at three different wave heights. The PTO coefficients are set to 
  $k_{\text{PTO}} = 1995.2$ N/m and  $b_{\text{PTO}}  =  80.64$  N$\cdot$ m/s. Fully-resolved WSI simulations are
  performed using $\rho_{\text{s}}/\rho_{\text{w}} = 0.9$, $d_{\text{s}} = 0.25$ m, $d = 0.65$ m, and $D = 0.16$ m.
}
\label{fig_heights}
\end{figure}

As a last parametric study of converter performance, we simulate the dynamic response of the submerged point absorber in regular waves 
of different heights. The buoy is subject to incident waves of height $\mathcal{H} = $ 0.01, 0.02 and 0.03 m having time periods in the range 
$\mathcal{T} \in [0.625 - 1.1]$ s. The absorption efficiency of the buoy for different wave heights is shown in Fig.~\ref{fig_heights}. 
As observed in the figure, the absorption efficiency decreases with increased wave heights. Waves with greater heights are more 
energetic and they result in an increased dynamic response of the submerged object. However,  the point absorber fails to absorb a significant 
portion of the available wave energy. One possible way to achieve an optimal performance for more energetic waves is to increase the size of \REVIEW{the power take-off unit and the WEC device~\cite{Falnes2002,Yu2018,Falnes2012}}.  However, bigger wave energy converters are more costly and therefore, a balance between cost and efficiency should be considered 
during the design stage of WECs~\cite{Falnes2012}.  

%%%%%%%%%%%%%%%%%%%%%%%%%%%%%%%%%
\section{Conclusions} \label{sec_conclusion}
In this study, we compared the dynamics of a 3-DOF cylindrical buoy  using potential flow and CFD models. The 
potential flow model is based on the time-domain Cummins equation, whereas the CFD model employs the 
\REVIEW{FD/BP method} --- a fully-Eulerian 
technique for modeling \REVIEW{FSI} problems. An advantage of FD/BP method over the FD/DLM method is its ability to 
incorporate external forces and torques in the equations of motion, and it enabled us to solve 
the coupled translational and rotational degrees of freedom of the buoy.

The comparison of the dynamics show that the Cummins model over-predicts the amplitude of heave and surge motions, whereas it results in an insignificant pitch of the buoy. Moreover, it does not capture the slow drift phenomena in the surge dynamics.  The CFD model was then used to study the wave absorption efficiency of the converter under varying PTO coefficients, mass density of the buoy, and incoming wave heights. It is demonstrated that the PTO coefficients predicted by the 
linear potential theory are sub-optimal for waves of moderate and high steepness. The wave absorption efficiency improves significantly when 
higher values of PTO damping coefficients are used. Simulations with different mass densities of the buoy show that converters 
with low mass density have an increased permanent load in their PTO and mooring lines. Moreover, the mass density also influences the 
range of resonance periods of the device. Finally, simulations with different wave heights show that at higher heights, the wave absorption 
efficiency of the converter decreases and a large portion of available wave power remains unabsorbed.    

%%%%%%%%%%%%%%%%%%%%%%%%%%%%%%

\section*{Acknowledgements}
A.P.S.B.~acknowledges support from NSF award OAC 1931368.  NSF XSEDE and SDSU Fermi compute resources are particularly acknowledged.

%%%%%%%%%%%%%%%%%%%%%%%%%%%%%%%%%

\section*{Data availability statement}
The data that support the findings of this study are openly available in IBAMR Github repository at \url{https://github.com/IBAMR/IBAMR}.

%%%%%%%%%%%%%%%%%%%%%%%%%%%%%%%%%
\bibliography{PointAbsorber}

\end{document}